\documentclass[%
superscriptaddress,
preprint,
preprintnumbers,
amsmath,amssymb,
aps,
prd,
floatfix,
notitlepage,
showpacs,
]{revtex4-2}
\usepackage{graphicx}
\usepackage{hyperref}
\usepackage{xcolor}
\usepackage{booktabs}
\usepackage{multirow}
\usepackage{siunitx}
\usepackage{parskip}
\newcommand{\Tr}{{\rm Tr}}
\newcommand{\tr}{{\rm tr}}

\newcommand{\Id}{\mbox{1\hspace{-1.2mm}I}}

\newcommand{\bea}{\begin{eqnarray}}
\newcommand{\eea}{\end{eqnarray}}
\newcommand{\BAN}{\begin{eqnarray*}}
\newcommand{\EAN}{\end{eqnarray*}}
\newcommand{\nn}{\nonumber\\}
\newcommand{\blue}{\color{blue}}

\begin{document}

\newcommand{\ASIOP}
{Institute of Physics, Academia Sinica, Taipei, Taiwan~11529, Republic of China}

\newcommand{\CTP}
{Center for Theoretical Physics, Department of Physics, \\ 
National Taiwan University, Taipei, Taiwan~10617, Republic of China}

\newcommand{\NTNU}
{Department of Physics, National Taiwan Normal University, \\ 
Taipei, Taiwan~11677, Republic of China}

\newcommand{\NCTS}
{Physics Division, National Center for Theoretical Sciences,  \\
Taipei, Taiwan~10617, Republic of China}

\preprint{NTUTH-23-505A}

\title{Symmetries of meson correlators in high-temperature QCD 
       with physical $(u/d, s, c)$ domain-wall quarks}

\author{Ting-Wai~Chiu}
\email{twchiu@phys.ntu.edu.tw}
\affiliation{\NTNU}
\affiliation{\ASIOP}
\affiliation{\NCTS}
\affiliation{\CTP}

\begin{abstract}

The correlation functions of meson interpolators 
in $N_f=2+1+1$ lattice QCD with optimal domain-wall quarks at the physical point 
are studied for six temperatures in the range $ T \sim $ 190-770 MeV. 
The meson interpolators include a complete set of Dirac bilinears,   
and each for six combinations of quark flavors. 
In this paper, we focus on the meson correlators of $u$ and $d$ quarks, 
and we discuss their implications for the effective restoration 
of $U(1)_A$ and $SU(2)_L \times SU(2)_R$ chiral symmetries,  
as well as the emergence of approximate $SU(2)_{CS}$ chiral spin symmetry. 

\end{abstract}

\maketitle

\section{Introduction}
\label{intro}

It is important to understand the nature of strongly interacting matter at high-temperatures, 
which is crucial for the mechanism of matter creation in the early Universe, as 
well as in relativisitic heavy ion collision experiments such as those at RHIC and LHC. 
A first step toward this goal is to find out the symmetries of QCD at high-temperatures,  
since the nature of matter is likely to be unveiled from its symmetries. 

At low temperatures $T < T_c $, quarks and gluons are confined in hadrons, 
and the chiral symmetry of QCD is spontaneously broken, 
with the nonzero chiral condensate ($\Sigma(T) \ne 0 $),   
\bea
\label{eq:Sigma_T}
\Sigma(T)=- \lim_{m_q \to 0} \lim_{V \to \infty} \frac{T}{V} 
            \int_0^{1/T} dt \int_V d^3 x \left< \Tr (D_c + m_q)^{-1} \right>.  
\eea 
Moreover, the $U(1)_A$ symmetry is explicitly broken by the chiral anomaly 
due to the quantum fluctuations of topologically nontrivial gauge fields. 

Since the quark mass explicitly breaks the $U(1)_A$ symmetry and the chiral symmetry, 
determining whether the $U(1)_A$ symmetry and the chiral symmetry 
are broken/restored at any $T$ should be performed in the massless limit.
Nevertheless, for QCD with physical $(u,d,s,c,b)$ quarks  
with nonzero quark masses, as the temperature $T$ is increased,   
the $SU(n_f)_L \times SU(n_f)_R $ chiral symmetry
is effectively restored successively from $n_f = 2 $ to 3, 4, and 5, say, 
as $ T \nearrow T_c^{u/d} \nearrow T_c^{s} \nearrow T_c^c \nearrow T_c^b $.  
Since the $SU(2)_L \times SU(2)_R$ chiral symmetry of physical $u$ and $d$ quarks 
is effectively restored at $ T \ge T_c^{u/d} $,  
its counterpart $T_c^0 $ in QCD with massless $(u,d,s,c,b)$ quarks is supposed to be 
at a lower temperature, i.e.,  $ T_c^{0} < T_c^{u/d} $.
Now, assuming the $U(1)_A$ symmetry in QCD with massless $(u,d,s,c,b)$ quarks is also effectively 
restored at $T_c^0$, it is unclear whether the $U(1)_A$ symmetry of $u$ and $d$ quarks in QCD with 
physical $(u,d,s,c,b)$ quarks is also effectively restored at $ T \ge T_c^{u/d} $, 
or at higher temperatures $ T \ge T_1^{u/d} \gtrsim T_c^{u/d} $.   
 
Since 1987\cite{Detar:1987kae}, 
there have been many lattice studies using spatial meson correlators (and their screening masses) 
to investigate the effective restoration of $U(1)_A$ and $ SU(2)_L \times SU(2)_R$ chiral symmetries
in high-temperature QCD, see, e.g., Ref.\cite{Bazavov:2019www} and references therein.  
In this paper, we will use the degeneracies of meson correlators of $u$ and $d$ quarks 
to determine the effective restoration or the emergence of any exact/approximate symmetries 
in high-temperature QCD, as discussed in Sec. \ref{symmetries}. 
For example, we use the degeneracy of meson correlators of vectors ($V_k \equiv \bar u \gamma_k d$) 
and axial-vectors ($A_k \equiv \bar u \gamma_5 \gamma_k d $)  
to determinate the effective restoration of the $SU(2)_L \times SU(2)_R $ chiral symmetry 
of $u$ and $d$ quarks, and the degeneracy of the meson correlators of the scalar 
($S \equiv \bar u d$) and the pseudoscalar ($P \equiv \bar u \gamma_5 d $) 
to determine the effective restoration of the $U(1)_A$ symmetry of $u$ and $d$ quarks. 

Now, the question is whether $U(1)_A$ and $SU(2)_L \times SU(2)_R$ are the only symmetries of 
QCD with physical $(u,d,s,c,b)$ quarks for $T \ge T_1^{u/d} \gtrsim T_c^{u/d} $, 
all the way up to the temperatures where the effective coupling among quarks and gluons 
becomes sufficiently weak (screened), and the quarks and gluons behave like deconfined particles 
forming the quark-gluon plasma.
In particular, it is interesting to find out whether there are any emergent symmetries 
which are manifested in observables (e.g., hadron correlators) but not in the QCD action.
Moreover, one may ask whether quarks are deconfined or confined inside these hadron-like objects 
for temperatures $ T \gtrsim T_c^{u/d}$.
In the latter case, the properties of these hadron-like objects would be quite different 
from those at $T < T_c^{u/d} $, since the chiral symmetry has been restored with $ \Sigma = 0 $. 

Recently, it has been observed that in $N_f=2$ lattice QCD with domain-wall fermions, 
at temperatures $T \sim$ 220-500 MeV $\sim$ (1.2-2.8)$T_c$ 
(where $T_c \sim 175$~MeV for $N_f=2$ lattice QCD), 
a larger symmetry group $SU(2)_{CS}$ [with $U(1)_A$ as a subgroup] 
\cite{Glozman:2014mka,Glozman:2015qva}
is approximately manifested in the multiplets of 
correlators of the $J=1$ meson interpolators \cite{Rohrhofer:2019qwq,Rohrhofer:2019qal}, 
as an approximate emergent symmetry in high-temperature QCD.  
This suggests the possible existence of hadron-like objects 
which are predominantly bound by the chromoelectric interactions into color singlets 
for a range of temperatures above $T_c$.
Now, the question is one of identifying the scenario of the emergence of approximate $SU(2)_{CS}$ 
chiral spin symmetry in QCD with dynamical light and heavy quarks. This motivates the present study.

In this paper, we study the temporal and spatial correlation functions of meson interpolators 
in $N_f=2+1+1$ lattice QCD with $(u, d, s, c)$ optimal domain-wall quarks at the physical point -   
on the $32^3 \times (16,12,10,8,6,4) $ lattices for temperatures 
in the range $ T \sim$ 190-770 MeV. 
The meson interpolators include a complete set of Dirac bilinears  
(scalar, pseudoscalar, vector, axial vector, tensor vector, and axial-tensor vector),     
and each for six combinations of quark flavors  
($\bar u d$, $\bar u s$, $\bar u c$, $\bar s c$, $\bar s s$, and $\bar c c $). 
We discuss the implications of these results for the effective restoration 
of the $SU(2)_L \times SU(2)_R $ and $U(1)_A$ chiral symmetries,  
as well as the emergence of approximate $SU(2)_{CS}$ chiral spin symmetry.  
In this paper, we focus on the meson correlators of $u$ and $d$ quarks.
The results of meson correlators with other flavor combinations  
($\bar u s$, $\bar u c$, $\bar s c$, $\bar s s$, and $\bar c c $) 
will be analyzed in a forthcoming paper \cite{chiu:2022ab}.

The outline of this paper is as follows: In Sec. \ref{symmetries}, 
we discuss the relationship between various symmetries 
[$SU(2)_L \times SU(2)_R$, $U(1)_A$, $SU(2)_{CS}$ and $SU(4)$]   
and the degeneracies of meson correlators.
In Sec. \ref{kappa}, the symmetry-breaking parameters 
for measuring various symmetries with the degeneracies of meson correlators are defined. 
In Sec. \ref{lattice}, the features of the gauge ensembles of $N_f=2+1+1$ lattice QCD  
at the physical point for this study are outlined.
The results of the temporal $t$ correlators for three temperatures in the range 
$T \simeq$ 190-310 MeV are presented in Sec. \ref{Ct_ud}, 
while those of the spatial $z$ correlators 
for six temperatures in the range $T \simeq 190-770$~MeV are presented in Sec. \ref{Cz_ud}.  
We discuss their implications    
for the effective restoration of $SU(2)_L \times SU(2)_R $ and $U(1)_A$ chiral symmetries,   
and the emergence of the approximate $SU(2)_{CS}$ chiral spin symmetry. 
We also compare our results with those 
in $N_f = 2$ lattice QCD \cite{Rohrhofer:2019qwq,Rohrhofer:2019qal}, 
as well as the noninteracting theory with free quarks.
In Sec. \ref{conclusion}, we conclude with some remarks.

\section{Symmetries and Meson Correlators} 
\label{symmetries}

In this section, we discuss the relationship between the symmetry 
and the degeneracy of the meson correlators in high-temperature QCD. 

The correlation function of meson interpolator $ \bar q_1 \Gamma q_2 $ 
is measured according to the formula
\bea
\label{eq:C_Gamma}
C_\Gamma(t,\vec{x}) = \left< (\bar q_1 \Gamma q_2)_x (\bar q_1 \Gamma q_2)_0^{\dagger}  \right >
= \left< \tr\left[ \Gamma (D_c + m_1)^{-1}_{0,x} \Gamma (D_c + m_2)^{-1}_{x,0} \right]  
  \right >_{\text{confs}}, 
\eea 
where $(D_c + m_q)^{-1} $ denotes the valence quark propagator with quark mass $ m_q $ in lattice QCD
with exact chiral symmetry, tr denotes the trace over the color and Dirac indices, and 
the brackets $\left< \cdots \right>_{\text{confs}} $ denote averaging over the gauge configurations.   
Here the label of a lattice site $x$ is understood to stand for 
$ (x_1, x_2, x_3, x_4) = (x, y, z, t)$, and the overall $\pm$ sign due to 
$ \gamma_4 \Gamma^\dagger \gamma_4 = \pm \Gamma $ has been suppressed. 

On a lattice of $N_x^3 \times N_t$ sites, the discrete Fourier transform of (\ref{eq:C_Gamma}) gives
\bea
\label{C_Gamma_p}
\widetilde{C}_\Gamma(t, \vec{p}, T) = \sum_{x_1, x_2, x_3} \exp(i \vec{p} \cdot \vec{x}) \ C_\Gamma(t, \vec{x}), 
\hspace{4mm} T = \frac{1}{N_t a}, 
\eea 
which is related to the spectral function $\rho_\Gamma(\omega, \vec{p}, T)$ 
through the integral transform, 
\bea
\label{eq:spectral_t}
\widetilde{C}_\Gamma(t,\vec{p},T) = \int_0^\infty \frac{d\omega}{2 \pi} \ 
\frac{\cosh\left[\omega\left(t-\frac{1}{2T}\right)\right]}{\sinh \left(\frac{\omega}{2T}\right)} \ 
\rho_\Gamma(\omega, \vec{p}, T). 
\eea 

The time-correlation function ($t$ correlator) of the meson interpolator $ \bar q_1 \Gamma q_2 $ is 
defined as  
\bea
\label{eq:C_Gamma_t}
C_\Gamma(t,T) = \sum_{x_1, x_2, x_3} C_{\Gamma}(t,\vec{x}),  
\eea 
which is equal to $ \widetilde{C}_\Gamma(t, \vec{p}=0, T) $, 
and is related to the spectral function at $ \vec{p} = 0 $.  

Alternatively, one can study the spatial correlation function in the $z$ direction ($z$ correlator) 
\bea
\label{eq:C_Gamma_z}
C_\Gamma(z,T) = \sum_{x_1, x_2, x_4} C_{\Gamma}(t,\vec{x}),  \hspace{4mm} T = \frac{1}{N_t a}, 
\eea 
which is related to the spectral function at $p_1 = p_2 = 0 $ 
through the integral transform
\bea
\label{eq:spectral_z}
C_\Gamma(z,T) = \int_0^\infty \frac{d\omega}{\pi \omega} 
                \int_{-\infty}^{+\infty} \frac{d p_3}{2 \pi} \exp( i p_3 z) \rho_\Gamma(\omega, p_3, T). 
\eea
If any symmetry manifests in the $z$ correlator, it should also  
appear in the spectral function $\rho(\omega, \vec{p}, T)$, 
since in thermal equilibrium, $\rho(\omega, \vec{p}, T) = \rho(\omega, |p|, T)$, isotropic 
in all directions of $\vec{p} $. 

In the following, it is understood that $C_\Gamma(t,T)$ is normalized by $ C_\Gamma(n_t=1,T)$, 
and similarly $C_\Gamma(z,T)$ is normalized by $ C_\Gamma(n_z=1,T)$. 

\subsection{Classification of meson interpolators}

The meson interpolators are classified 
according to their transformation properties as listed in Table \ref{tab:bilinear}. 
The $\Gamma$ matricies are given for the $t$ correlators in the second column,  
and the $z$ correlators in the third column. 
Note that $V_4$ and $A_4$ are omitted for the $t$ correlators,   
since $C_{V_4}(t)$ and $C_{A_4}(t)$ do not propagate in the $t$ direction
when the chiral symmetry of $u$ and $d$ quarks is effectively restored for $T > T_c$. 
Similarly, $V_3$ and $A_3$ are omitted for the $z$ correlators.

\begin{table}[ht]
\centering
\caption{The classification of meson interpolators $\bar q_1 \Gamma q_2 $, 
and their names and notations. The $\Gamma$ matricies in the second column are 
for the $t$ correlators, while those in the third column for the $z$ correlators.} 
\setlength{\tabcolsep}{4pt}
\vspace{2mm}
\begin{tabular}{c c c}
\hline
\hline
Name and notation & $\Gamma$ (for $t$ correlators) &  $\Gamma$ (for $z$ correlators)  \\
\hline
\multicolumn{1}{l}{Scalar ($S$)}                  
&  $\Id$           &  $\Id$                         \\
\multicolumn{1}{l}{Pseudocalar ($P$)}             
&  $\gamma_5$      &  $\gamma_5$                    \\
\multicolumn{1}{l}{Vector ($V_k$)}                
&  $\gamma_k \ (k=1,2,3)$  &  $\gamma_k \ (k=1,2,4)$                    \\
\multicolumn{1}{l}{Axial vector ($A_k$)}          
&  $\gamma_5 \gamma_k \ (k=1,2,3)$   &  $\gamma_5 \gamma_k \ (k=1,2,4)$           \\
\multicolumn{1}{l}{Tensor vector ($T_k$)}         
&  $\gamma_4 \gamma_k \ (k=1,2,3)$   &  $\gamma_3 \gamma_k \ (k=1,2,4)$           \\
\multicolumn{1}{l}{Axial-tensor vector ($X_k$)}   
&  $\gamma_5 \gamma_4 \gamma_k \ (k=1,2,3)$  &  $\gamma_5 \gamma_3 \gamma_k \ (k=1,2,4)$  \\
\hline
\hline
\end{tabular}
\label{tab:bilinear}
\end{table}

For the vector meson correlators, the rotational symmetry in the continuum
is reduced to the discrete permutation symmetry on the lattice.
For the $t$ correlators, the rotational symmetry becomes the $S_3$ symmetry 
of the $x$, $y$, and $z$ components,      
which gives $ C_{V_1} = C_{V_2} = C_{V_3} $, $ C_{A_1} = C_{A_2} = C_{A_3} $, 
$ C_{T_1} = C_{T_2} = C_{T_3} $, and $ C_{X_1} = C_{X_2} = C_{X_3} $. 
For the $z$ correlators, it becomes the $S_2$ symmetry of the $x$ and $y$ components,   
which gives $ C_{V_1} = C_{V_2} $, $ C_{A_1} = C_{A_2} $, 
$ C_{T_1} = C_{T_2} $, and $ C_{X_1} = C_{X_2} $. 

\subsection{$U(1)_A$ symmetry}

For the scalar ($S$) and the pseudoscalar ($P$) bilinears, their correlators
can be transformed into each other by the global $U(1)_A$ transformations 
\bea
q(x) \rightarrow \exp(i \gamma_5 \theta) q(x), \hspace{4mm}
\bar{q}(x) \rightarrow \bar{q}(x) \gamma_4 \exp(-i \gamma_5 \theta) \gamma_4.
\eea 
Similarly, for the tensor vector ($T_k$) and the axial-tensor vector ($X_k$), their 
correlators can be transformed into each other by the global $ U(1)_A $ transformations.
If $U(1)_A$ is effectively restored for $ T \gtrsim T_1^q$ (where $T_1^q$ depends on the 
masses of $q_1$ and $q_2$), the correlators of scalar ($S$) and pseudoscalar ($P$) are degenerate,  
and also those of tensor vectors ($T_k$) and axial-tensor vectors ($X_k$), i.e., 
\BAN
&& C_S(t) = C_P(t); \hspace{4mm} C_{T_k}(t) = C_{X_k}(t), \hspace{4mm} k=1,2,3,  \\
&& C_S(z) = C_P(z); \hspace{4mm} C_{T_k}(z) = C_{X_k}(z), \hspace{4mm} k=1,2,4.
\EAN 
Thus the effective restoration of the $U(1)_A$ symmetry is equivalent to the 
emergence of two multiplets
\bea
\label{eq:U1A}
(S,P); \ (\{ T_k \}, \{ X_k \}),  
\eea 
where $k=1,2,3 $ for $t$ correlators and $k=1,2,4$ for $z$ correlators.

\subsection{$SU(2)_L \times SU(2)_R $ flavor chiral symmetry}
 
For the $SU(2) $ flavor doublet $ q = (q_1, q_2)^T $, we consider the vector bilinears ($V_k$) 
\BAN 
\bar q(x) \gamma_k \frac{\tau_\pm}{2} q(x),  \hspace{4mm} \tau_{\pm} = \tau_1 \pm i \tau_2,   
\EAN
where $\{ \tau_1, \tau_2, \tau_3 \}$ are Pauli matrices, and  
$\{\tau_i/2, i=1,2,3\}$ are the generators of the $SU(2)$ group in the flavor space. 
Similarly, the axial-vector bilinears ($A_k$)  
can be written as 
\BAN 
\bar q(x) \gamma_5 \gamma_k \frac{\tau_\pm}{2} q(x).  
\EAN
The correlators of vector and axial-vector bilinears 
can be transformed into each other by the flavor nonsinglet axial rotations
\bea
q(x) \rightarrow \exp\left( i\gamma_5 \frac{\vec{\tau}}{2} \cdot \vec{\theta} \right) q(x), 
\hspace{4mm}
\bar{q}(x) \rightarrow \bar{q}(x) \gamma_4 
\exp\left(-i\gamma_5 \frac{\vec{\tau}}{2} \cdot \vec{\theta} \right) \gamma_4. 
\eea
If the $ SU(2)_L \times SU(2)_R $ chiral symmetry of the flavor doublet is effectively restored 
for $ T \gtrsim T_c^q$ (where $T_c^q$ depends on the masses of $q_1$ and $q_2$), 
the correlators of the vector bilinears ($V_k$) and the axial-vector bilinears ($A_k$) are degenerate, 
i.e., $ C_{V_k} = C_{A_k} $.  
Thus the effective restoration of $ SU(2)_L \times SU(2)_R $ chiral symmetry 
is equivalent to the emergence of the multiplet
\bea
\label{eq:SU2xSU2}
(\{ V_k \}, \{ A_k \} ),  
\eea 
where $k=1,2,3 $ for $t$ correlators and $k=1,2,4$ for $z$ correlators.

\subsection{$SU(2)_{CS}$ chiral spin symmetry}

The $SU(2)_{CS}$ chiral spin transformations \cite{Glozman:2014mka,Glozman:2015qva} are defined by 
\bea
\label{eq:q_SU2_CS}
q(x) \rightarrow \exp\left(i\frac{\vec{\Sigma}_\mu}{2} \cdot \vec{\theta} \right) q(x), \hspace{4mm} 
\bar{q}(x) \rightarrow \bar{q}(x) 
\gamma_4 \exp\left(-i \frac{\vec{\Sigma}_\mu}{2} \cdot \vec{\theta} \right) \gamma_4,  
\hspace{4mm} \mu=1,2,3,4, 
\eea   
where $ \vec{\Sigma}_\mu = \{ \gamma_\mu, i \gamma_\mu \gamma_5, \gamma_5  \} $, 
and $\vec{\theta}$ are global parameters. 
The choice of $\mu$ for a given observable is fixed by the requirement that 
the $ SU(2)_{CS} $ transformations do not mix operators with different spin. 

The QCD Lagrangian is not invariant under $SU(2)_{CS}$ transformations, but only 
the chromoelectic part of the quark-gluon interaction, and also  
the color charge $ Q^a = \int d^4 x q^\dagger(x) T^a q(x) $.  
In a given reference frame (e.g., the rest frame of the medium), 
the quark-gluon interaction in the QCD Lagrangian can be decomposed 
into the temporal and spatial parts,  
\BAN
\bar q(x) \left\{ \gamma_4 [\partial_4 + ig T^a A_4^a(x)]  
               + \sum_{k=1,2,3} \gamma_k [\partial_k + i g T^a A_k^a (x) ] \right\} q(x),  
\EAN
where the chromoelectric interaction term $ i g q^\dagger(x) T^a A_4^a(x) q(x) $ 
is invariant under the $SU(2)_{CS}$ transformations, while the chromomagnetic interaction 
and the kinetic terms break the $SU(2)_{CS}$ symmetry. 
If the $SU(2)_{CS}$ chiral spin symmetry turns out to be exact for a range of temperatures 
in high-temperature QCD, then the quarks cannot behave like free fermions at these temperatures 
since the latter break the $SU(2)_{CS}$ symmetry. 
Consequently, it is likely that there are hadron-like objects which are predominantly bound by 
the chromoelectric interactions into color singlets. 
On the other hand, if $SU(2)_{CS}$ is an approximate emergent symmetry, 
then the chromomagnetic interactions could also play some role in forming these hadron-like objects, 
and the dominance of the chromoelectric interactions depends on 
to what extent the $SU(2)_{CS} $ symmetry emerges as an exact symmetry.

In the following, we discuss the $SU(2)_{CS}$ multiplets of vector meson correlators, 
which are generated by the $SU(2)_{CS} $ transformations.  

For the $t$ correlators, the choice of $\mu=4$ satisfies the requirement  
that the $ SU(2)_{CS} $ transformations do not mix operators with different spin. 
Then the $SU(2)_{CS} \times S_3 $ transformations generate one triplet and one nonet: 
\bea
\label{eq:SU2CS_S3_k4}
(A_1, A_2, A_3); \ (V_1, V_2, V_3, T_1, T_2, T_3, X_1, X_2, X_3).  
\eea

For $ T \gtrsim T_c^q$, the $ SU(2)_L \times SU(2)_R $ chiral symmetry 
of the flavor doublet $ (q_1, q_2) $ is effectively restored (i.e., $ C_{A_k} = C_{V_k}, k=1,2,3$), 
and then the triplet and the nonet are degenerate into a single multiplet:   
\bea
\label{eq:SU2CS_S3_SU2A_k4}
(A_1, A_2, A_3, V_1, V_2, V_3, T_1, T_2, T_3, X_1, X_2, X_3).  
\eea
This suggests the possibility of a larger symmetry group $SU(4) $ 
for $T > T_c^q$ which contains $ SU(2)_L \times SU(2)_R \times SU(2)_{CS} $ as a subgroup. 
For the full $SU(4) \times S_3 $ symmetry, the multiplet in (\ref{eq:SU2CS_S3_SU2A_k4})
is enlarged to include the flavor-singlet partners of $V_k$, $T_k$ and $X_k$,  
while the flavor-singlet partners of $A_1$, $A_2$ and $A_3$ are $SU(4)$ singlets, i.e., 
\small
\bea
\label{eq:SU4_S3_k4}
\hspace{-4mm} (A_1^0, A_2^0, A_3^0); (V_1, V_2, V_3, A_1, A_2, A_3, T_1, T_2, T_3, X_1, X_2, X_3,   
 V_1^0, V_2^0, V_3^0, T_1^0, T_2^0, T_3^0, X_1^0, X_2^0, X_3^0), \hspace{4mm} 
\eea
\normalsize
where the superscript ``0" denotes the flavor singlet. 

For the $z$ correlators, $\mu=1$ and $\mu=2$ each satisfies the requirement 
that the $ SU(2)_{CS} $ transformations do not mix operators with different spin.  
Then, the $SU(2)_{CS} \times S_2 $ transformations with $\mu=1$ and $\mu=2$ together 
generate the following multiplets:
\bea
\label{eq:SU2CS_S2_a}
(V_1, V_2); \ (A_1, A_2, T_4, X_4),  \\
V_4; \ (A_4, T_1, T_2, X_1, X_2).  
\label{eq:SU2CS_S2_b}
\eea
For $ T \gtrsim T_c^q $, the $ SU(2)_L \times SU(2)_R $ chiral symmetry of the $(q_1, q_2)$ doublet 
is effectively restored, and the multiplets 
in Eqs. (\ref{eq:SU2CS_S2_a}) and (\ref{eq:SU2CS_S2_b}) become two sextets:    
\bea
\label{eq:SU2CS_S2_SU2A_a}
(V_1, V_2, A_1, A_2, T_4, X_4),  \\
(V_4, A_4, T_1, T_2, X_1, X_2).  
\label{eq:SU2CS_S2_SU2A_b}
\eea
This suggests the possibility of a larger symmetry group $ SU(4) $ for $T > T_c^q$ 
which contains $ SU(2)_L \times SU(2)_R \times SU(2)_{CS} $ as a subgroup. 
For the full $SU(4) \times S_2$ symmetry, each of the multiplets 
in Eqs. (\ref{eq:SU2CS_S2_SU2A_a}) and (\ref{eq:SU2CS_S2_SU2A_b}) is enlarged 
to include the flavor-singlet partners of $A_k$, $T_k$ and $X_k$, 
while the flavor-singlet partners of $V_1$, $V_2$, and $V_4$ are $SU(4)$ singlets, i.e., 
\bea
\label{eq:SU4_S2_a}
(V_1^0, V_2^0); \ (V_1, V_2, A_1, A_2, T_4, X_4, A_1^0, A_2^0, T_4^0, X_4^0),   \\
V_4^0; \ (V_4, A_4, T_1, T_2, X_1, X_2, A_4^0, T_1^0, T_2^0, X_1^0, X_2^0).  
\label{eq:SU4_S2_b}  
\eea

To investigate the full $SU(4)$ symmetry, it is necessary to examine the  
degeneracies of the correlators in the multiplets of Eqs. (\ref{eq:SU4_S3_k4}), 
(\ref{eq:SU4_S2_a}), and (\ref{eq:SU4_S2_b}) which involve the flavor singlets. 
Since the evaluations of the correlators of flavor singlets require the 
disconnected diagrams which have been omitted in this work,  
we are not in a position to determine the emergence of the full $SU(4)$ symmetry, 
even if its subgroup $SU(2)_L \times SU(2)_R \times SU_{CS}(2) $ is manifested approximately
due to the effective restoration of $SU(2)_L \times SU(2)_R $ chiral symmetry
and the emergence of approximate $SU(2)_{CS}$ chiral spin symmetry. 
Nevertheless, the splittings between the correlators of the flavor singlet 
and the nonsinglet of $u$ and $d$ quarks 
are usually very small compared to the correlators of the nonsinglet. 
Thus we can envision that 
the flavor singlets in Eqs. (\ref{eq:SU4_S3_k4}), (\ref{eq:SU4_S2_a}), and (\ref{eq:SU4_S2_b}) 
would be approximately degenerate with all members in the multiplet. 
In order to justify this, computing the correlators of flavor singlets is indispensible.   
 
To investigate the manifestation of various symmetries from the degeneracies of 
the $t$ correlators and the $z$ correlators of vector mesons, 
in view of the $S_3$ and $S_2$ symmetries, 
it suffices to focus on the ``1" components of the vector meson correlators
(i.e., $C_{V_1}$, $C_{A_1}$, $C_{T_1}$, $C_{X_1}$, and their flavor-singlet partners),  
while all ``2" and ``3" components can be suppressed.  
With this convention, the multiplets of $SU(2)_{CS}$ in Eqs. (\ref{eq:SU2CS_S3_k4}),  
(\ref{eq:SU2CS_S2_a}), and (\ref{eq:SU2CS_S2_b}) can be abbreviated as: 
\bea
\label{eq:SU2CS_t}
\mbox{$t$ correlators}: \hspace{2mm} && (A_1); \ (V_1, T_1, X_1), \\ 
\label{eq:SU2CS_z_a}
\mbox{$z$ correlators}: \hspace{2mm} && (V_1); \ (A_1, T_4, X_4), \\
                                     && (V_4); \ (A_4, T_1, X_1),     
\label{eq:SU2CS_z_b}
\eea       
and the degeneracies in the above triplets signal 
the emergence of $SU(2)_{CS}$ chiral spin symmetry. 
Similarly, the $SU(4)$ multiplets in Eqs.
(\ref{eq:SU4_S3_k4}), (\ref{eq:SU4_S2_a}), and (\ref{eq:SU4_S2_b}) can be abbreviated as: 
\bea
\label{eq:SU4_t}
\mbox{$t$ correlators}: \hspace{2mm} && A_1^0; \ (A_1, V_1, T_1, X_1, V_1^0, T_1^0, X_1^0), \\ 
\label{eq:SU4_z_a}
\mbox{$z$ correlators}: \hspace{2mm} && V_1^0; \ (V_1, A_1, T_4, X_4, A_1^0, T_4^0, X_4^0),   \\
                                     && V_4^0; \ (V_4, A_4, T_1, X_1, A_4^0, T_1^0, X_1^0),  
\label{eq:SU4_z_b}
\eea       
and the degeneracies in the above multiplets signal the emergence of $SU(4)$ symmetry. 

For $ T > T_1^q \gtrsim T_c^q$, the $SU(2)_L \times SU(2)_R$ and $U(1)_A$ chiral symmetries are 
effectively restored, and $ C_{V_k} = C_{A_k} $, $ C_{T_k} = C_{X_k} $, $ C_{V_k}^0 = C_{A_k}^0 $, 
and $ C_{T_k}^0 = C_{X_k}^0 $. 
Thus, to examine the $SU(2)_{CS}$ symmetry, one only needs to check the degeneracy of 
$t$ correlators of $( V_1, T_1 ) $ in Eq. (\ref{eq:SU2CS_t}),  
the degeneracy of $z$ correlators of $ (A_1, T_4 )$ in Eq. (\ref{eq:SU2CS_z_a}), 
and the degeneracy of $z$ correlators of $ (A_4, T_1) $ in (\ref{eq:SU2CS_z_b}). 
Meanwhile, for the $SU(4)$ symmetry, it only needs to check the degenerancy of the $t$ correlators 
of $( V_1, T_1, V_1^0, T_1^0 ) $ in Eq. (\ref{eq:SU4_t}),  
the degeneracy of the $z$ correlators of $ (A_1, T_4, A_1^0, T_4^0 )$ in Eq. (\ref{eq:SU4_z_a}), 
and also of $ (A_4, T_1, A_4^0, T_1^0 )$ in Eq. (\ref{eq:SU4_z_b}).

\section{Symmetry-Breaking Parameters}
\label{kappa}

In order to give a quantitative measure for the manifestation of symmetries from the degeneracy
of temporal/spatial correlators, we consider the symmetry-breaking parameters as follows.
To this end, we write the meson correlators as functions of dimensionaless variables 
\bea
\label{eq:tT}
&& tT = (n_t a)/(N_t a) = n_t/N_t, \\
&& zT = (n_z a)/(N_t a) = n_z/N_t, 
\label{eq:zT}
\eea
where $ T $ is the temperature.    

\subsection{$U(1)_A$ and $SU(2)_L \times SU(2)_R$ symmetry-breaking parameters}

For the $U(1)_A$ symmetry, its breaking in the pseudoscalar ($P$) and scalar ($S$) channels
can be measured by
\bea
\label{eq:k_PS_t}
&& \kappa_{PS}(tT) = 1-\frac{C_S(tT)}{C_P(tT)}, \hspace{4mm}  n_t > 1,  \\
&& \kappa_{PS}(zT) = 1-\frac{C_S(zT)}{C_P(zT)}, \hspace{4mm}  n_z > 1,
\label{eq:k_PS_z}
\eea
where $C_{S} $ and $C_{P} $ are normalized correlators 
(with normalization equal to 1 at $n_t =1$ or $n_z=1$).
If $ C_P $ and $ C_S $ are exactly degenerate at $T$, then $\kappa_{PS} = 0 $ for any $tT$ ($zT$),
and the $U(1)_A$ symmetry is effectively restored at $T$.
On the other hand, if there is any discrepancy between $C_P$ and $C_S$ at any $tT$ ($zT$),
then $\kappa_{PS} $ is nonzero at this $tT$ ($zT$), and this suggests that $U(1)_A$ 
is not completely restored at $T$.
Obviously, this criterion is more stringent than
the equality of the thermal masses from the temporal correlators as well as 
the screening masses from the spatial correlators.
Similarly, the $U(1)_A $ symmetry breaking 
in the channels of tensor vectors ($T_k$) and axial-tensor vectors ($X_k$) can be measured by
\bea
\label{eq:k_TX_t}
&& \kappa_{TX}(tT) = 1-\frac{C_{X_k}(tT)}{C_{T_k}(tT)}, \hspace{4mm} n_t > 1, \hspace{4mm} (k=1,2,3), \\ 
&& \kappa_{TX}(zT) = 1-\frac{C_{X_k}(zT)}{C_{T_k}(zT)}, \hspace{4mm} n_z > 1, \hspace{4mm} (k=1,2,4). 
\label{eq:k_TX_z}
\eea
Due to the $S_3$ symmetry of the $t$ correlators, it suffices only to examine the $k=1$ component 
in Eq. (\ref{eq:k_TX_t}). Similarly, due to the $S_2$ symmetry of the $z$ correlators, 
one only needs to examine the $k=1$ and $k=4$ components of Eq. (\ref{eq:k_TX_z}). 
In practice, there is no difference 
between $k=1$ and $k=4$ components (up to the statistical uncertainties); 
thus, the $k=4$ component is suppressed in the following.    

By the same token, the breaking of $SU(2)_L \times SU(2)_R $ chiral symmetry can be measured by
\bea
\label{eq:k_VA_t}
&& \kappa_{VA}(tT) = 1-\frac{C_{A_k}(tT)}{C_{V_k}(tT)}, \hspace{4mm} n_t > 1, \hspace{4mm} (k=1,2,3), \\ 
&& \kappa_{VA}(zT) = 1-\frac{C_{A_k}(zT)}{C_{V_k}(zT)}, \hspace{4mm} n_z > 1, \hspace{4mm} (k=1,2,4).  
\label{eq:k_VA_z}
\eea
If $ C_{A_k} $ and $ C_{V_k} $ are degenerate, then $\kappa_{VA}= 0 $ for any $tT$ ($zT$), 
and the $SU(2)_L \times SU(2)_R $ chiral symmetry is effectively restored. 
Following the above discussion for $\kappa_{TX}$, 
the components of $k=2,3,4$ in Eqs. (\ref{eq:k_VA_t}) and (\ref{eq:k_VA_z}) are suppressed.

\subsection{$SU(2)_{CS}$ symmetry-breaking and fading parameters}

For the $t$ correlators, the $SU(2)_{CS} $ symmetry breaking can be measured by  
the splitting of $V_1$ and $T_1$ in the multiplet [Eq. (\ref{eq:SU2CS_t})]:  
\bea
\label{eq:k_AT_t}
\kappa_{AT}(tT) = \frac{C_{V_1}(tT)}{C_{T_1}(tT)} - 1, \hspace{4mm} n_t > 1,
\eea
where $V_1$ and $T_1$ are connected by the $SU(2)_{CS}$ transformations.
In general, the splitting between $V_1(tT)$ and $T_1(tT)$ 
is a monotonic decreasing function of $T$ for a fixed $tT$, and so is $\kappa_{AT}(tT) $. 

As the temperature $T$ is increased, 
the separation between the multiplets of $SU(2)_{CS}$ and $U(1)_A$ is decreased.
Therefore, at sufficiently high-temperatures, the $SU(2)_{CS} \times SU(2)_L \times SU(2)_R$ multiplet  
$M_1 = (A_1, V_1, T_1, X_1) $ and the $U(1)_A$ multiplet $M_0 = (P, S)$ merge together, 
and then the approximate $SU(2)_{CS}$ symmetry becomes washed out, 
and only the $U(1)_A \times SU(2)_L \times SU(2)_R $ chiral symmetry remains.
The fading of the approximate $SU(2)_{CS}$ symmetry  
can be measured by the ratio of the splitting between $V_1$ and $T_1$ in the $M_1$ multiplet
to the separation of $M_1$ and $M_0$ multiplets:
\bea
\label{eq:kappa_t}
\kappa(tT) = \frac{C_{V_1}(tT) - C_{T_1}(tT)}{C_{M_0}(tT) - C_{M_1}(tT)}, \hspace{4mm} n_t > 1,
\eea
where
\BAN
&& C_{M_0}(tT) = \frac{1}{2} \left[ C_P(tT) + C_S(tT) \right],  \\
&& C_{M_1}(tT) = \frac{1}{4} \left[ C_{A_1}(tT) + C_{V_1}(tT) + C_{T_1}(tT) + C_{X_1}(tT) \right].
\EAN
In general, $\kappa(tT) $ is a monotonic increasing function of $T$ for a fixed $tT$. 
If $\kappa(tT) \ll 1 $ for a range of $T$, then  
the approximate $SU(2)_{CS}$ symmetry is well defined for this window of $T$.
On the other hand, if $ \kappa(tT) \gtrsim 1$ for $T > T_f$, 
then the approximate $SU(2)_{CS}$ symmetry becomes washed out,
and only the $U(1)_A \times SU(2)_L \times SU(2)_R $ chiral symmetry remains. 

Thus, to determine to what extent the $SU(2)_{CS}$ symmetry
is manifested in the $t$ correlators, it is necessary to examine whether 
both $\kappa(tT)$ and $\kappa_{AT}(tT)$ are sufficiently small. 
For a fixed $tT$, the condition
\bea
\label{eq:SU2_CS_crit_t}
\left(~|\kappa_{AT}(tT)| < \epsilon_{CS}~\right)~\land ~\left(~|\kappa(tT)| < \epsilon_{CS} ~\right)
\eea
serves as a criterion for the emergence of approximate $SU(2)_{CS}$ symmetry in the $t$ correlators, 
where $ \epsilon_{CS} $ specifies the precision of $SU(2)_{CS}$ symmetry.
Once $\epsilon_{CS}$ is given, the range of temperatures satisfying Eq. (\ref{eq:SU2_CS_crit_t}) 
can be determined for a fixed $tT$. Roughly speaking, if there exists a window of temperatures 
satisfying Eq. (\ref{eq:SU2_CS_crit_t}) with $\epsilon_{CS} \le 0.01 $, 
then the $SU(2)_{CS}$ symmetry can be regarded as an exact symmetry emerging in this window. 
Here the upper bound 0.01 is estimated based on the maximum values of 
$\kappa_{PS}(tT)$, $\kappa_{TX}(tT)$ and $\kappa_{VA}(tT)$ among all values of $T$ and $tT$ 
[Iin this study, as given in Sec. \ref{Ct_ud_A}.
On the other hand, if no temperatures satisfying (\ref{eq:SU2_CS_crit_t}) exist 
with $ \epsilon_{CS} < 0.50 $, then the $SU(2)_{CS}$ symmetry can be regarded as not 
emerging in this theory e.g., the noninteracting theory with free fermions on the lattice. 
Otherwise, $ 0.01 < \epsilon_{CS} \le 0.5$, and the $SU(2)_{CS}$ symmetry can be regarded as 
an approximate emergent symmetry in this window. 

Next, we turn to the $SU(2)_{CS}$ symmetry-breaking and fading parameters for the $z$ correlators.
Note that at sufficiently high-temperatures, the $U(1)_A$ multiplet $M_0 = (P, S)$ and the 
$SU(2)_{CS} \times SU(2)_L \times SU(2)_R$ multiplet  
$M_2=(V_1, A_1, T_4, X_4)$ merge together,  
and then the approximate $SU(2)_{CS}$ symmetry becomes washed out, 
and only the $U(1)_A \times SU(2)_L \times SU(2)_R $ chiral symmetry remains.  
On the other hand, the $SU(2)_{CS} \times SU(2)_L \times SU(2)_R$ multiplet  
$ M_4 = (V_4, A_4, T_1, X_1) $ never merges with 
$M_0$ and $M_2$ even in the limit $T \to \infty$ 
(i.e., the noninteracting theory with free quarks), 
which can be seen from Eqs. (\ref{eq:M02_free}) and (\ref{eq:M4_free}).
Thus, the multiplet $M_4$ is irrelevant to the fading of the approximate $SU(2)_{CS} $ symmetry.  

Now it is straightforward to transcribe Eqs. (\ref{eq:k_AT_t})$-$(\ref{eq:SU2_CS_crit_t}) 
to their counterparts for the $z$ correlators. 
This gives the $SU(2)_{CS}$ symmetry-breaking and fading parameters    
\bea
\label{eq:k_AT_z}
&& \kappa_{AT}(zT) = \frac{C_{A_1}(zT)}{C_{T_4}(zT)} - 1, \hspace{4mm} n_z > 1, \\ 
&& \kappa(zT) = \frac{C_{A_1}(zT) - C_{T_4}(zT)}{C_{M_0}(zT) - C_{M_2}(zT)}, \hspace{4mm} n_z > 1, 
\label{eq:kappa_z}
\eea
where
\BAN
&& C_{M_0}(zT) \equiv \frac{1}{2} \left[ C_P(zT) + C_S(zT) \right],  \\
&& C_{M_2}(zT) \equiv \frac{1}{4} \left[ C_{V_1}(zT) + C_{A_1}(zT) + C_{T_4}(zT) + C_{X_4}(zT) \right],   
\EAN
and the criterion for the emergence of approximate $SU(2)_{CS}$ symmetry in the $z$ correlators is   
\bea
\label{eq:SU2_CS_crit_z}
\left(~|\kappa_{AT}(zT)| < \epsilon_{CS}~\right)~\land ~\left(~|\kappa(zT)| < \epsilon_{CS} ~\right),
\eea 
where $\epsilon_{CS} $ is not necessarily equal to that in Eq. (\ref{eq:SU2_CS_crit_t}). 
In general, for a fixed $zT$, $\kappa_{AT}(zT) $ is a monotonic decreasing function of $T$,  
while $\kappa(zT) $ is a monotonic increasing function of $T$. 
Once $\epsilon_{CS}$ is given, the range of temperatures satisfying Eq. (\ref{eq:SU2_CS_crit_z}) 
can be determined for a fixed $zT$. 
Note that even for the same $\epsilon_{CS}$ and $ tT = zT $, 
the window satisfying Eq. (\ref{eq:SU2_CS_crit_t}) 
is most likely different from that satisfying Eq. (\ref{eq:SU2_CS_crit_z}). 
Nevertheless, the classification of the emergent $SU(2)_{CS}$ symmetry as an 
(exact, approximate, nonexisting) symmetry according to 
$ \epsilon_{CS} = \left(\le 0.01,\ (0.01, 0.5],\ > 0.5 \right)$ 
can be used in both cases.  
Here, the upper bound 0.01 for exact $SU(2)_{CS}$ symmetry is estimated based on 
the maximum values of $\kappa_{PS}(zT)$, $\kappa_{TX}(zT)$ and $\kappa_{VA}(zT)$ 
among all values of $T$ and $zT$ in this study, as given in Sec. \ref{Cz_ud_A}.


Finally, we note that the $\kappa(tT)$ defined in Ref.\cite{Rohrhofer:2019qal} 
for the $t$ correlators can be written as
\bea
\label{eq:kappa_Nf2_t}
\kappa(tT) = -\frac{C_{V_1}(tT) - C_{T_1}(tT)}{C_S(tT) - C_{V_1}(tT)}, 
\eea
where the denominator is different from that in Eq. (\ref{eq:kappa_t}). 
However, for $ T > T_1 \gtrsim T_c $,  
with the effective restoration of $U(1)_A$ and $SU(2)_L \times SU(2)_R$ of $u$ and $d$ quarks,     
then $C_P = C_S$, $ C_{V_1} = C_{A_1}$, and $C_{T_1} = C_{X_1}$.  
Thus the difference between the denominators of Eqs. (\ref{eq:kappa_Nf2_t}) and (\ref{eq:kappa_t}) 
is equal to $ [C_{T_1}(tT) - C_{V_1}(tT) ]/2 $,  
which is negligible comparing with the denominator $[C_S(tT) - C_{V_1}(tT)]$ itself.  
Thus the discrepancy due to two different definitions of $ \kappa(tT) $ 
in (\ref{eq:kappa_Nf2_t}) and (\ref{eq:kappa_t}) is negligible 
for the meson correlators of $u$ and $d$ quarks, except for an overall minus sign.    

Moreover, the $\kappa(zT)$ defined in Ref.\cite{Rohrhofer:2019qwq} for the $z$ correlators 
can be written as 
\bea
\label{eq:kappa_Nf2_z}
\kappa(zT) = \left| \frac{C_{A_1}(zT) - C_{T_4}(zT)}{C_S(zT) - C_{A_1}(zT)} \right|, 
\eea
where the denominator is different from that in Eq. (\ref{eq:kappa_z}). 
Again, for $ T > T_1 \gtrsim T_c$, with     
$C_P = C_S$, $ C_{V_1} = C_{A_1}$, and $C_{T_4} = C_{X_4}$,  
the difference between the denominators of (\ref{eq:kappa_Nf2_z}) and (\ref{eq:kappa_z})  
is equal to $ [C_{T_4}(zT) - C_{A_1}(zT) ]/2 $,  
which is negligible compared with the denominator $[C_S(zT)-C_{A_1}(zT)]$ itself.   
Thus, the discrepancy due to two different definitions of $ \kappa(zT) $ 
in Eqs. (\ref{eq:kappa_Nf2_z}) and (\ref{eq:kappa_z}) is negligible 
for the meson correlators of $u$ and $d$ quarks.

\section{Gauge ensembles} 
\label{lattice}

The gauge ensembles in this study are generated by hybrid Monte Carlo (HMC) simulation 
of lattice QCD with $N_f=2+1+1$ optimal domain-wall quarks \cite{Chiu:2002ir} at the physical point,
on the $32^3 \times (16,12,10,8,6,4)$ lattices, 
with the plaquette gauge action at $\beta = 6/g^2 = \{6.20, 6.18 \} $.
This set of ensembles are generated with the same actions \cite{Chiu:2011bm,Chen:2014hyy}
and algorithms as their counterparts 
on the $64^3 \times (20,16,12,10,8,6)$ lattices \cite{Chen:2022fid}, 
but with one-eighth of the spatial volume. 
The simulations were performed on a GPU cluster of 32 nodes (64 GPUs)
with various Nvidia GPUs consisting of GTX-970/1060/1070/1080 and TITAN-X. 
The initial thermalization of each ensemble was performed in one node 
with one GPU or two GPUs with peer-to-peer communication via the PCIe bus. 
The initial thermalization of each ensemble was performed in one node with $1-2$ GPUs.
After thermalization, a set of gauge configurations were sampled and distributed
to $16-32$ simulation units, and each unit ($1-2$ GPUs) 
performed an independent stream of HMC simulation. 
For each HMC stream, one configuration was sampled for every five trajectories. 
Finally collecting all sampled configurations from all HMC streams 
gives the total number of configurations of each ensemble. 
The lattice parameters and statistics of the gauge ensembles for computing the meson correlators
in this study are summarized in Table \ref{tab:6_ensembles}.
The temperatures of these six ensembles are in the range $\sim$ 190-770 MeV, all above 
the pseudocritical temperature $ T_c \sim 150 $~MeV.

\begin{table}[h!]
\begin{center}
\caption{The lattice parameters and statistics of the six gauge ensembles 
for computing the meson correlators.
The last three columns are the residual masses of $u/d$, $s$, and $c$ quarks.}
\setlength{\tabcolsep}{3pt}
\vspace{-2mm}
\begin{tabular}{|cccccccccccc|}
\hline 
    $\beta$
  & $a$[fm]
  & $ N_x $
  & $ N_t $
  & $ m_{u/d} a $
  & $ m_{s} a $
  & $ m_{c} a $
  & $T$[MeV]
  & $N_{\rm confs}$
  & $ (m_{u/d} a)_{\rm res} $
  & $ (m_{s} a)_{\rm res} $
  & $ (m_{c} a)_{\rm res} $
\\
\hline
\hline
6.20 & 0.0641 & 32 & 16 & 0.00125 & 0.040 & 0.550 & 193 & 583 
& $ 1.9(2) \times 10^{-5}$ & $1.5(2) \times 10^{-5}$ & $ 4.3(7) \times 10^{-6} $\\
6.18 & 0.0685 & 32 & 12 & 0.00180 & 0.058 & 0.626 & 240 & 781 
& $1.9(2) \times 10^{-5}$ & $1.6(1) \times 10^{-5}$ & $ 3.8(5) \times 10^{-6} $\\
6.20 & 0.0641 & 32 & 10 & 0.00125 & 0.040 & 0.550 & 307 & 481 
& $5.7(7) \times 10^{-6}$ & $5.1(6) \times 10^{-6}$ & $ 1.4(2) \times 10^{-6} $\\
6.20 & 0.0641 & 32 & 8  & 0.00125 & 0.040 & 0.550 & 384 & 468 
& $6.3(9) \times 10^{-6}$ & $6.0(7) \times 10^{-6}$ & $ 3.0(9) \times 10^{-6} $\\
6.20 & 0.0641 & 32 & 6  & 0.00125 & 0.040 & 0.550 & 512 & 431 
& $5.8(9) \times 10^{-6}$ & $5.6(8) \times 10^{-6}$ & $ 3.4(7) \times 10^{-6} $\\
6.20 & 0.0641 & 32 & 4  & 0.00125 & 0.040 & 0.550 & 768 & 991 
& $1.2(2) \times 10^{-6}$ & $1.2(2) \times 10^{-6}$ & $ 1.2(2) \times 10^{-6} $\\
\hline
\end{tabular}
\label{tab:6_ensembles}
\end{center}
\end{table}

The lattice spacing and the $(u/d, s, c)$ quark masses are determined on the 
the $ 32^3 \times 64 $ lattices, with the number of configurations  
$(221, 292)$ for $\beta = (6.18, 6.20)$ respectively.
The lattice spacing is determined using the Wilson flow 
\cite{Narayanan:2006rf,Luscher:2010iy}
with the condition $ \{ t^2 \langle E(t) \rangle \} |_{t=t_0} = 0.3 $
and the input $ \sqrt{t_0} = 0.1416(8) $~fm 
\cite{Bazavov:2015yea}.
The physical $(u/d, s, c)$ quark masses are obtained by tuning their masses such that 
the masses of the lowest-lying states extracted from 
the time-correlation functions of the meson operators
$ \{ \bar u \gamma_5 d, \bar s \gamma_i s, \bar c \gamma_i c \} $  
are in good agreement with the physical masses of 
$\pi^{\pm}(140) $, $\phi(1020)$, and $J/\psi(3097)$. 

The chiral symmetry breaking due to finite $N_s=16$ (in the fifth dimension)
can be measured by the residual mass of each quark flavor \cite{Chen:2012jya}, 
as given in the last three columns of Table \ref{tab:6_ensembles}.
The residual masses of $(u/d, s, c)$ quarks are
less than ($1.5\%, 0.04\%, 0.001\%$) of their bare masses, 
amounting to less than (0.06, 0.05, 0.02) MeV/$c^2$, respectively. 
This asserts that the chiral symmetry is well preserved such that the deviation
of the bare quark mass $m_q$ is sufficiently small in the
effective 4D Dirac operator of optimal DWF, for both light and heavy quarks.
In other words, the chiral symmetry in the simulations are sufficiently precise
to guarantee that the hadronic observables (e.g., meson correlators) 
can be evaluated to high precision, with the associated uncertainty
much less than those due to statistics and other systematics.

\begin{figure}[!ht]
  \centering
  \caption{
   The left panels are the normalized $t$ correlators of $ \bar u \Gamma d $
   in $N_f=2+1+1$ lattice QCD at the physical point for $T=(193, 240, 307)$~MeV,
   while the right panels are the symmetry-breaking parameters corresponding to the left panels. 
  }
  \begin{tabular}{@{}c@{}c@{}}
  \includegraphics[width=7.0cm,clip=true]{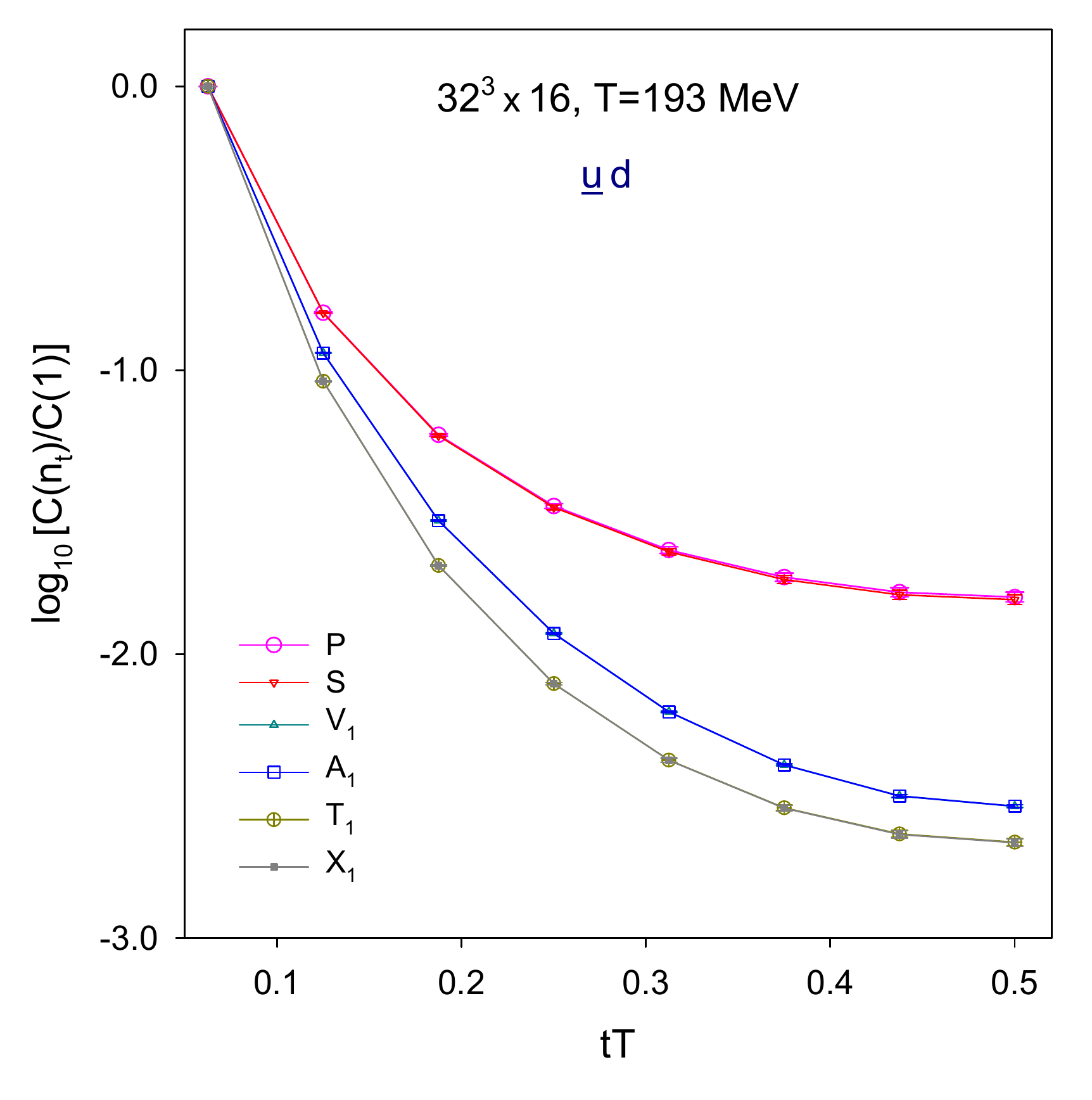}
&
  \includegraphics[width=7.0cm,clip=true]{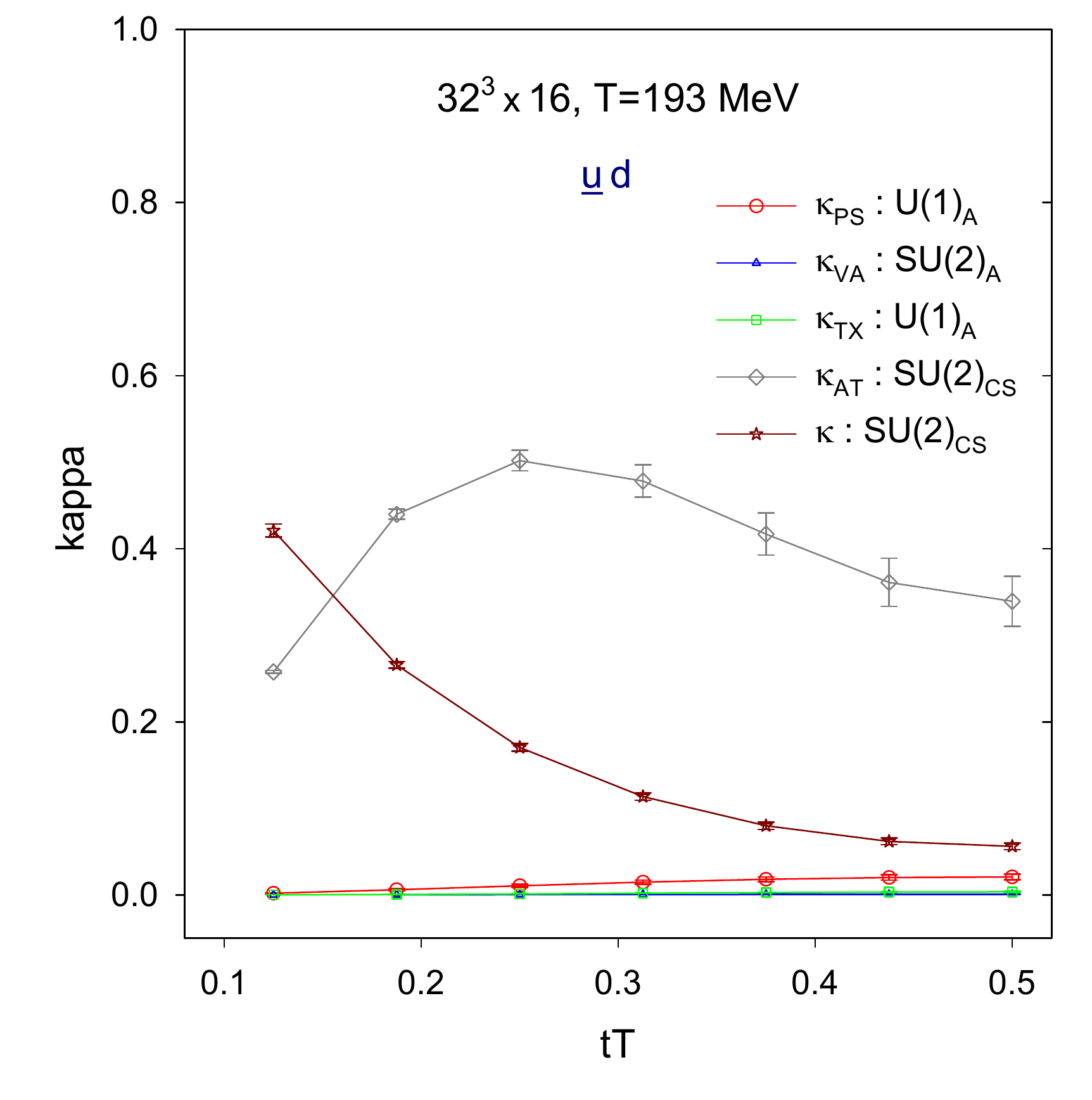}
\\
  \includegraphics[width=7.0cm,clip=true]{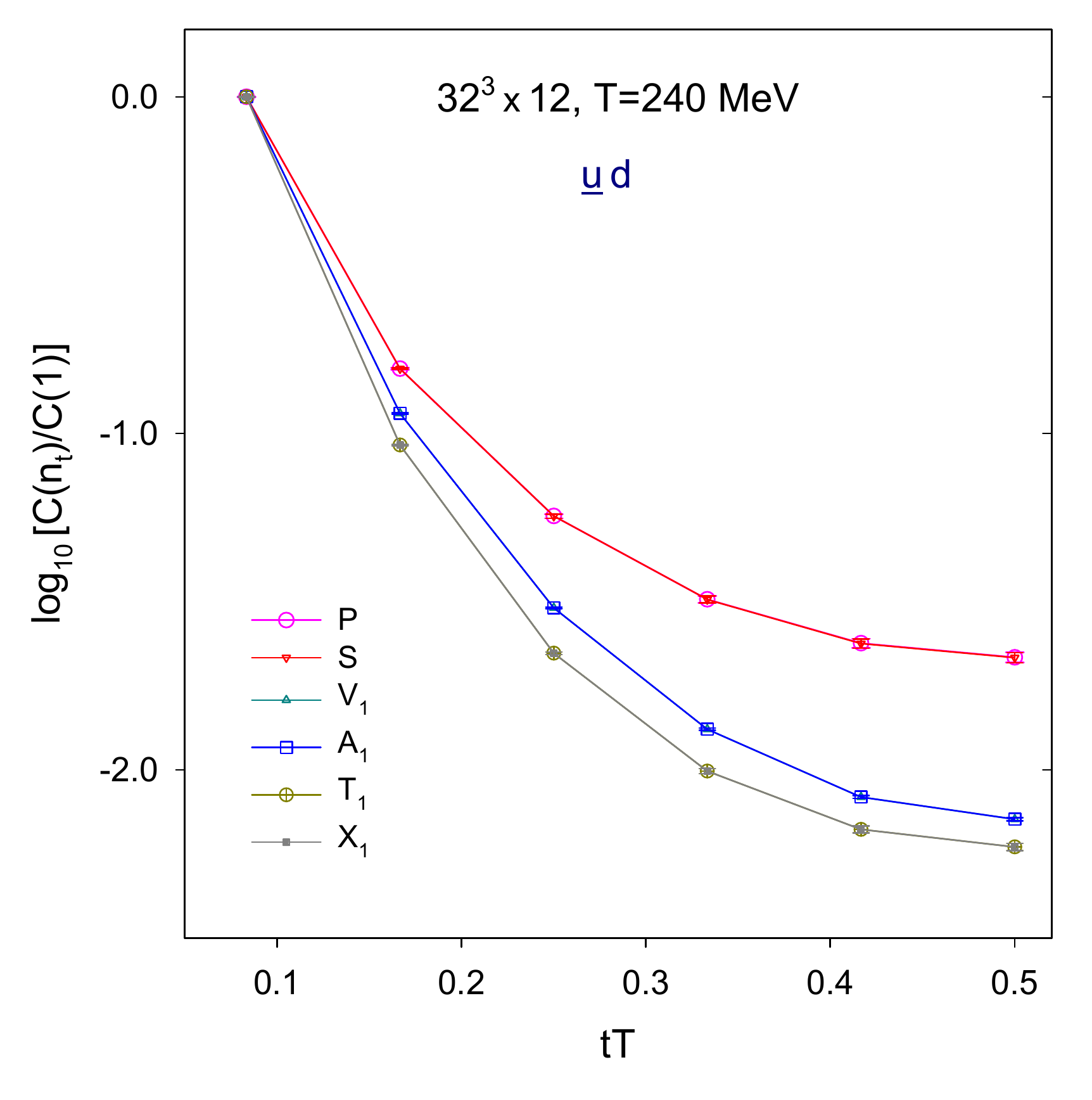}
&
  \includegraphics[width=7.0cm,clip=true]{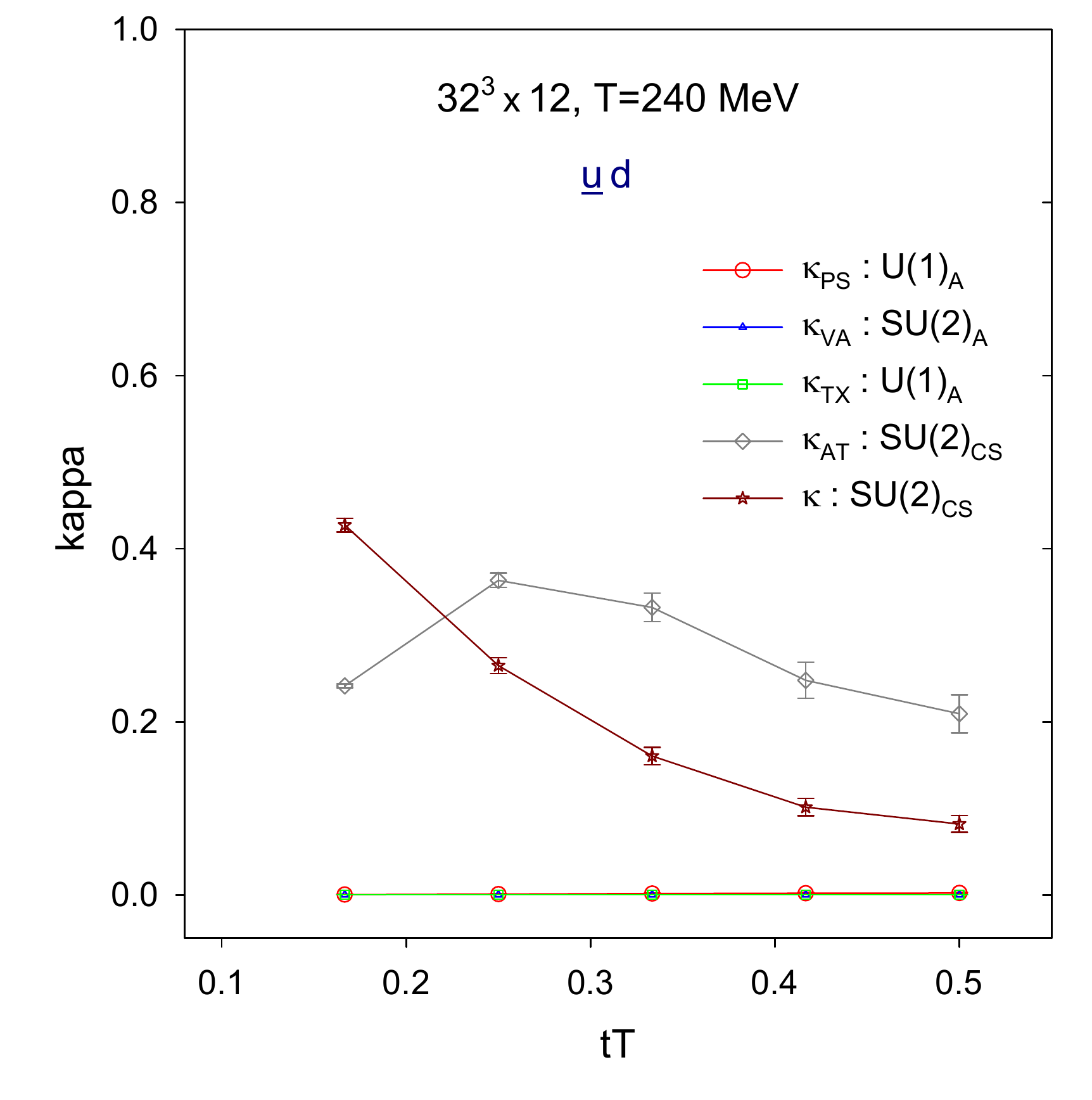}
\\
  \includegraphics[width=7.0cm,clip=true]{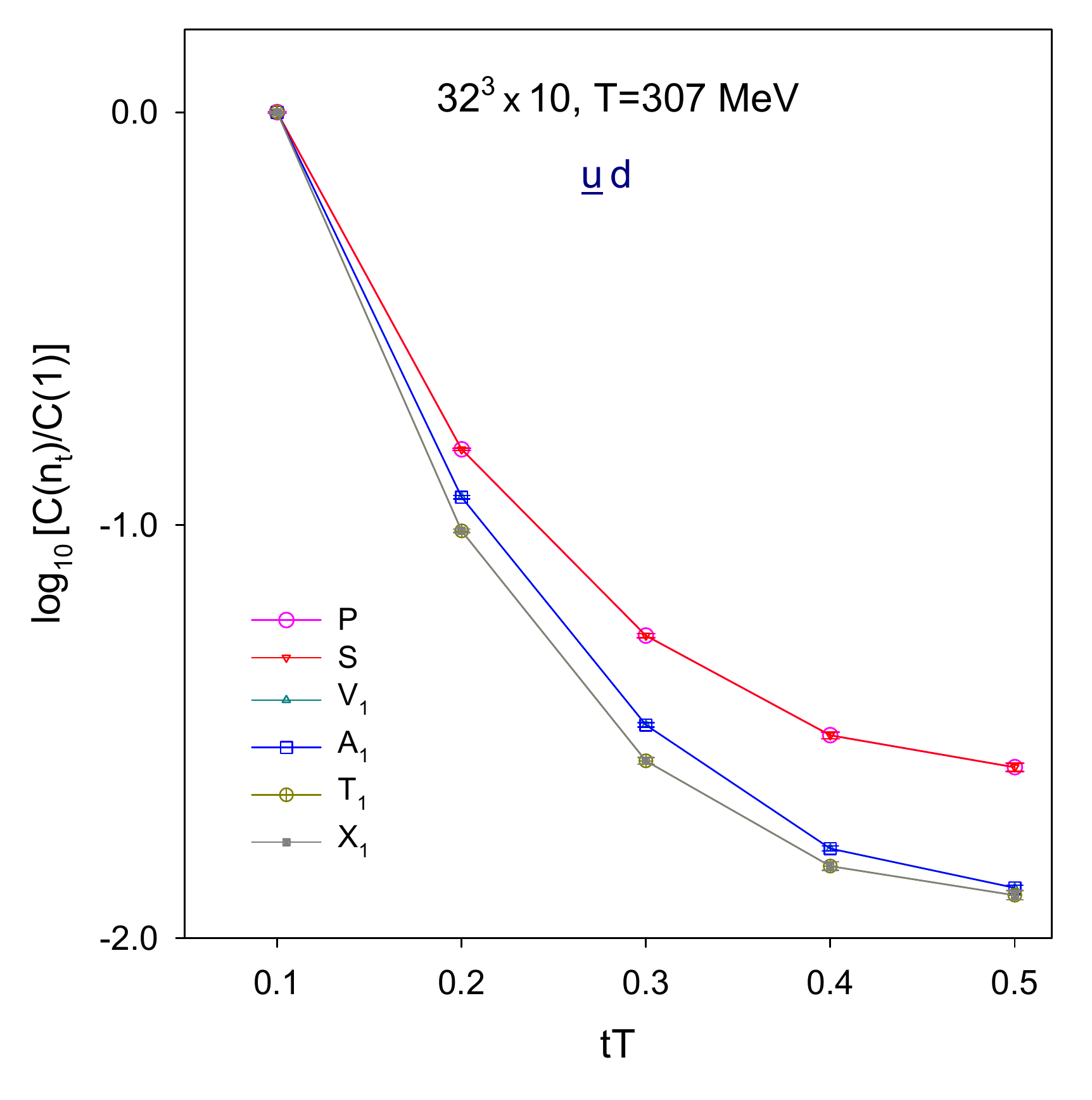}
&
  \includegraphics[width=7.0cm,clip=true]{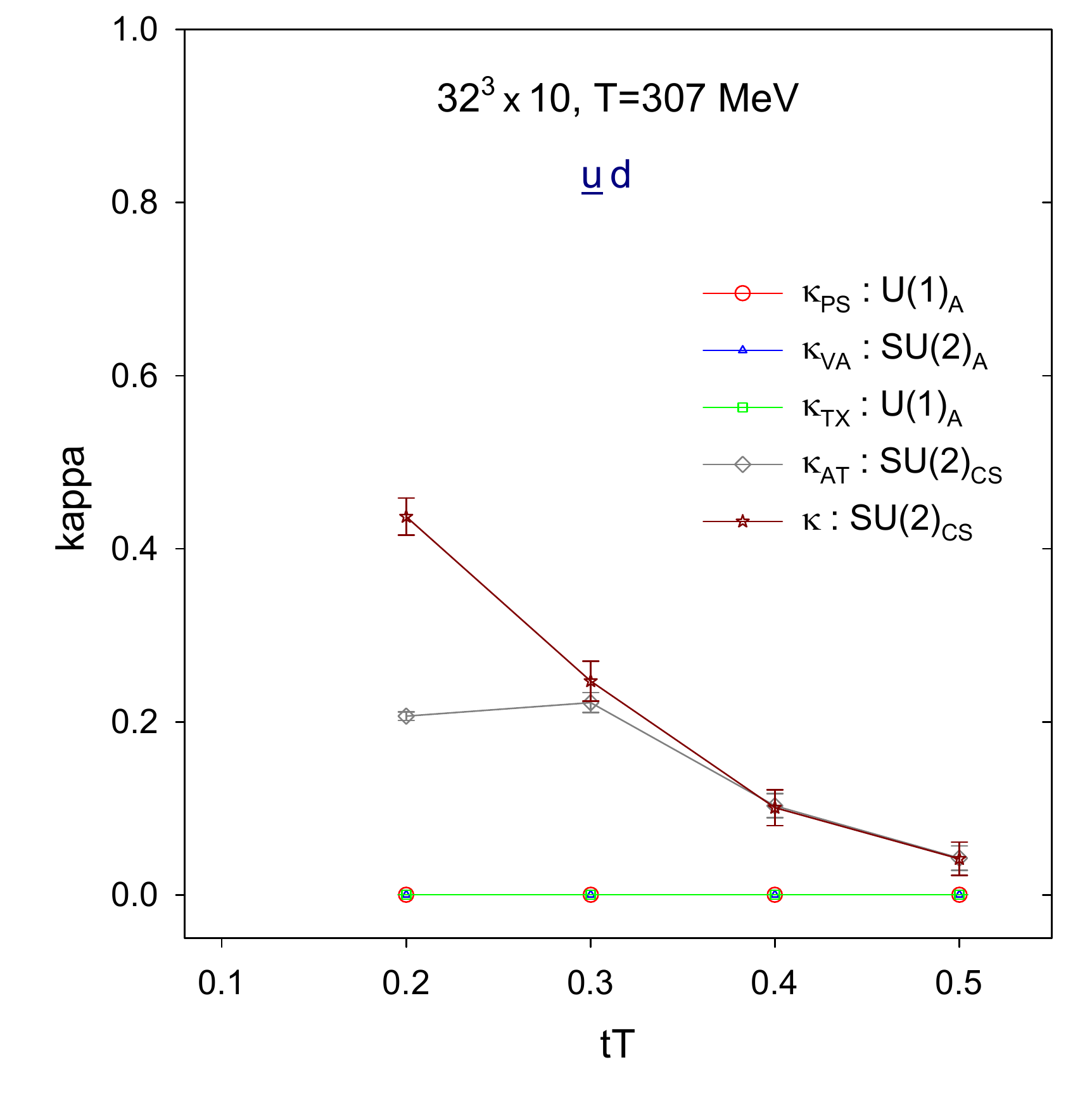}
\end{tabular}
\label{fig:Ct_ud}
\end{figure}

\section{Temporal correlators of $ \bar{u} \Gamma d $} 
\label{Ct_ud}

\subsection{Results of $N_f=2+1+1$ lattice QCD}
\label{Ct_ud_A}

In the left panels of Fig. \ref{fig:Ct_ud}, the temporal correlators 
of $\bar u \Gamma d $ are plotted as a function of the dimensionless variable $tT$ 
[Eq. \ref{eq:tT}].
Each panel displays the normalized $t$ correlators (with the normalization equal to 1 at $n_t=1$)  
for all meson interpolators (see Table \ref{tab:bilinear}),  
for $n_t$ from $1$ to $N_t/2$.
Due to the degeneracy (the $S_3$ symmetry) of the ``1," ``2," and ``3" components 
in the $t$ correlators of $J=1$ mesons, only the ``1" components are plotted in Fig. \ref{fig:Ct_ud}.

For the three temperatures in the range $T \sim $ 190-310 MeV, the $U(1)_A$ symmetry 
seems to be effectively restored, as shown by the degeneracies $C_P(t)=C_S(t)$   
and $C_{T_1}(t) = C_{X_1}(t)$. 
Moreover, the $SU(2)_L \times SU(2)_R$ chiral symmetry is also effectively restored,  
as shown by the degeneracy $ C_{V_1}(t) = C_{A_1}(t) $. 

Due to the effective restoration of $U(1)_A$ and $SU(2)_L \times SU(2)_R $ chiral symmetries, 
in each left panel of Fig. \ref{fig:Ct_ud},  
there emerge three distinct multiplets: $(P, S)$, $(V_1, A_1 )$, and $(T_1, X_1)$. 
They appear in the order 
\bea 
\label{eq:Ct_ud}
C_{P,S} > C_{V_1,A_1} > C_{T_1,X_1}, \hspace{2mm} {\rm for} \ n_t >1, 
\eea 
which is consistent with $N_f=2+1+1$ lattice QCD at $T< T_c \sim 150$~MeV.  

As the temperature $T$ is increased from $193$~MeV to $307$~MeV, 
the multiplets $(V_1, A_1)$ and $(T_1, X_1 )$ tend to merge together to form 
a single multiplet $ M_1 = ( A_1, V_1, T_1, X_1 ) $,  
in agreement with the $SU(2)_{CS} $ multiplets [Eq. \ref{eq:SU2CS_t}] 
and the $SU(4)$ multiplet [Eq. \ref{eq:SU4_t}]. 
This suggests the emergence of approximate $SU(2)_{CS}$ and $SU(4)$ symmetries. 
Moreover, we observe that the separation between $M_1$ and the $U(1)_A$ multiplet $ M_0 = (P, S) $ 
becomes smaller and smaller as $T$ is increased from 193~MeV to 307~MeV. 
Therefore, at sufficiently high-temperatures above 307~MeV, say, $ T \ge T_f $,  
$M_1$ and $M_0$ would merge together, and then the approximate $SU(2)_{CS}$ and $SU(4)$ symmetries 
become washed out, and only the $U(1)_A \times SU(2)_L \times SU(2)_R $ chiral symmetry remains. 
In other words, the approximate $SU(2)_{CS}$ and $SU(4)$ symmetries  
can only appear in a range of temperatures above $T_c$, 
say $T_c < T_{cs} \lesssim T \lesssim T_f $, 
where $T_{cs} $ and $T_f$ depend on the $\epsilon_{CS}$ in the criterion (\ref{eq:SU2_CS_crit_t}) 
for the emergence of approximate $SU(2)_{CS}$ symmetry in the $t$ correlators.

Next we examine the symmetries in the temporal correlators  
with the symmetry-breaking parameters as defined by 
$ \kappa_{PS} $ [Eq. \ref{eq:k_PS_t}], $\kappa_{TX}$ [Eq. \ref{eq:k_TX_t}], 
$\kappa_{VA}$ [Eq. \ref{eq:k_VA_t}], $\kappa_{AT}$ [Eq. \ref{eq:k_AT_t}] 
and $\kappa$ [Eq. \ref{eq:kappa_t}] in Sec. \ref{kappa}.
In Fig. \ref{fig:Ct_ud}, the symmetry-breaking parameters 
are plotted in the right panels,  
with one-to-one correspondence to the $t$ correlators in the left panels. 

For all three temperatures in the range $T \sim $ 190-310 MeV, 
the $SU(2)_L \times SU(2)_R$ chiral symmetry is effectively restored 
with the maximum value of $ \kappa_{VA} $ equal to $5.2(8) \times 10^{-4} $ 
at $ T \sim 193 $~MeV and $tT = 0.5$. 

For the $U(1)_A$ symmetry, there are tiny breakings at $T = 193 $~MeV and $tT=0.5$
with the maximum value of $ \kappa_{TX} $ equal to $ 3.5(5) \times 10^{-3} $, while 
that of $\kappa_{PS}$ is equal to $2.1(3) \times 10^{-2} $.   
This seems to suggest that the effective restoration of $U(1)_A$ symmetry 
occurs at temperatures higher than 193 MeV. 
To confirm or refute this requires us to determine $\kappa_{PS}$ and $\kappa_{TX}$ 
in the continuum limit, which is beyond the scope of this paper. 

%
%

\begin{figure}[!ht]
\centering
  \caption{
    The $SU(2)_{CS}$ symmetry-breaking and fading parameters $(\kappa_{AT}, \kappa)$ 
    in $N_f=2+1+1$ lattice QCD at the physical point, 
    for $tT=(0.5, 0.25)$ and $T = (193, 240, 307)$~MeV. 
  }
\begin{tabular}{@{}c@{}c@{}}
  \includegraphics[width=8.0cm,clip=true]{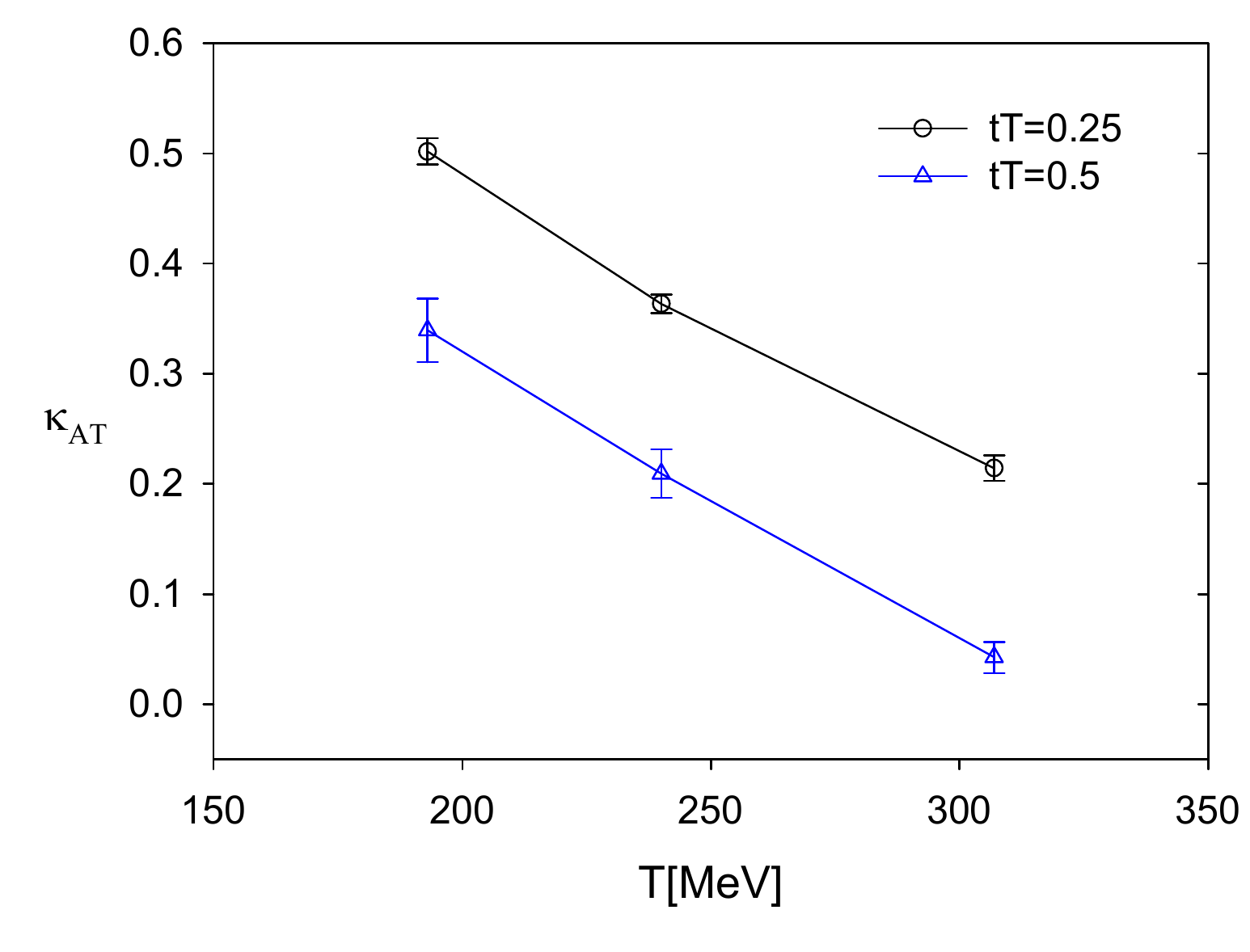}
&
  \includegraphics[width=8.0cm,clip=true]{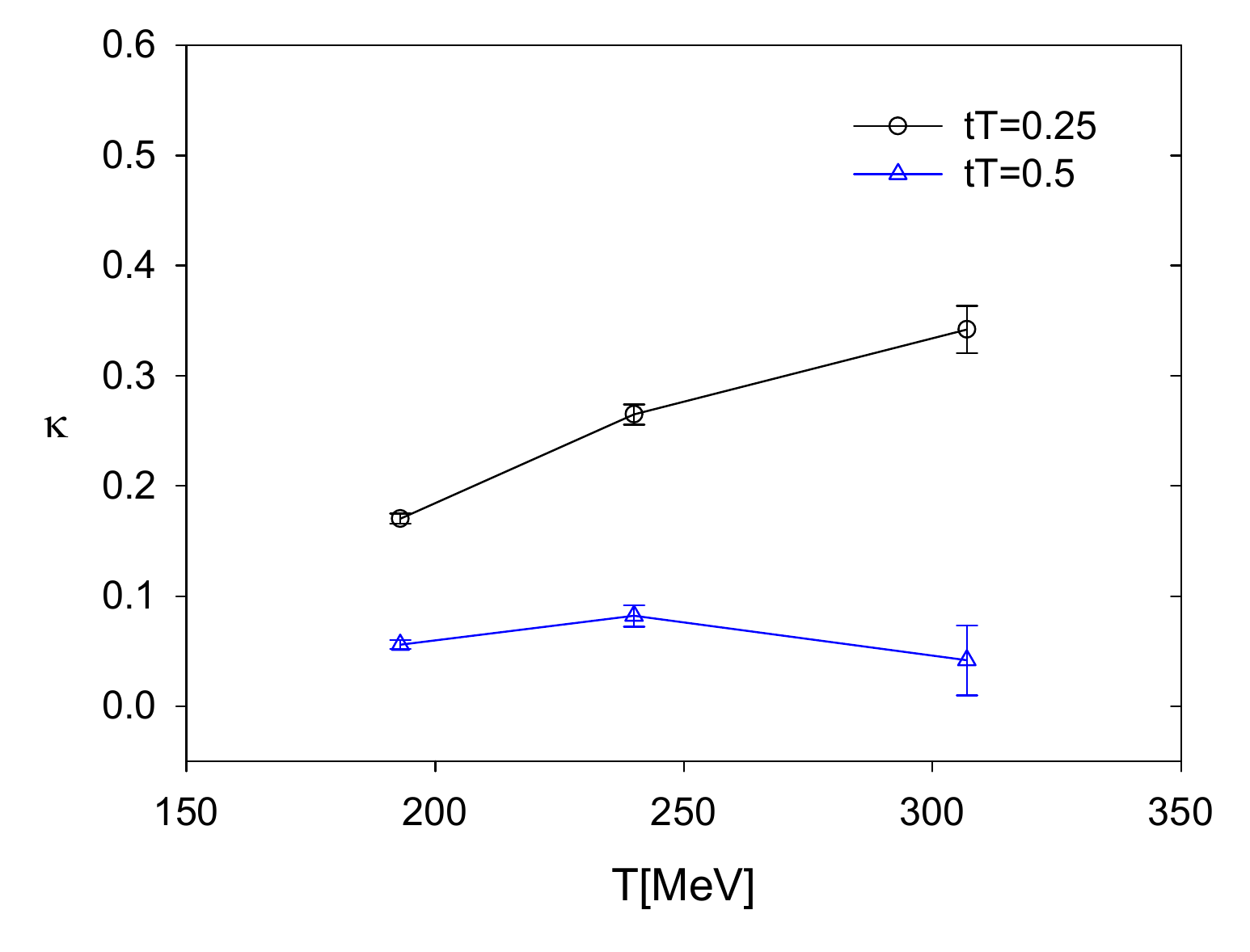}
\end{tabular}
\label{fig:Kat_K_ud_t}
\end{figure}

\begin{table}[!ht]
\caption{The approximate range of temperatures satisfying the criterion [Eq. \ref{eq:SU2_CS_crit_t}] 
with $\epsilon_{CS} = (0.20, 0.10, 0.05, 0.03) $ for $tT=(0.5, 0.25)$.  
In the second column ($tT=0.5$), $T_x$, $T_y$, and $T_z$ have yet to be determined.} 
\centering
\setlength{\tabcolsep}{4pt}
\vspace{2mm}
\begin{tabular}{|ccc|}
\hline
  $\epsilon_{CS}$
  & $tT=0.5$
  & $tT=0.25$ \\
\hline
\hline
0.20 &  $\sim 244$~MeV$-T_x$    & NULL   \\ 
0.10 &  $\sim 280$~MeV$-T_y$    & NULL   \\
0.05 &  $\sim 304$~MeV$-T_z$    & NULL   \\
0.03 &  NULL                    & NULL   \\
\hline
\end{tabular}
\label{tab:Nf2p1p1_ud_t}    
\end{table}

For the $SU(2)_{CS}$ chiral spin symmetry, it turns out to be a rather approximate symmetry 
in comparison with the $U(1)_A$ and $SU(2)_L \times SU(2)_R $ chiral symmetries, as shown 
in the right panels of Fig. \ref{fig:Ct_ud}. 
Also, this can be seen by plotting $\kappa_{AT}$ and $ \kappa $ versus the temperature $T$, 
for $tT = 0.50$ and $tT=0.25$, as shown in Fig. \ref{fig:Kat_K_ud_t}.
Here, the data points at $T=307$~MeV ($N_t=10$) for $tT = 0.25$ are obtained by 
interpolation between $tT=0.2$ and $tT=0.3$. 
For $tT=0.5$, $\kappa_{AT}$ is decreased from $0.34(3)$ to $0.21(2)$ to $0.04(2)$ 
as $T$ is increased from 193 MeV to 307 MeV, 
while $ \kappa$ is changed from $0.056(4)$ to $0.08(1)$ to $0.04(3)$. 
The last data point of $\kappa$ at $T=307$~MeV looks exceptional. 
Presumably, for any fixed $tT$, $\kappa$ is a monotonic increasing function of $T$. 
It is unknown why the data of $\kappa$ at $T=307$~MeV are not a clean cut.   
It could be just due to the finite-size effects of the small $N_t = 10$ in the temporal direction.     
Further investigations are needed to clarify this. 
For $tT=0.25$, $\kappa_{AT}$ is decreased from $0.50(1)$ to $0.36(1)$ to $0.21(1)$ 
as $T$ is increased from 193 MeV to 307 MeV, 
while $ \kappa$ is increased from $0.170(4)$ to $0.26(1)$ to $0.34(2)$.

Now, using the data of $\kappa_{AT}$ and $\kappa$ as plotted in Fig. \ref{fig:Kat_K_ud_t} 
and the criterion in Eq. (\ref{eq:SU2_CS_crit_t}), the ranges of temperatures 
for the emergence of approximate $SU(2)_{CS}$ symmetry can be determined, as tabulated 
in Table \ref{tab:Nf2p1p1_ud_t}, for $tT=(0.5, 0.25)$ 
and $\epsilon_{CS} = (0.20, 0.10, 0.05, 0.03)$.

For $tT=0.5$ in the second column of Table \ref{tab:Nf2p1p1_ud_t}, 
the lower bound of $T$ is increased as the $\epsilon_{CS}$ is decreased,  
and then at $\epsilon_{CS} = 0.03$, the window is shrunk to zero. 
The upper bounds of the window for $\epsilon_{CS} = (0.20, 0.10, 0.05)$ are 
$T_x$, $T_y$ and $T_z$, which have yet to be determined. 
The fact that the window is shrunk to zero for $\epsilon_{CS} \le 0.03 $ 
implies that the $SU(2)_{CS}$ symmetry of the temporal correlators of $u$ and $d$ quarks 
in $N_f=2+1+1$ lattice QCD 
is at most an approximate emergent symmetry, which never becomes an exact symmetry, 
unlike the $U(1)_A \times SU(2)_L \times SU(2)_R $ chiral symmetry, 
which is effectively restored as an exact symmetry for $T > T_1 \gtrsim T_c $.

For $tT = 0.25$ in the third column of Table \ref{tab:Nf2p1p1_ud_t}, 
there are no temperatures satisfying 
the criterion [Eq. \ref{eq:SU2_CS_crit_t}] with $\epsilon_{CS} \le 0.20$. 
Note that the $t$ correlators at $tT=0.25$ (with a small $t$) have large contributions 
from the excited states, thus they may not suitable 
for the criterion of [Eq. \ref{eq:SU2_CS_crit_t}].

\begin{figure}[!ht]
  \centering
  \caption{ 
 The left panels are $t$ correlators of $\bar u \Gamma d $ constructed by the free-quark propagators 
 (see text for details).
    The right panels are the symmetry-breaking parameters 
    ($\kappa_{PS}$, $\kappa_{TX}$, $\kappa_{VA}$, $\kappa_{AT}$ and $\kappa$) 
    corresponding to the $t$ correlators in the left panels. 
  }
\begin{tabular}{@{}c@{}c@{}}
  \includegraphics[width=7.0cm,clip=true]{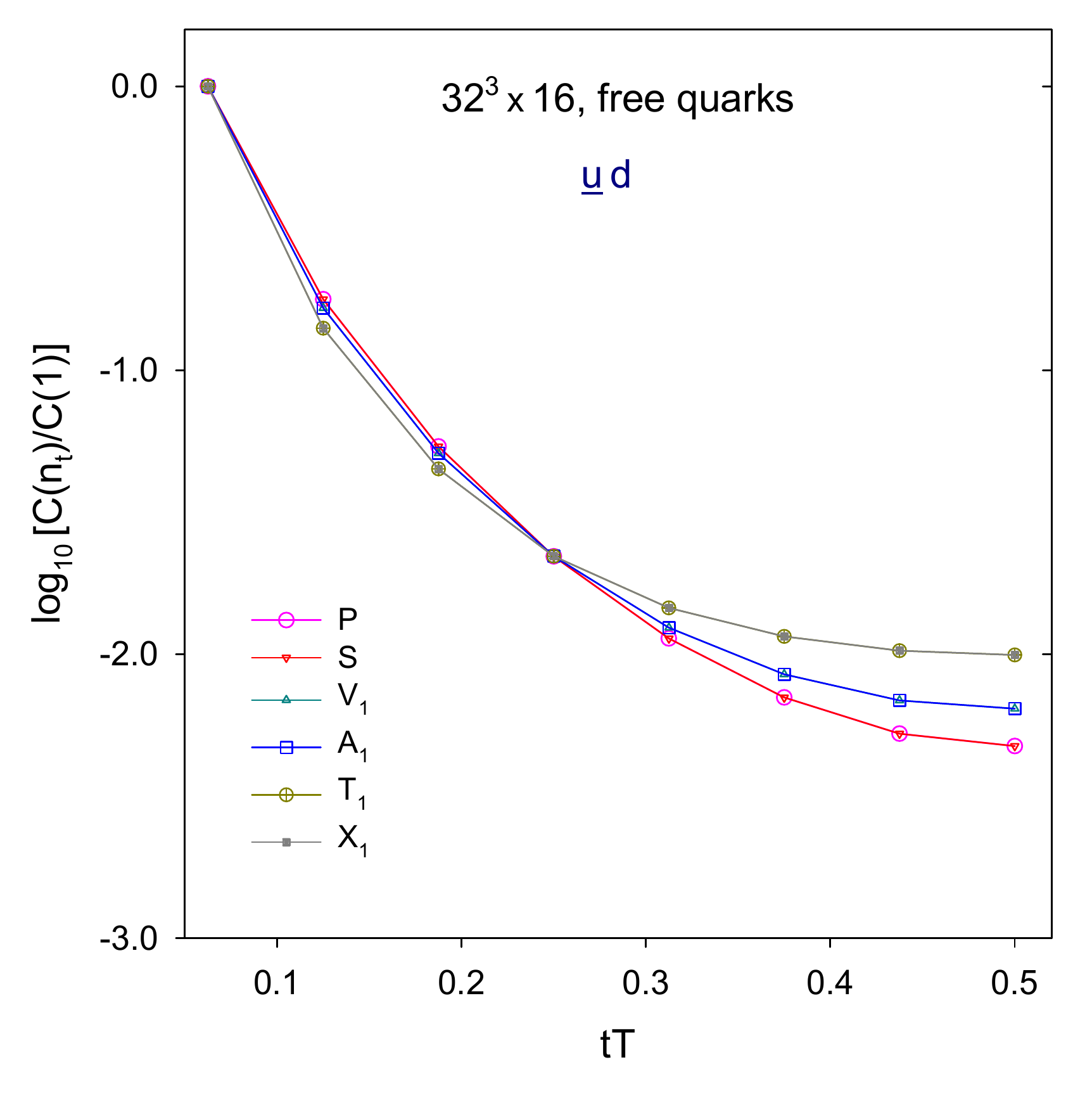}
&
  \includegraphics[width=7.0cm,clip=true]{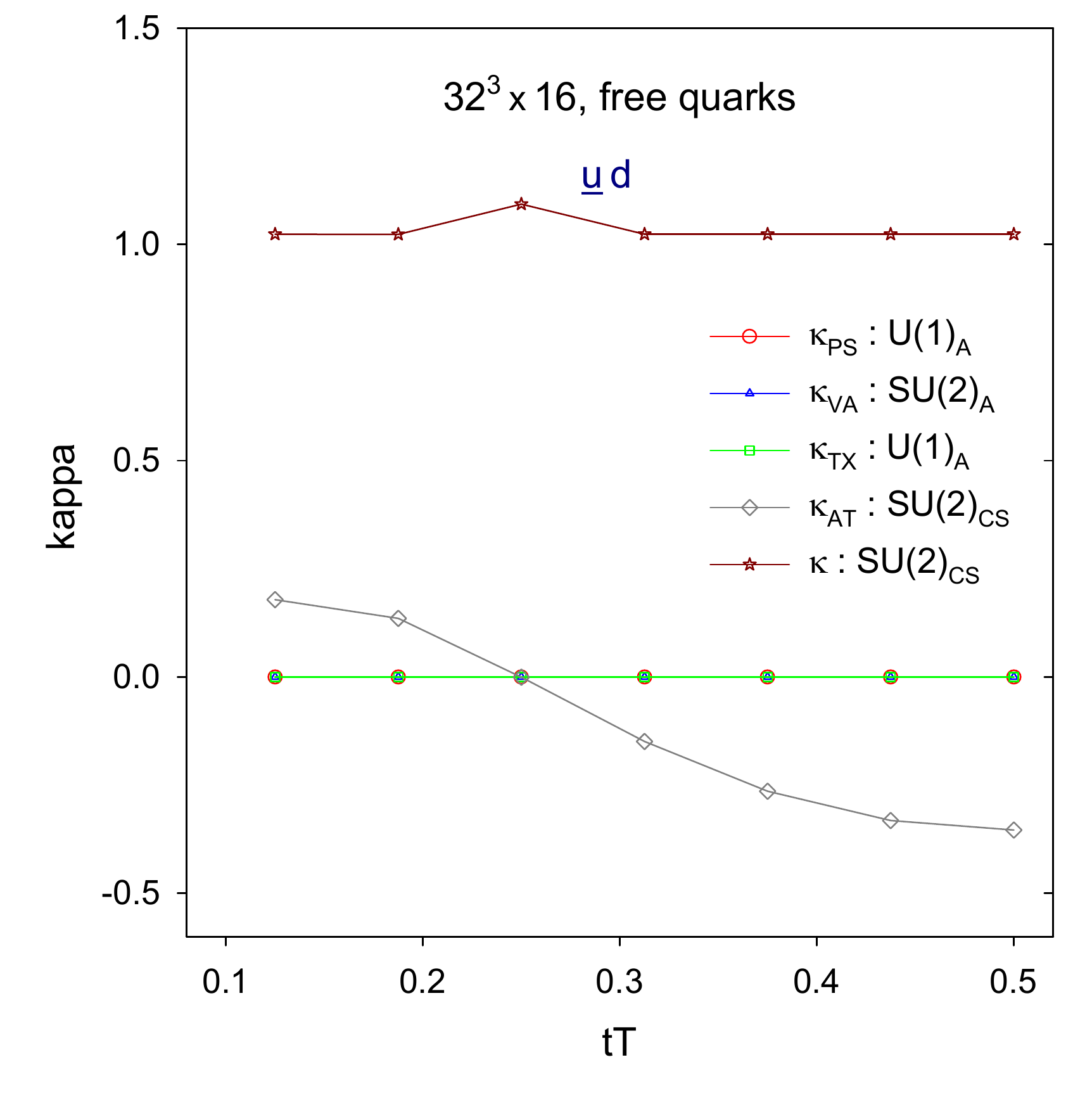}
\\
  \includegraphics[width=7.0cm,clip=true]{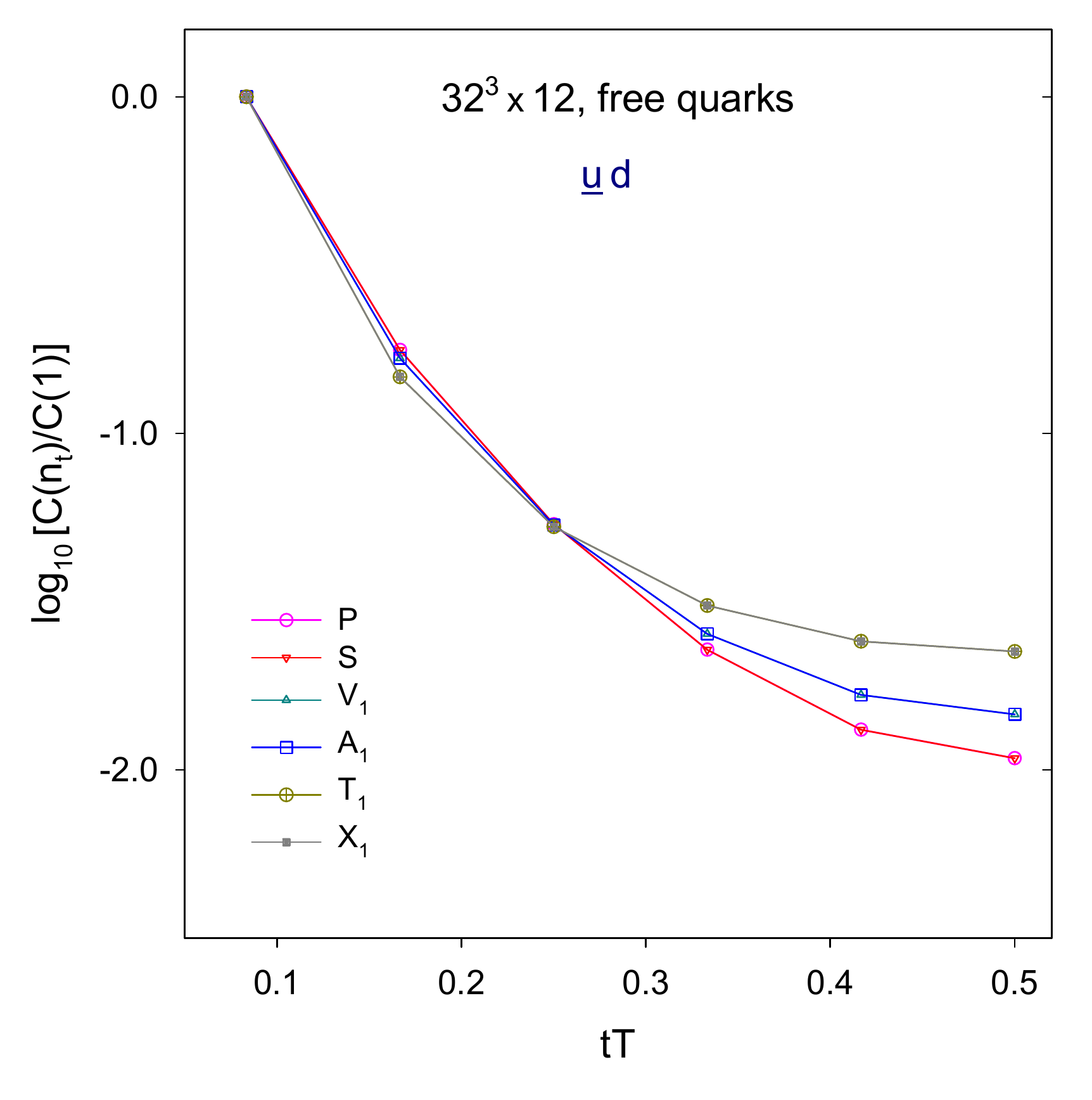}
&
  \includegraphics[width=7.0cm,clip=true]{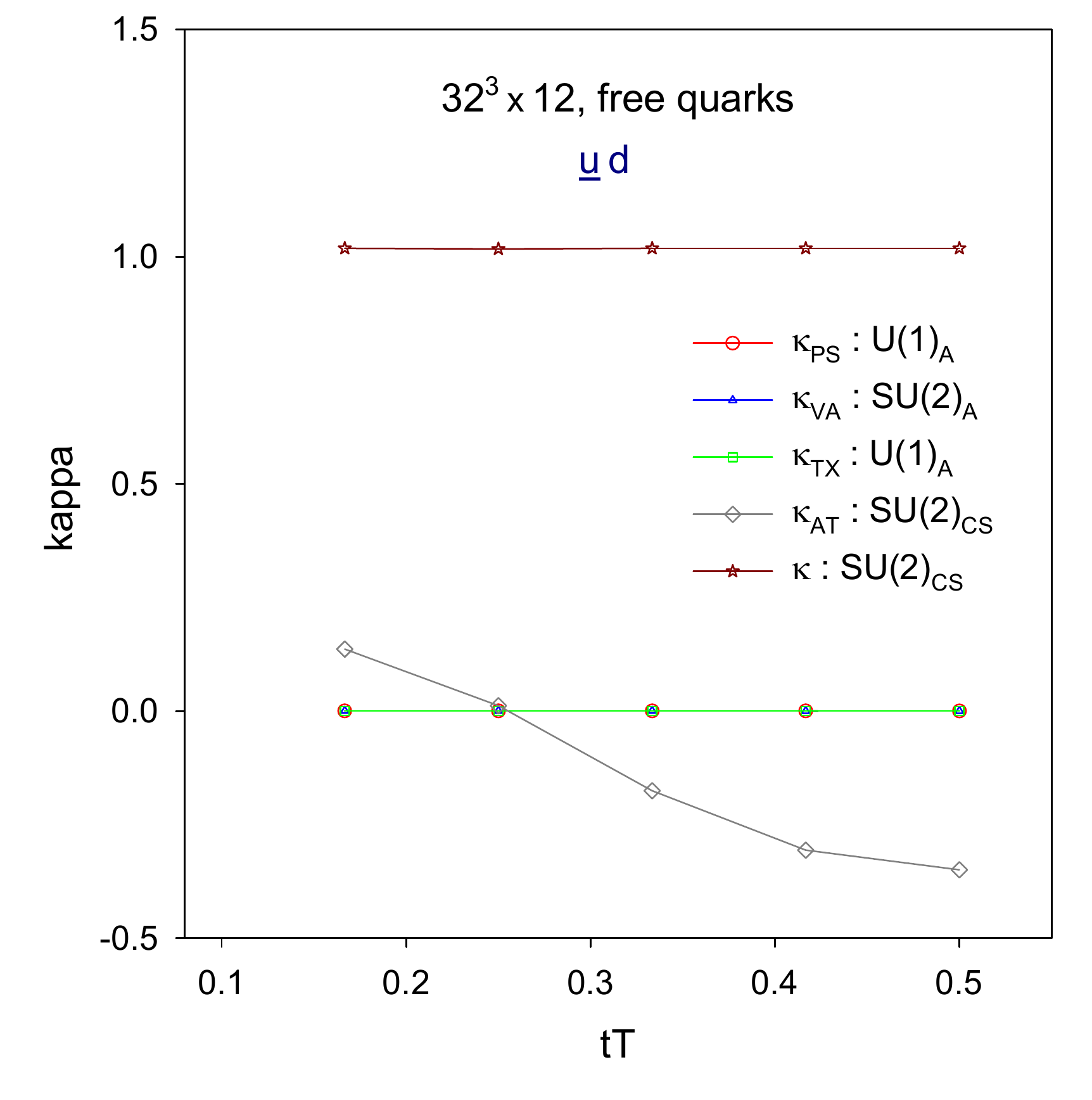}
\\
  \includegraphics[width=7.0cm,clip=true]{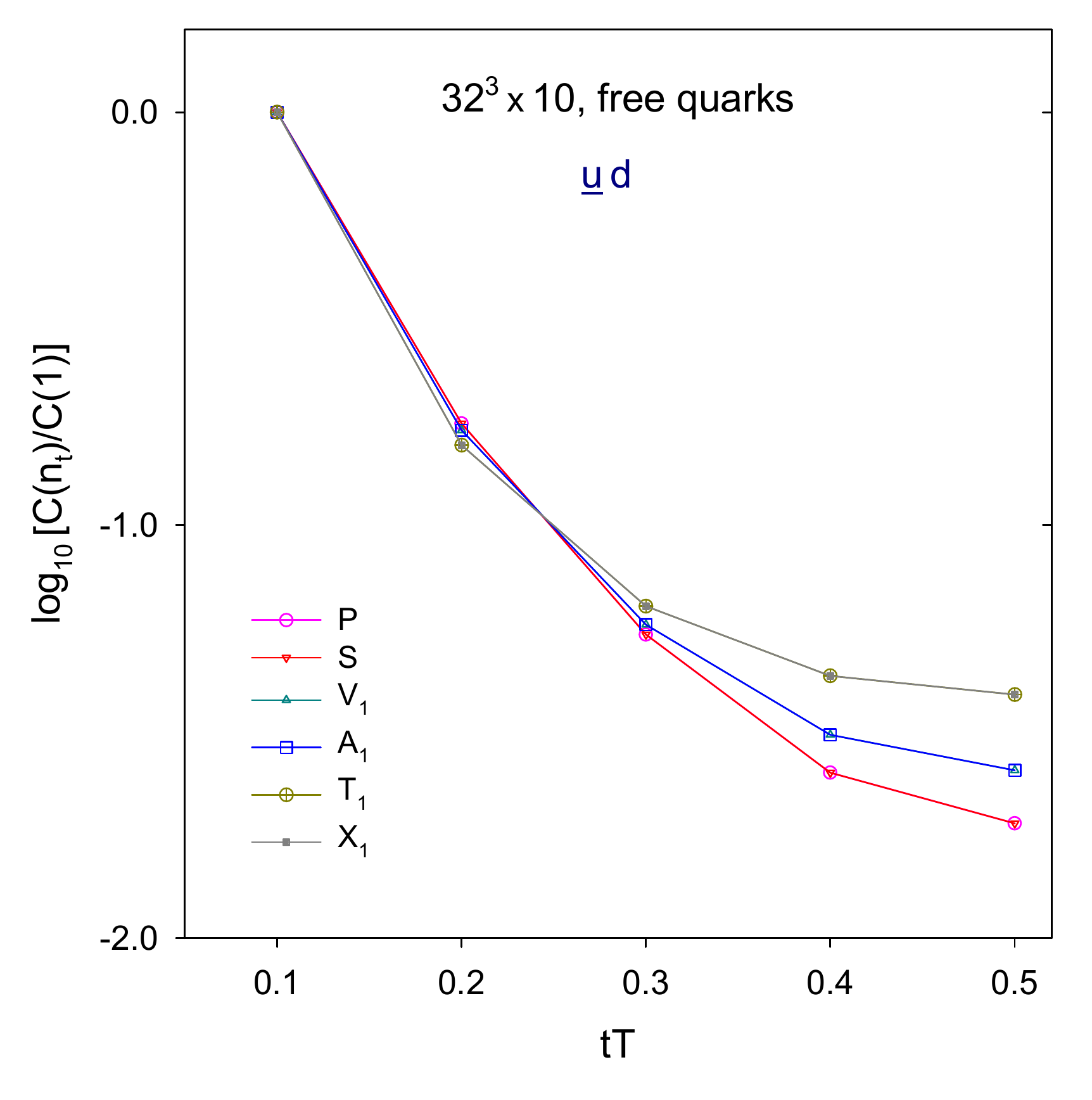}
&
  \includegraphics[width=7.0cm,clip=true]{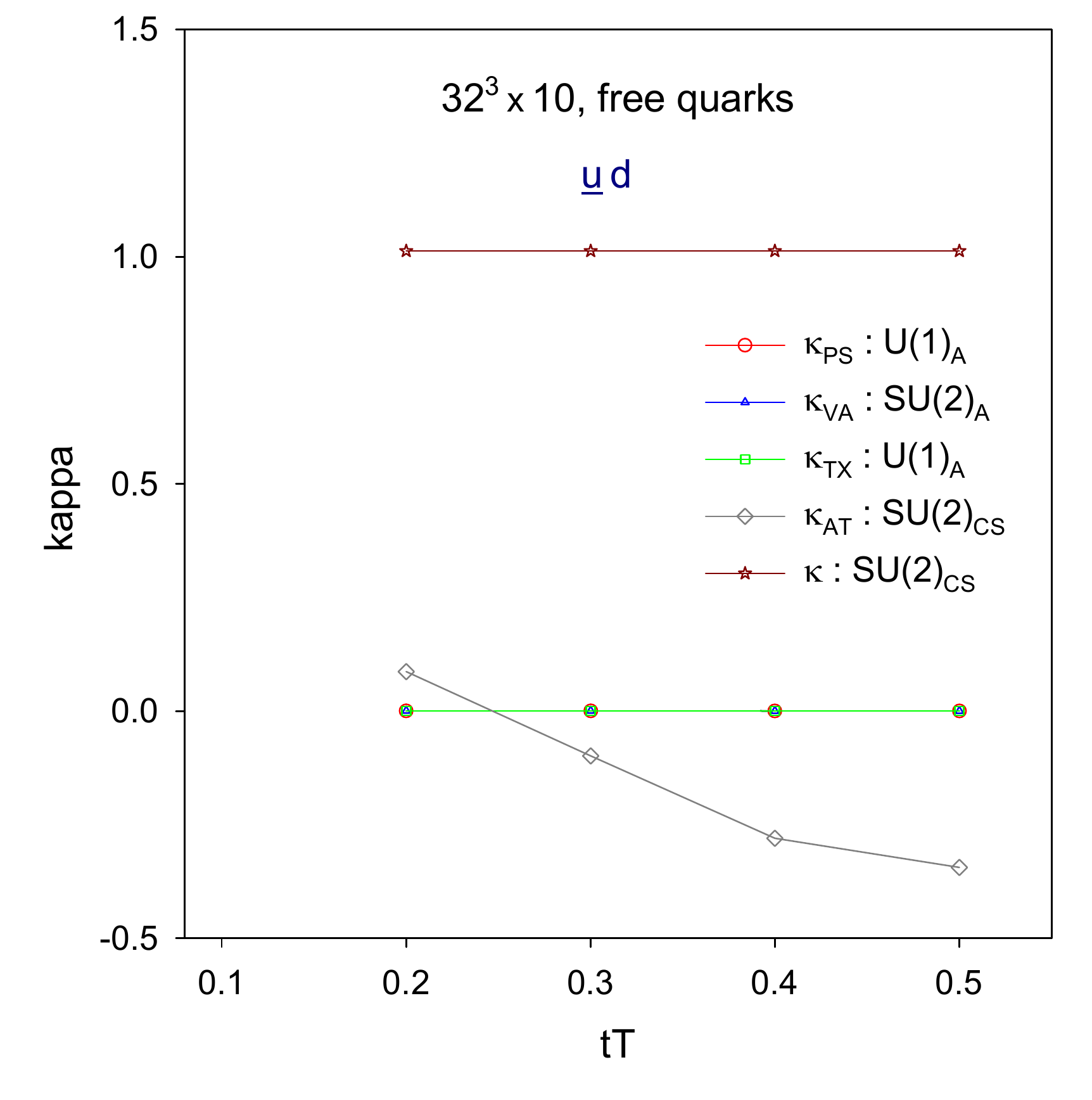}
\end{tabular}
  \label{fig:Ct_ud_free}
\end{figure}

\subsection{Comparison with the noninteracting theory}
\label{free_t}

The $t$ correlators of $\bar u \Gamma d$ constructed with 
free-quark propagators are plotted in the left panels of Fig. \ref{fig:Ct_ud_free}.
The free-quark propagators are computed with the same boundary conditions, the same lattice size,   
and the same $u/d$ quark masses as those in $N_f=2+1+1$ QCD, but  
with all link variables equal to the identity matrix.
Note that the lattice spacing $a$ and the temperature $T=1/(N_t a) $
are not defined for the free quarks. 
Thus, the label $tT$ of the horizontal axis in Fig. \ref{fig:Ct_ud_free}
should be regarded as $ tT = n_t/N_t $. 
In the following, the temperature $T$ for all quantities with free quarks 
is always understood to be the corresponding temperature $T=1/(N_t a) $ in 
$N_f = 2+1+1$ lattice QCD with the same $N_t$.

In the left panels of Fig. \ref{fig:Ct_ud_free}, for all three lattice sizes 
$32^3 \times (16,12,10) $, 
the $U(1)_A \times SU(2)_L \times SU(2)_R$ chiral symmetry is almost exact in spite of the 
nonzero $u/d$ quark masses, as shown by the degeneracies $C_P(t) = C_S(t)$,  
$C_{T_1}(t) = C_{X_1}(t)$, and $ C_{V_1}(t) = C_{A_1}(t)$.  
Consequently, it appears that there are only three distinct $t$ correlators in each left panel 
of Fig. \ref{fig:Ct_ud_free}. They are in the order of 
\bea
\label{eq:Ct_ud_free_A} 
&& C_{P,S}({\rm free}) > C_{V_1, A_1}({\rm free}) > C_{T_1,X_1}({\rm free}), 
\hspace{2mm} {\rm for} \ 1 < n_t < N_t/4, \\
&& C_{P,S}({\rm free}) < C_{V_1, A_1}({\rm free}) < C_{T_1,X_1}({\rm free}), 
\hspace{2mm} {\rm for} \ n_t \ge N_t/4, 
\label{eq:Ct_ud_free_B} 
\eea 
which is different from the order [Eq. \ref{eq:Ct_ud}] of the $N_f=2+1+1$ lattice QCD 
at $ T \sim$ 190-310 MeV. 

Next we examine the symmetries in the $t$ correlators of free quarks  
with the symmetry-breaking parameters as defined in Sec.~\ref{kappa}.
In the right panels of Fig.~\ref{fig:Ct_ud_free}, 
the symmetry-breaking parameters are plotted versus $tT = n_t/N_t$ for $N_t = (16, 12, 10)$. 
For $U(1)_A$ and $SU(2)_L \times SU(2)_R$ chiral symmetries, 
$\kappa_{PS} \simeq \kappa_{TX} \simeq \kappa_{VA} < 10^{-6}$,  
which shows that the $U(1)_A \times SU(2)_L \times SU(2)_R$ chiral symmetry is almost exact 
in the noninteracting theory with free quarks, in spite of the nonzero $u/d$ quark masses.
For the $SU(2)_{CS}$ symmetry, the symmetry-breaking and fading parameters
$\kappa_{AT}(tT)$ and $\kappa(tT)$ are much larger than those 
($\kappa_{PS}$, $ \kappa_{TX}$, and $\kappa_{VA} $) 
of $U(1)_A$ and $SU(2)_L \times SU(2)_R$ chiral symmetries. 
Since $ \kappa(tT) \gtrsim 1 $ for any $tT=n_t/N_t$ and $N_t$, there does not exist any $N_t$  
satisfying the criterion [Eq.~\ref{eq:SU2_CS_crit_t}] with $ \epsilon_{CS} < 1 $.
Thus the $SU(2)_{CS}$ symmetry does not emerge in the noninteracting theory on a lattice, 
in contrast to the $N_f=2+1+1$ lattice QCD at the physical point, 
with the emergence of approximate $SU(2)_{CS}$ symmetry in the windows 
as tabulated in Table~\ref{tab:Nf2p1p1_ud_t}.  
This implies that $u$ and $d$ quarks at these temperatures 
must be dynamically very different from the free or quasifree fermions.
If the deconfined quarks in high-temperature QCD behave like free or quasifree fermions,   
then the $u$ and $d$ quarks in $N_f=2+1+1$ lattice QCD at the temperatures with 
approximate $SU(2)_{CS}$ emergent symmetry are likely to be confined inside hadron-like objects, 
which are predominantly bound by the chromoelectric interactions into color singlets.
Moreover, since the emergent $SU(2)_{CS}$ symmetry is not an exact symmetry,           
the role of chromomagnetic interactions in forming these hadron-like objects cannot be neglected.

\subsection{Comparison with the $N_f=2$ lattice QCD}

In Ref.~\cite{Rohrhofer:2019qal}, the symmetries of temporal correlators of $\bar u \Gamma d$ 
were studied in $N_f=2$ lattice QCD at $T = 220$~MeV with M\"obius domain-wall fermions, 
on the $48^3 \times 12$ lattice with lattice spacing $a=0.075$~fm. 

Comparing the $t$ correlators of $N_f=2+1+1 $ lattice QCD at $ T = 240 $~MeV 
(in the middle-left panel of Fig.~\ref{fig:Ct_ud}) with those of $N_f=2 $ lattice QCD 
at $ T = 220 $~MeV \cite{Rohrhofer:2019qal}, 
we see that in both cases, the order of Eq. (\ref{eq:Ct_ud}) is satisfied, 
and $U_A(1) $ and $SU(2)_L \times SU(2)_R $ chiral symmetries are effectively restored. 
However, the $SU(2)_{CS} $ symmetry breakings in $N_f=2+1+1$ lattice QCD are larger than those in  
$N_f=2$ lattice QCD. This can be seen from the approximately degenerate multiplets 
$(V_1, A_1)$ and $(T_1, X_1)$ in the middle-left panel of Fig.~\ref{fig:Ct_ud} 
versus the highly degenerate multiplets $(V_1, A_1)$ and $(T_1, X_1)$ 
in the right panel of Fig. 2 in Ref. \cite{Rohrhofer:2019qal}.
Consequently, the values of $\kappa_{AT}$ (\ref{eq:k_AT_t}) 
of $N_f=2+1+1$ lattice QCD (as shown on the middle-right panel of Fig. \ref{fig:Ct_ud}) 
are larger than their counterparts of $N_f=2$ lattice QCD  
(which are not shown explicitly in Ref.~\cite{Rohrhofer:2019qal}).

\begin{figure}[!ht]
  \centering
  \caption{
    Comparision of the $SU(2)_{CS}$ symmetry-fading parameter $\kappa$ 
    between $N_f = 2+1+1$ lattice QCD at $T=(193, 240)$~MeV (this work) 
    and $N_f=2$ lattice QCD at $T=220$~MeV \cite{Rohrhofer:2019qal}.
  }
  \includegraphics[width=9.0cm,clip=true]{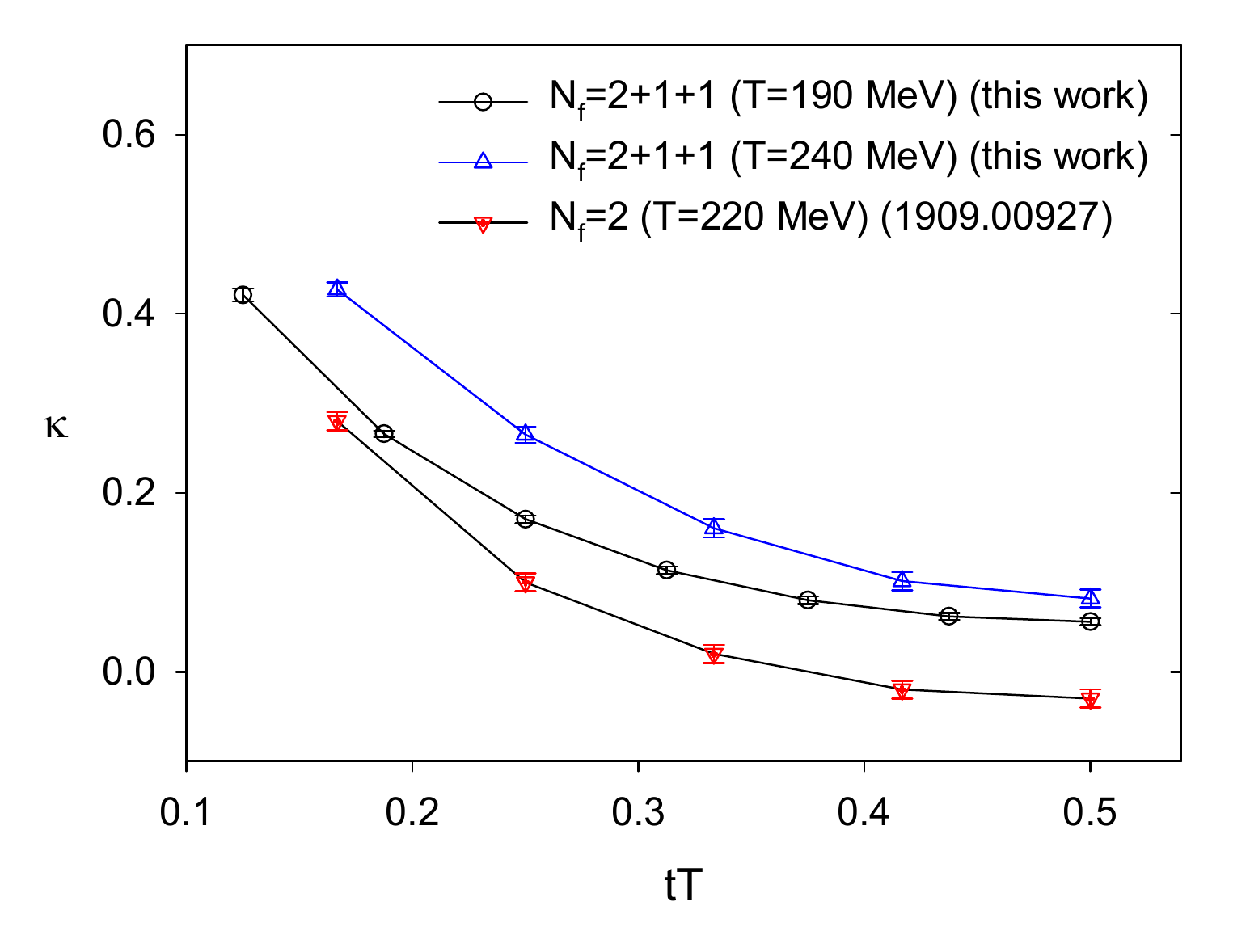}
  \label{fig:K_compare_t}
\end{figure}

Next, we compare the $ SU(2)_{CS} $ symmetry-fading parameter $\kappa$ (\ref{eq:kappa_t}) 
between $N_f=2+1+1$ and $N_f=2$ lattice QCD. 
In Fig.~\ref{fig:K_compare_t}, 
the values of $\kappa$ are plotted for $N_f=2$ lattice QCD at $T = 220$~MeV  
(which are read off from Fig.~3 of Ref.~\cite{Rohrhofer:2019qal},  
 after multiplying $(-1)$ due to different definitions of $\kappa$),  
and also for $N_f = 2+1+1$ lattice QCD at $T = (193, 240)$~MeV
(same as the values of $\kappa$ in the right panels of Fig.~\ref{fig:Ct_ud}).
Evidently, the $\kappa$ values of $N_f=2+1+1$ lattice QCD at $T =(193, 240)$~MeV 
are larger than those of $N_f=2$ lattice QCD at $T = 220$~MeV.

\section{Spatial correlators of $ \bar u \Gamma d $} 
\label{Cz_ud}

\subsection{\bf The issue of unphysical meson states and its resolution} 
\label{issues}

\begin{figure}[!ht]
  \centering 
  \caption{The normalized $z$ correlators of meson interpolators $\bar u \Gamma d$ 
           on the $32^3 \times 16 $ lattice at $T = 193 $~MeV (left panel), 
           and their counterparts constructed with the free-quark propagators (right panel). 
           The quark propagators are computed with periodic boundary conditions   
           in the $(x,y,z)$-directions and an antiperiodic boundary condition in the $t$ direction.  
  }
  \begin{tabular}{@{}c@{}c@{}}
  \includegraphics[width=7.5cm,clip=true]{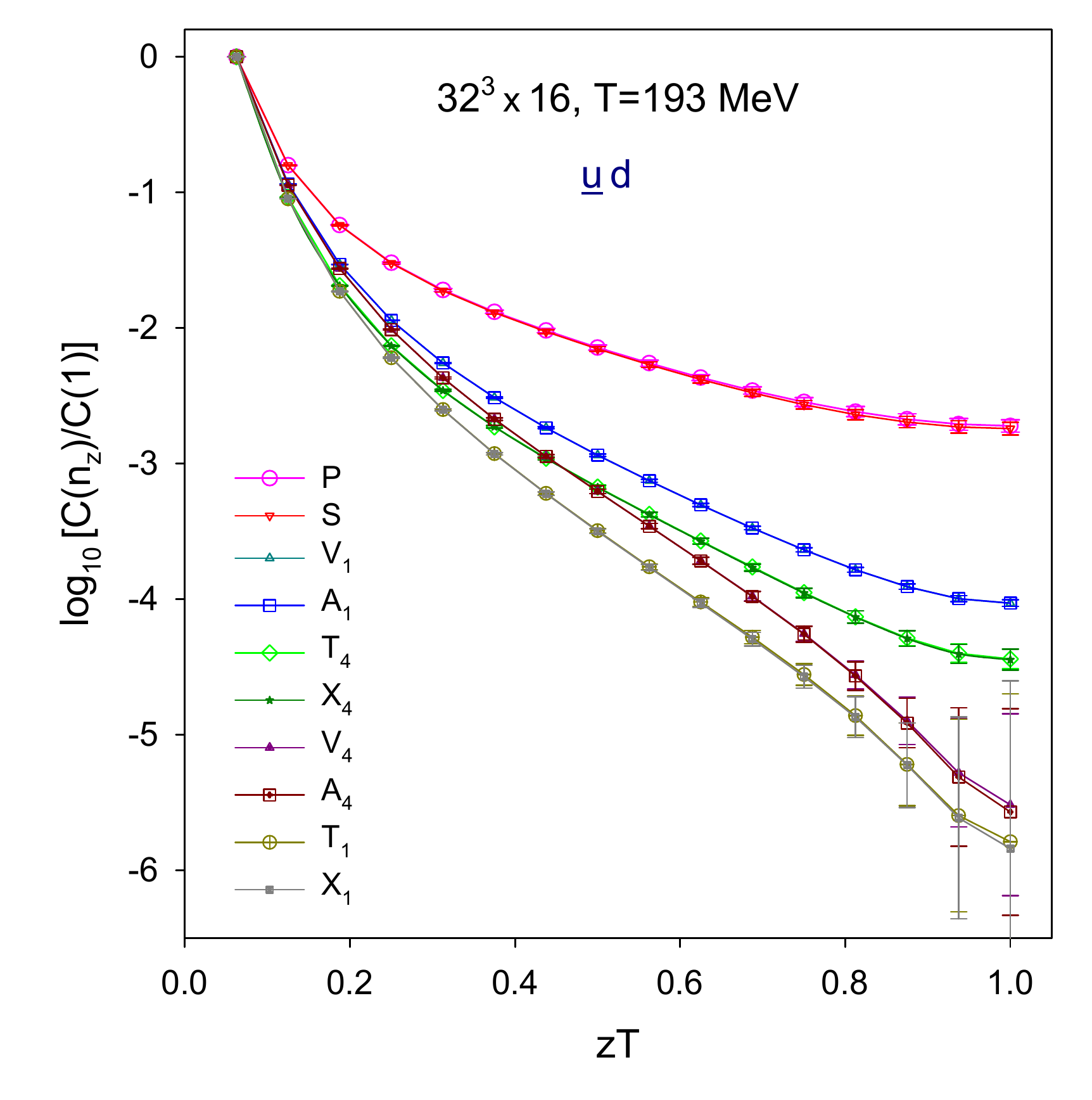}
&
  \includegraphics[width=7.5cm,clip=true]{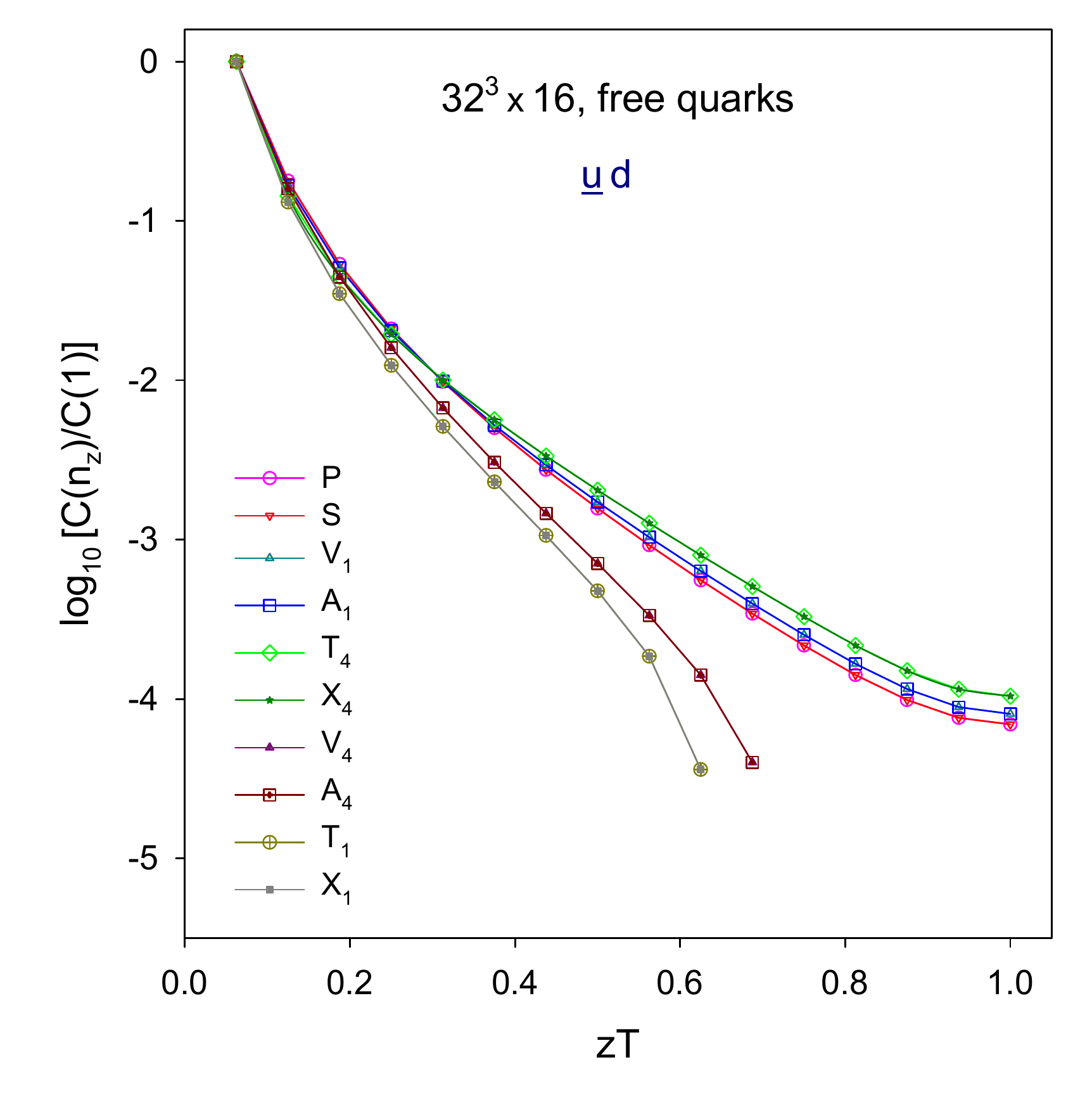} 
  \end{tabular}
\label{fig:Cz_l32t16_b620_ud}
\end{figure}

In Fig. \ref{fig:Cz_l32t16_b620_ud}, the normalized $z$ correlators 
of $ \bar u \Gamma d $ (see Table \ref{tab:bilinear}) 
at $T$ = 193 MeV are plotted in the left panel, while their counterparts  
constructed with the-free quark propagators are plotted in the right panel. 
Here, the normalized $z$ correlators are plotted as a function of the 
dimensionless variable $zT$ (\ref{eq:zT}).
Due to the degeneracy (the $S_2$ symmetry) of the ``1" and ``2" components 
in the $z$ correlators of vector meson interpolators, only the ``1" components are plotted.

We note that $C_{V_4,A_4} $ and $C_{T_1, X_1}$ at large distances with $n_z \ge 12 $ 
are seriously distorted by the contribution of unphysical meson states,  
which have the opposite sign from physical meson states.  Consequently, 
the cancellation between the contributions of the physical and the unphysical meson states 
produces large statistical errors for $C_{V_4,A_4} $ and $C_{T_1, X_1}$ at $n_z \ge 12$. 
In the case of free quarks, the issue of unphysical meson states 
is even more serious, as shown in the right panel of Fig. \ref{fig:Cz_l32t16_b620_ud}, 
in which $C_{V_4,A_4} < 0 $  for $ n_z \ge 12 $, and $C_{T_1, X_1} < 0 $ for $ n_z \ge 11 $. 
The issue due to the unphysical meson states is also visible in the meson spatial correlators 
of $N_f=2$ lattice QCD \cite{Rohrhofer:2019qwq}, and it was discussed in Ref. \cite{Glozman:2020qvo}.

The unphysical meson states are essentially due to the superposition of $+\hat{z}$ (forward) and 
$-\hat{z}$ (backward) running quark propagators, which are nothing but the finite size effects. 
%
%
Since the unphysical meson states change sign if the boundary condition in the $z$ direction 
is changed from periodic to antiperiodic, this  
leads to the following prescription for eliminating the contribution of 
unphysical meson states to the spatial $z$ correlators. 

\begin{figure}[!ht]
  \centering 
  \caption{The contribution of unphysical meson states to the $z$ correlators 
           in Fig. \ref{fig:Cz_l32t16_b620_ud} are eliminated with the proposed precription. 
           Here, each $z$ correlator is the average of two $z$ correlators constructed 
           from two sets of quark propagators with periodic and antiperiodic boundary conditions 
           in the $z$ direction. (See text for details.)
  }
  \begin{tabular}{@{}c@{}c@{}}
  \includegraphics[width=7.5cm,clip=true]{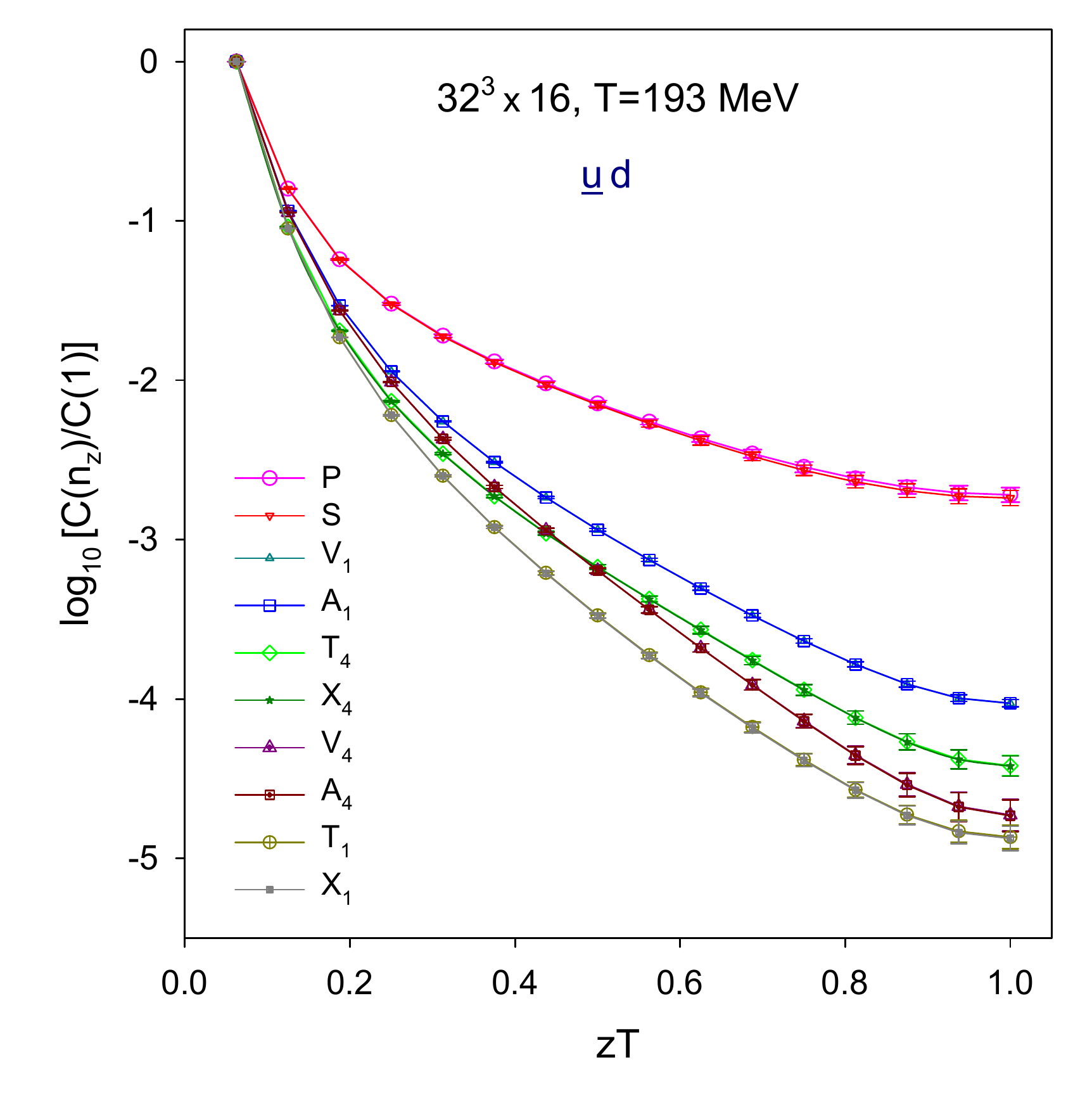}
&
  \includegraphics[width=7.5cm,clip=true]{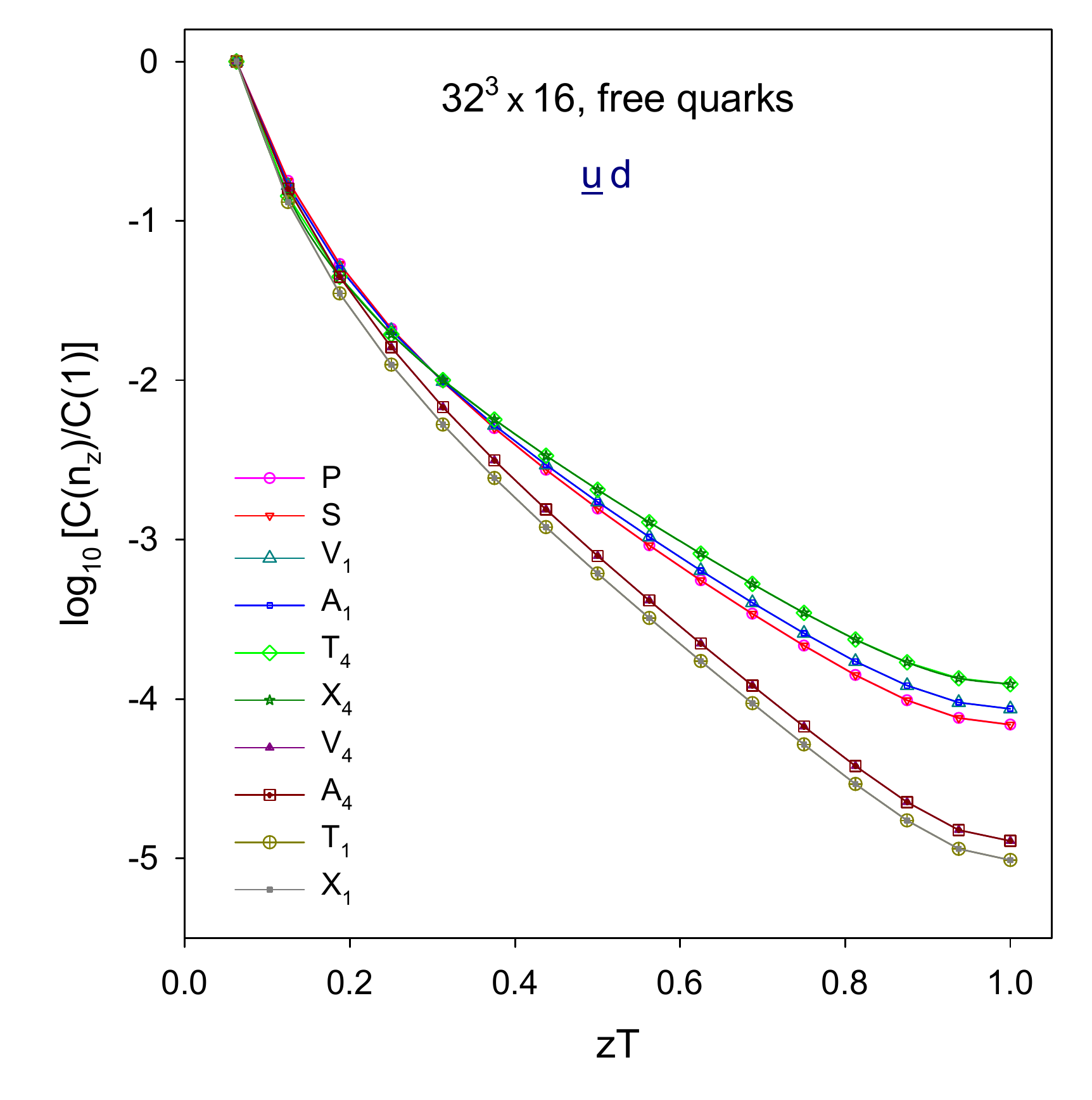} 
  \end{tabular}
\label{fig:Cz_l32t16_b620_ud_av4}
\end{figure}

First, we compute two sets of quark propagators with periodic and antiperiodic
boundary conditions in the $z$ direction, while their boundary conditions in $(x,y,t)$ directions 
are the same: i.e., periodic in the $(x,y)$ directions, and antiperiodic in the $t$ direction.  
Each set of quark propagators are used to construct the $z$ correlators independently, 
and finally taking the average of these two $z$ correlators. 
Then, the contribution of unphysical meson states to the $z$ correlators 
can be cancelled configuration by configuration, 
up to the numerical precision of the quark propagators. 
Using this prescription, the averaged $z$ correlators of $ \bar u \Gamma d $
at $T = 193 $~MeV are plotted in the left panel of Fig. \ref{fig:Cz_l32t16_b620_ud_av4}, 
while their counterparts constructed with the free quark propagators are plotted in the right panel. 
Evidently, the contributions of unphysical meson states are eliminated 
in both $N_f=2+1+1$ lattice QCD and the noninteracting theory with free quarks. 
Note that there is another viable prescription for eliminating the unphysical meson states, 
which will be discussed in Sec. \ref{conclusion}.

\subsection{Results of $N_f=2+1+1$ lattice QCD}
\label{Cz_ud_A}

\begin{figure}[!ht]
  \centering
  \caption{
  The spatial $z$ correlators of $ \bar u \Gamma d $ in $N_f=2+1+1$ lattice QCD for 
  $T \simeq 190-770 $~MeV.
  }
  \begin{tabular}{@{}c@{}c@{}}
  \includegraphics[width=7.0cm,clip=true]{Cz_l32t16_b620_ud.pdf}
&
  \includegraphics[width=7.0cm,clip=true]{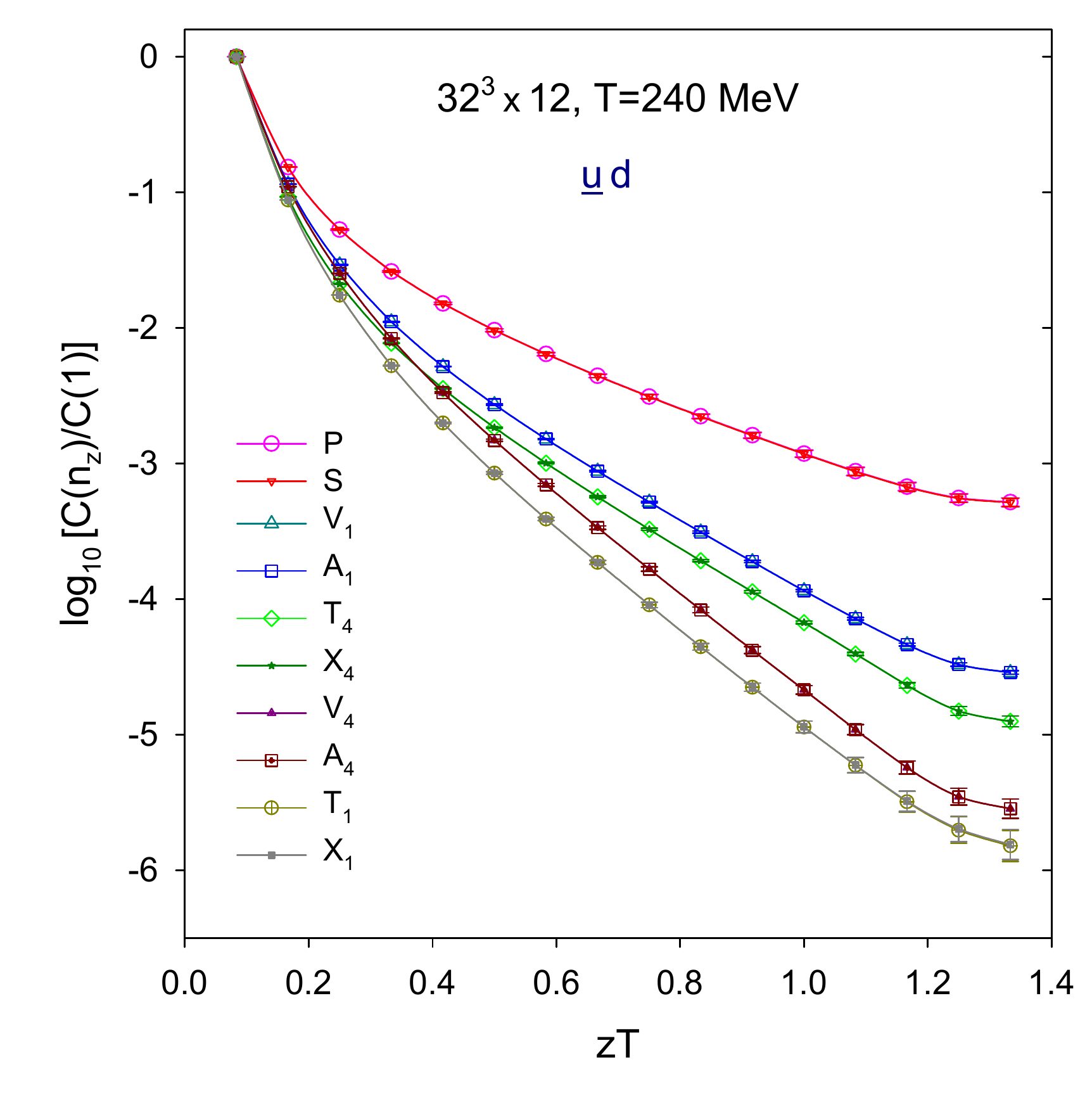}
\\
  \includegraphics[width=7.0cm,clip=true]{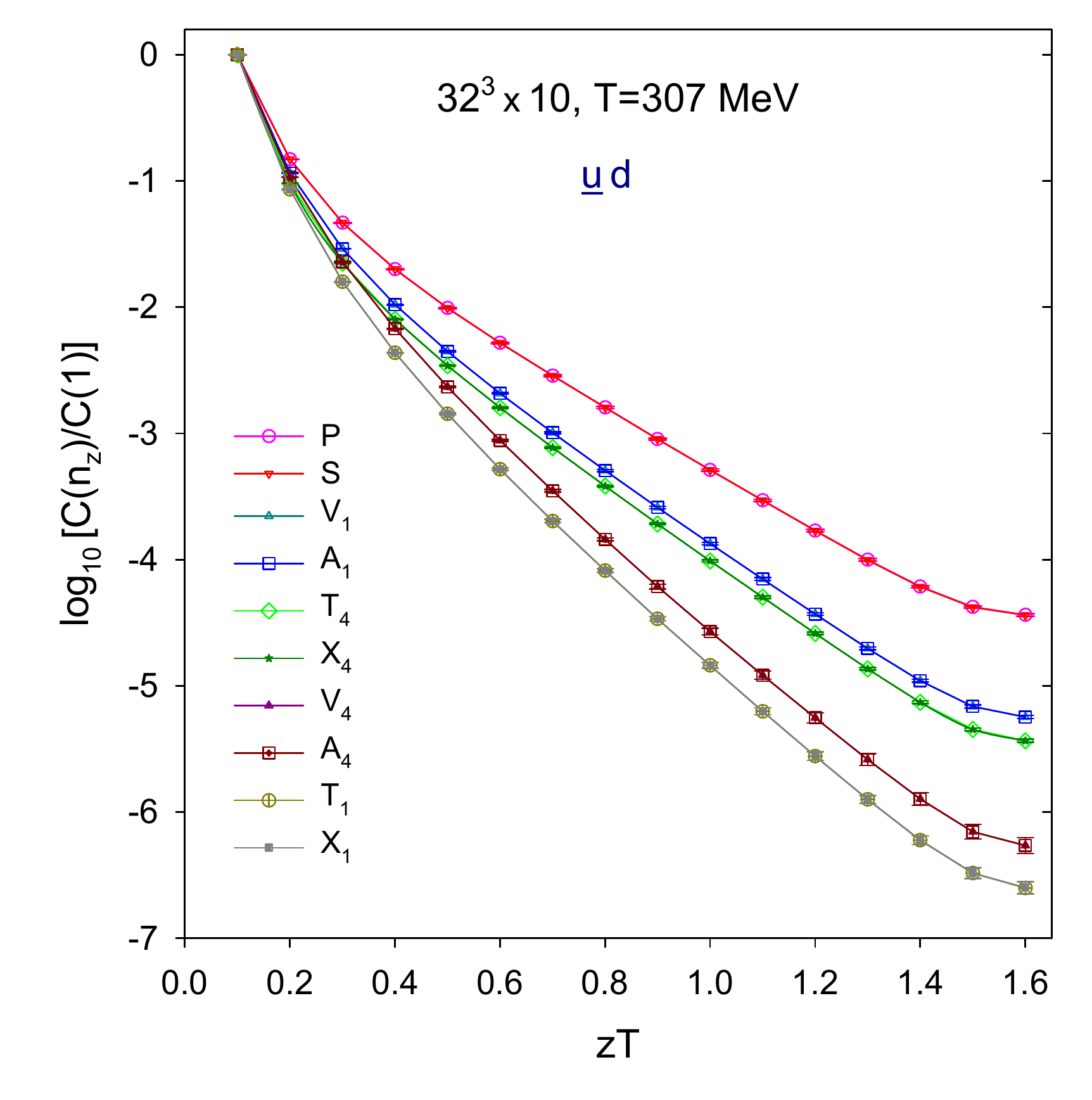}
&
  \includegraphics[width=7.0cm,clip=true]{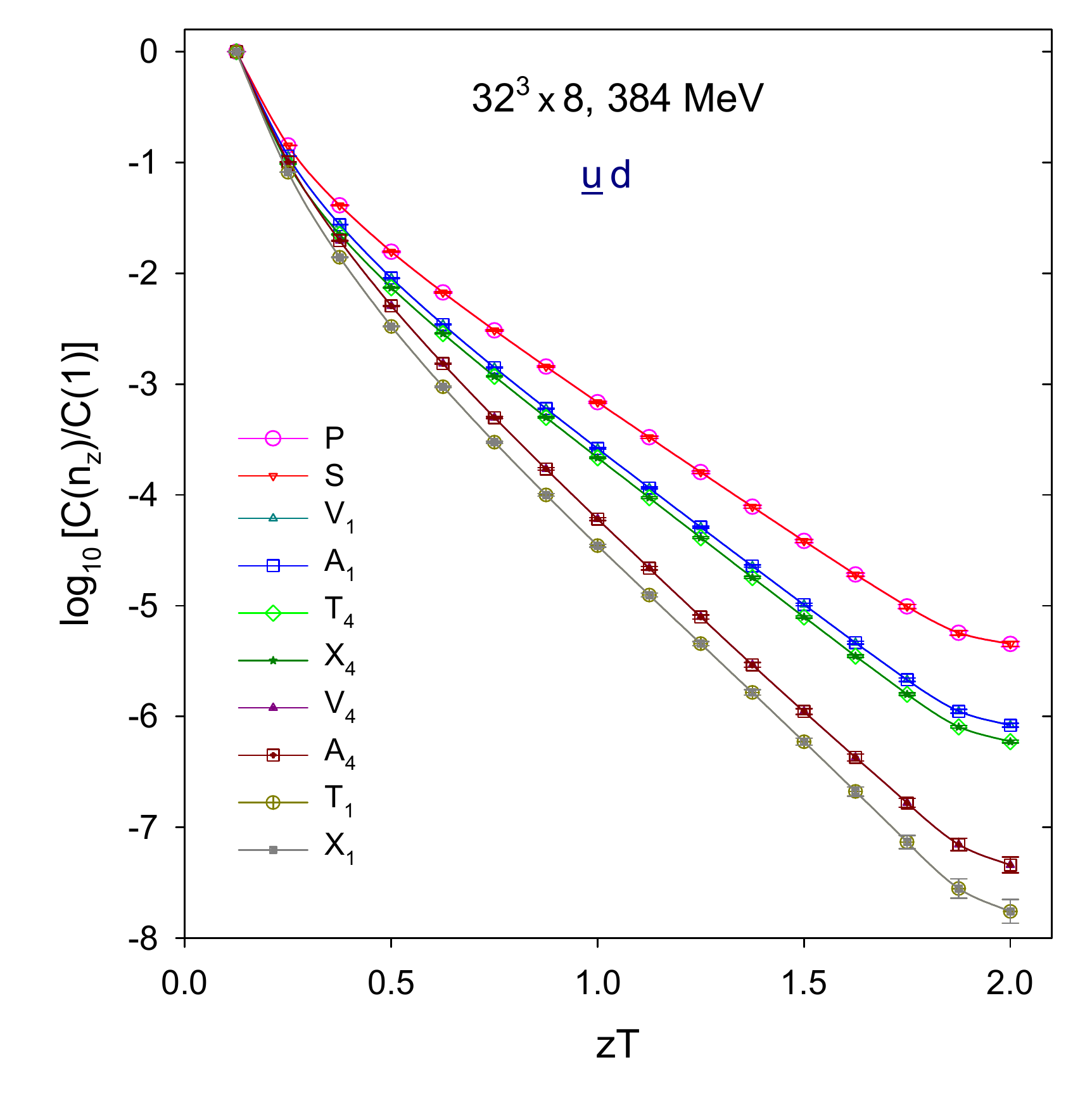}
\\
  \includegraphics[width=7.0cm,clip=true]{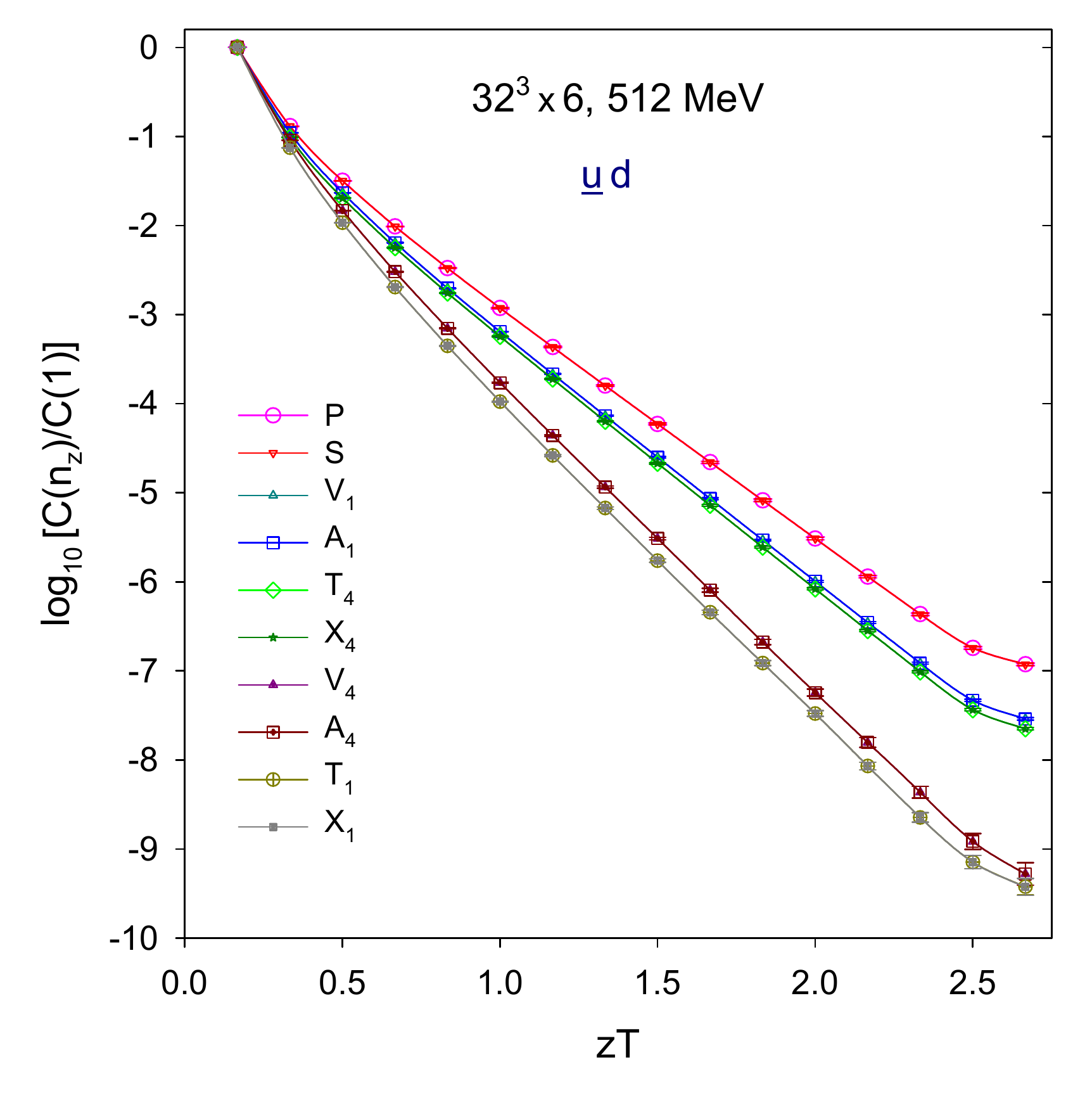}
&
  \includegraphics[width=7.0cm,clip=true]{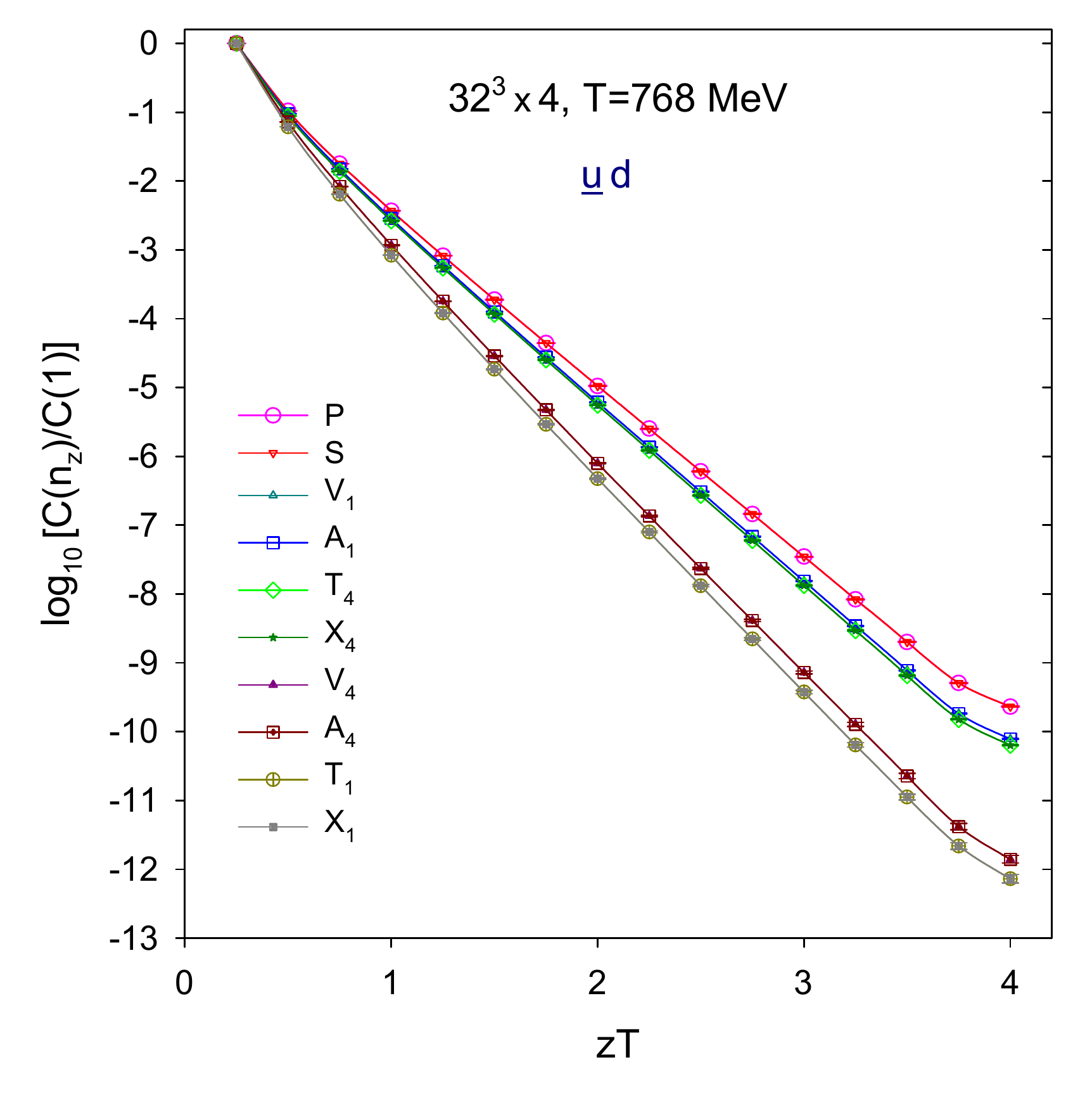}
\\
  \end{tabular}
\label{fig:Cz_ud}
\end{figure}

In the following section, for the spatial $z$ correlators, we always use the average 
of two $z$ correlators constructed from two sets of quark propagators 
with periodic and antiperiodic boundary conditions in the $z$ direction.
In each panel of Fig. \ref{fig:Cz_ud}, the normalized $z$ correlators of $\bar u \Gamma d $ 
(see Table \ref{tab:bilinear}) are plotted as a function of the dimensionless variable 
$zT$ [Eq. \ref{eq:zT}].
Due to the degeneracy (the $S_2$ symmetry) of the ``1" and ``2" components 
in the $z$ correlators of vector mesons, only the ``1" components are plotted.

For all six temperatures in the range $T \sim 190-770$~MeV, 
the $SU(2)_L \times SU(2)_R$ chiral symmetry is effectively restored,   
as manifested in the degeneracy: $ C_{V_1}(z) = C_{A_1}(z)$. 
Moreover, the $U(1)_A$ symmetry is effectively restored,  
as manifested in the degeneracies $C_P(z) = C_S(z)$ 
(except for the small breakings at large $z$ at $T = 193$~MeV), 
and $C_{T_1}(z) = C_{X_1}(z)$.   

Due to the effective restoration of $U(1)_A$ and $SU(2)_L \times SU(2)_R$ chiral symmetries, 
it appears that there are only five distinct $z$ correlators in each panel of Fig. \ref{fig:Cz_ud}.  
They are in the order 
\bea 
\label{eq:Cz_ud}
C_{P, S} > C_{V_1, A_1} > C_{T_4, X_4} > C_{V_4, A_4} > C_{T_1, X_1}, \ {\rm for} \ n_z \ge 7.   
\eea
Note that there is a ``level crossing" in the channels of $(T_4, X_4)$  and $ (V_4, A_4)$ 
at $ T = 193 $~MeV: namely,   
$ C_{T_4, X_4} < C_{V_4, A_4} $ for $ 1 < n_z \le 7 $, while        
$ C_{T_4, X_4} > C_{V_4, A_4} $ for $ n_z > 7 $.       

As the temperature is increased from 193 MeV to 768 MeV, 
we see the emergence of three distinct multiplets,   
\BAN 
&& M_0 = ( P, S ), \\
&& M_2 = ( V_1, A_1, T_4, X_4 ), \\
&& M_4 = ( V_4, A_4, T_1, X_1 ), 
\EAN  
which become more pronounced at higher temperatures.
Note that the emergence of the multiplets $M_2$ and $M_4$ is     
in agreement with the $SU(2)_{CS}$ multiplets [Eqs. (\ref{eq:SU2CS_S2_a}) and (\ref{eq:SU2CS_S2_b})] 
and the $SU(4)$ multiplets [Eq. (\ref{eq:SU4_S2_a}) and (\ref{eq:SU4_S2_b})]. 
This suggests the emergence of the approximate $SU(2)_{CS}$  
and $SU(4)$ symmetries for $T \sim $ 380-770 MeV. 
Moreover, the splitting between the multiplets $M_2$ and $M_0$ 
is decreased as the temperature is increased.
Thus, at sufficiently high-temperatures, 
$M_2$ and $M_0$ would merge together to form a single multiplet,  
and then the approximate $SU(2)_{CS}$ and $SU(4)$ symmetries become washed out, and only the 
$U(1)_A \times SU(2)_L \times SU(2)_R $ chiral symmetry remains. 
In other words, the approximate $SU(2)_{CS}$ and $SU(4)$ symmetries  
can only appear in a range of temperatures above $T_c$, 
say $T_c < T_{cs} \lesssim T \lesssim T_f $, 
where $T_{cs} $ and $T_f$ depend on the $\epsilon_{CS}$ in the criterion 
[Eq. (\ref{eq:SU2_CS_crit_z})] 
for the emergence of approximate $SU(2)_{CS}$ symmetry in the $z$ correlators.

Note that the multiplet $M_4$ never merges with the multiplets $M_0$ and $M_2$, 
even in the limit $T \to \infty $ (the noninteracting theory with free quarks). 
This can be seen as follows.  
In the noninteracting theory, the $z$ correlators of $M_4$ have different asymptotic behaviors 
from those of $M_0$ and $M_2$, namely
\bea
\label{eq:M02_free}
\lim_{z \to \infty}  C_{P, S, V_1, A_1, T_4, X_4} (z) &\to&  c_0 \frac{e^{-2 \pi  z T }}{z}, \\   
\lim_{z \to \infty}  C_{V_4, A_4, T_1, X_1} (z)  &\to&  c_4 \frac{e^{- 2 \pi z T}}{z^2}, 
\label{eq:M4_free}
\eea 
where $c_0$ and $c_4$ are fixed by the normalization $C_\Gamma (n_z = 1) = 1$.
Evidently, Eq.~(\ref{eq:M4_free}) never merges with Eq.~(\ref{eq:M02_free}), 
which can be easily seen by plotting $\log [C_\Gamma(z) ]$ versus $z$.  
Thus, turning on the QCD interactions must make $M_4$ further apart from 
$M_0$ and $M_2$.

\begin{figure}[!ht]
  \centering
  \caption{
   The symmetry-breaking parameters of $z$ correlators of $ \bar u \Gamma d $ 
   in $N_f=2+1+1$ lattice QCD for six temperatures in the range $ \sim 190-770$~MeV.
  }
  \begin{tabular}{@{}c@{}c@{}}
  \includegraphics[width=7.0cm,clip=true]{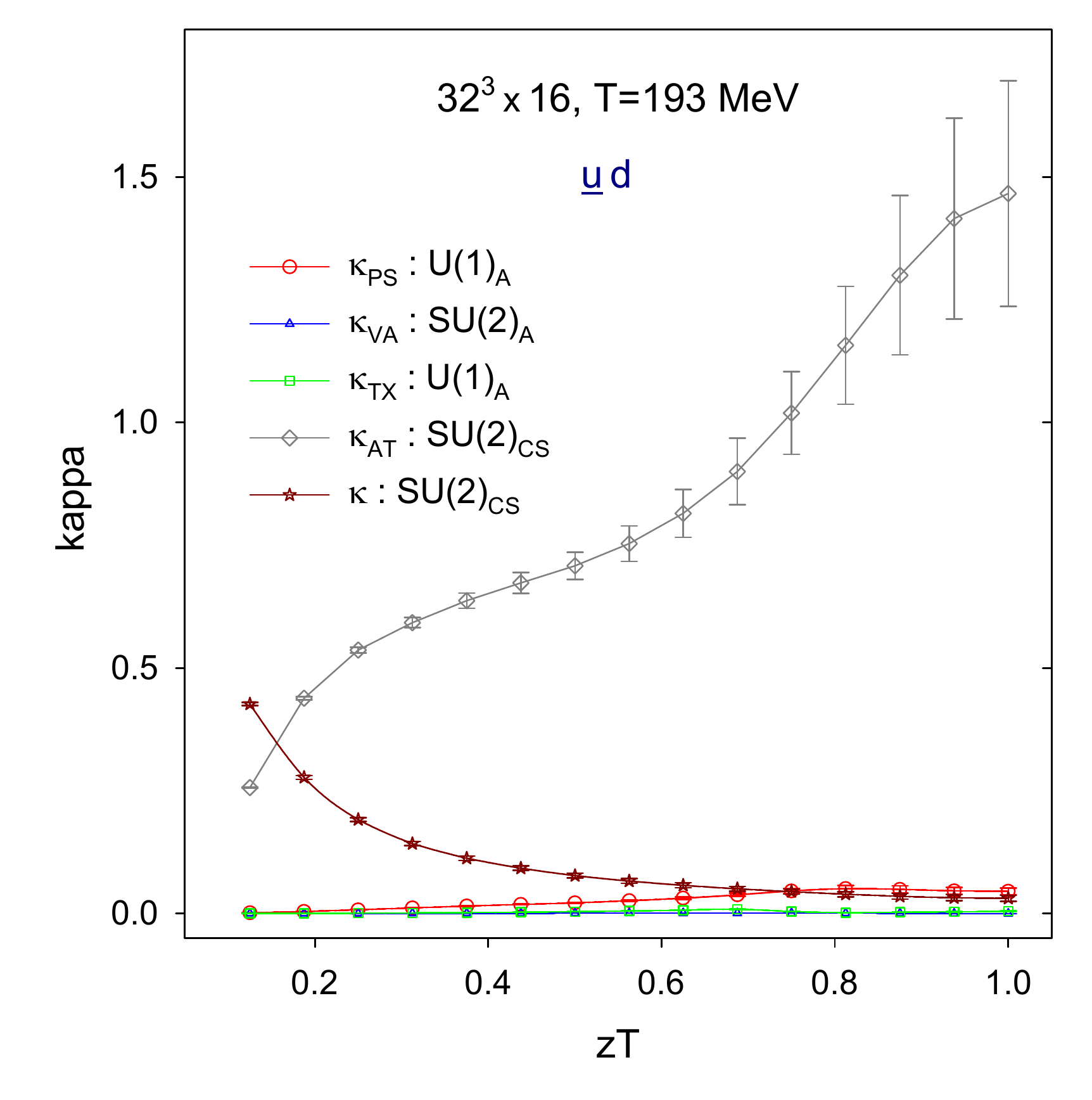}
&
  \includegraphics[width=7.0cm,clip=true]{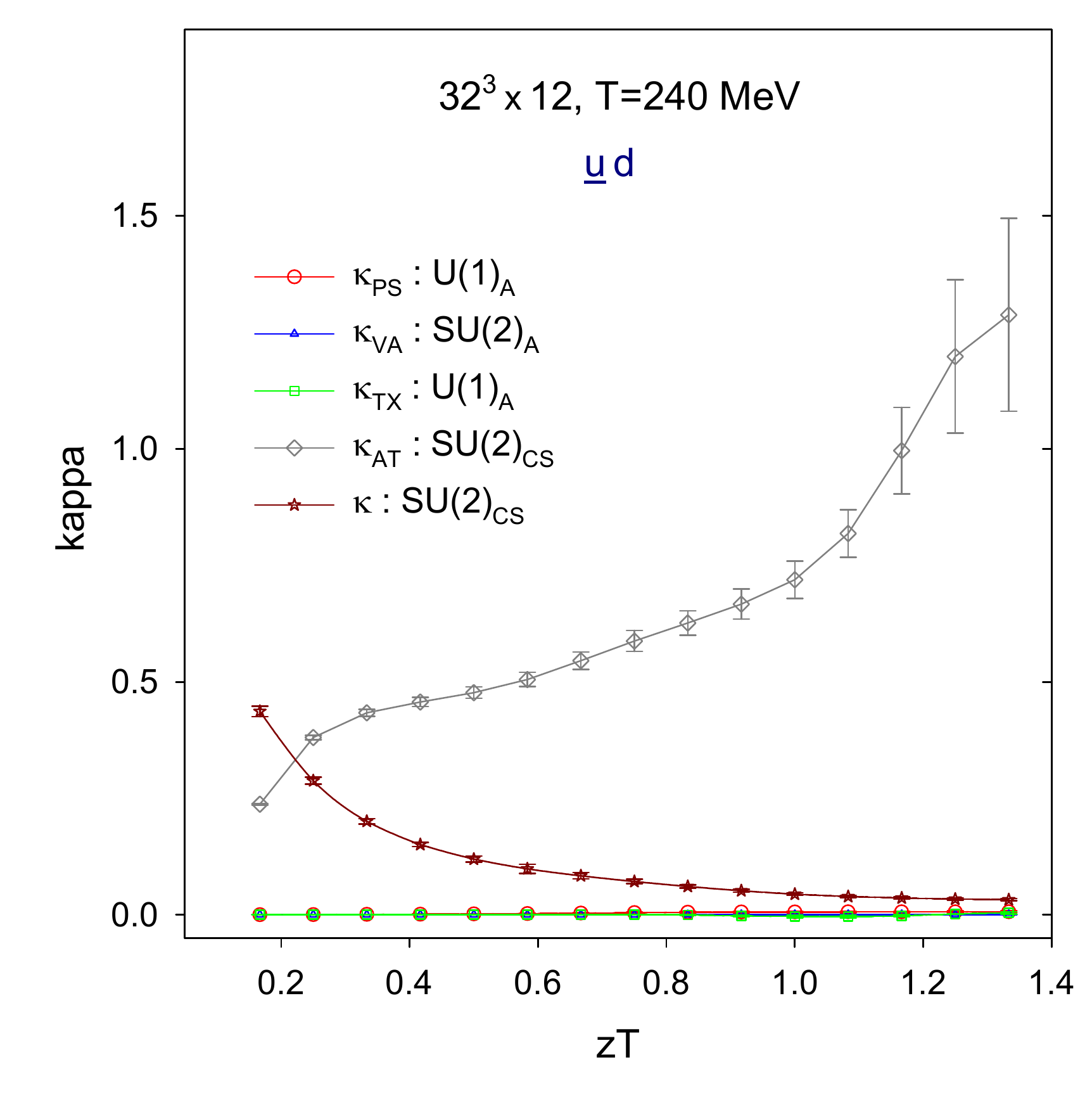}
\\
  \includegraphics[width=7.0cm,clip=true]{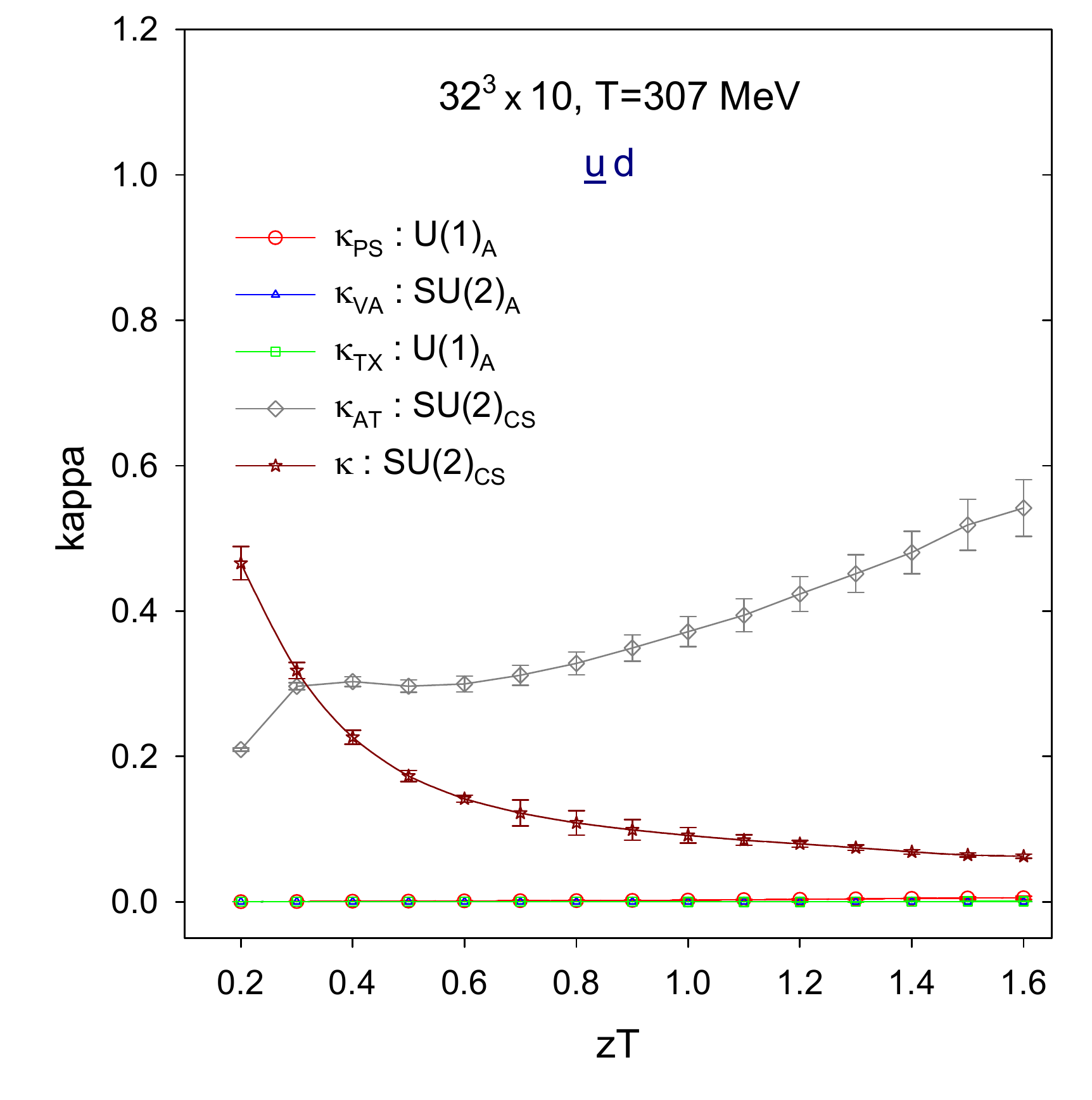}
&
  \includegraphics[width=7.0cm,clip=true]{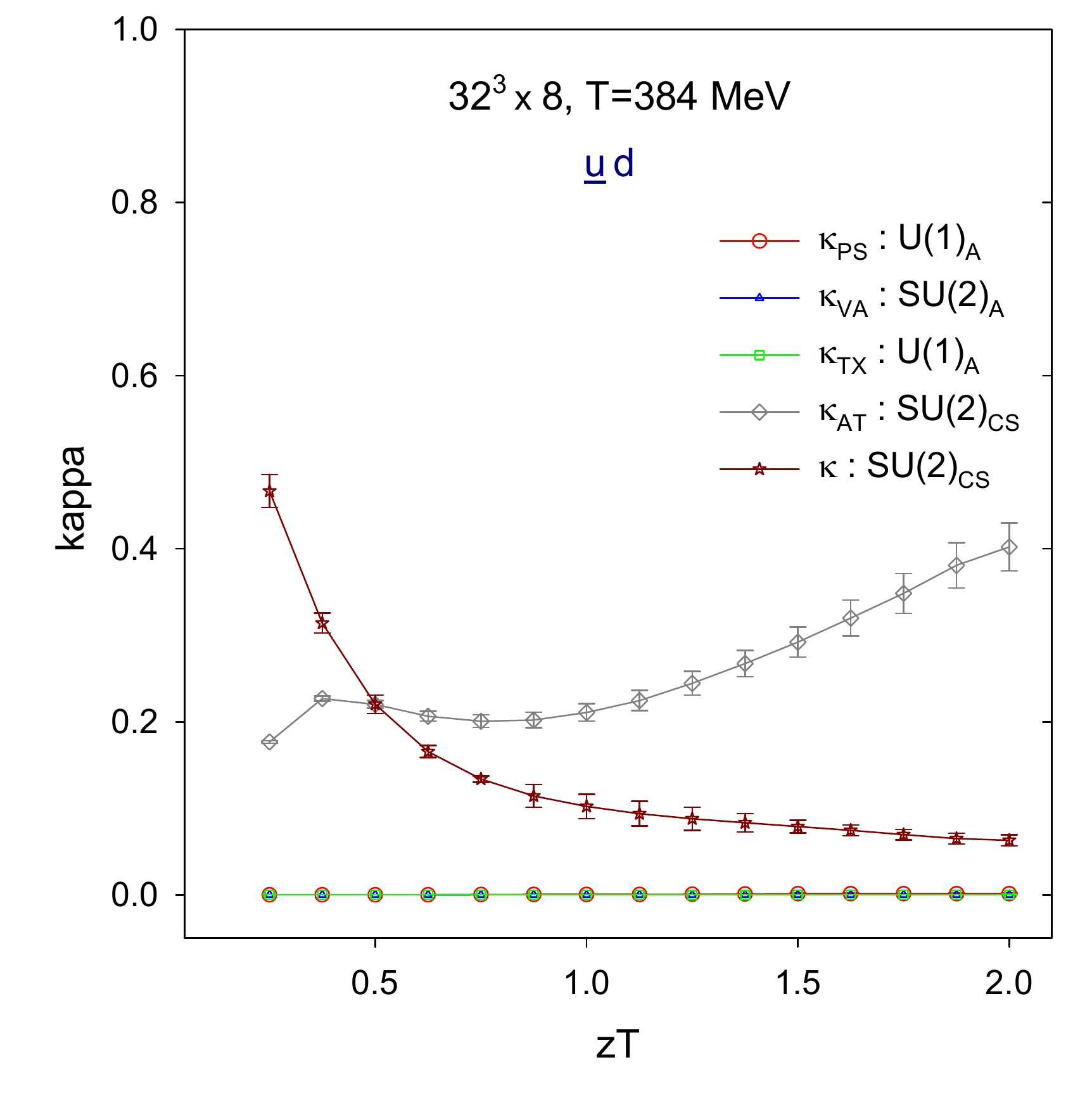}
\\
  \includegraphics[width=7.0cm,clip=true]{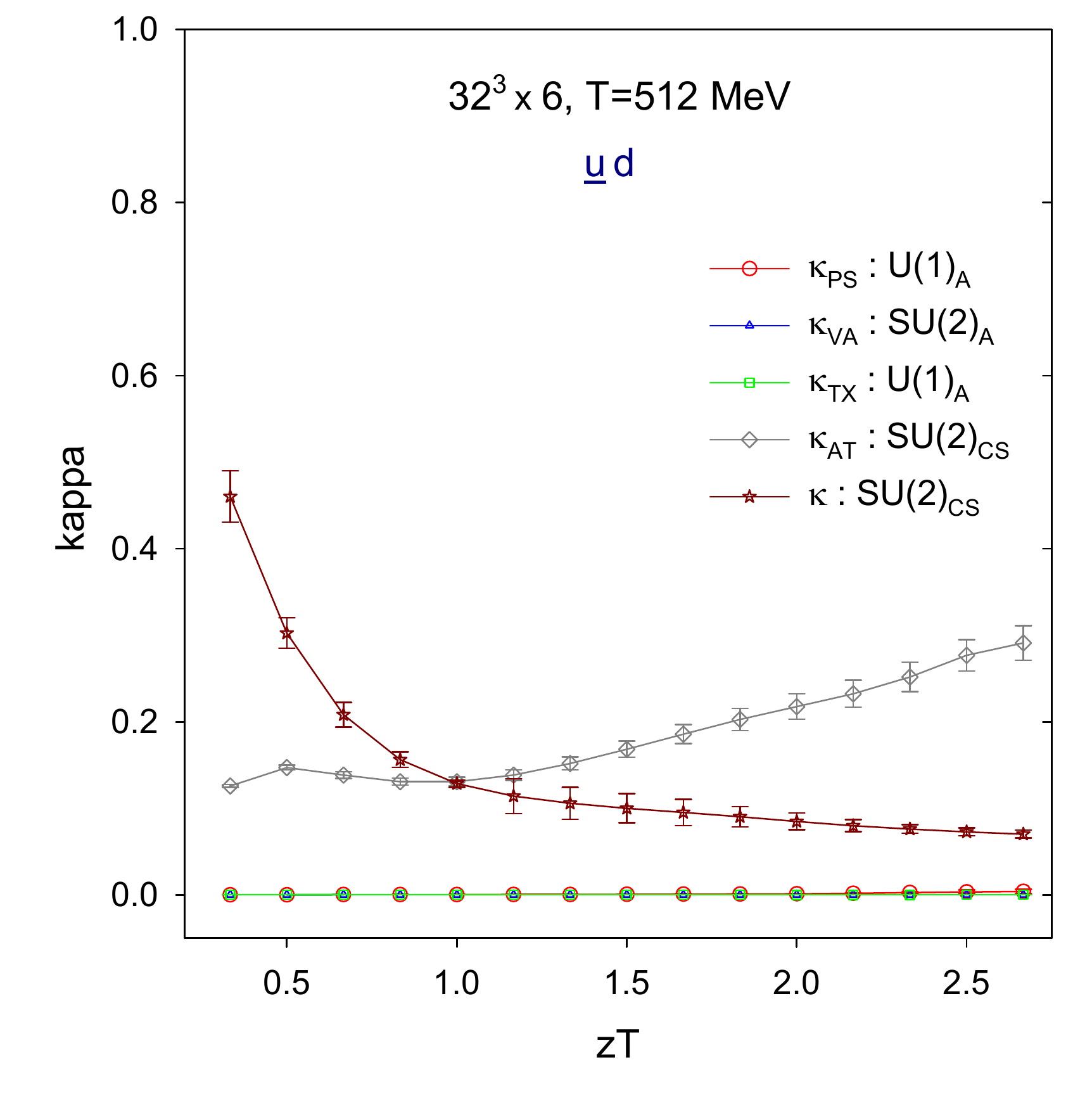}
&
  \includegraphics[width=7.0cm,clip=true]{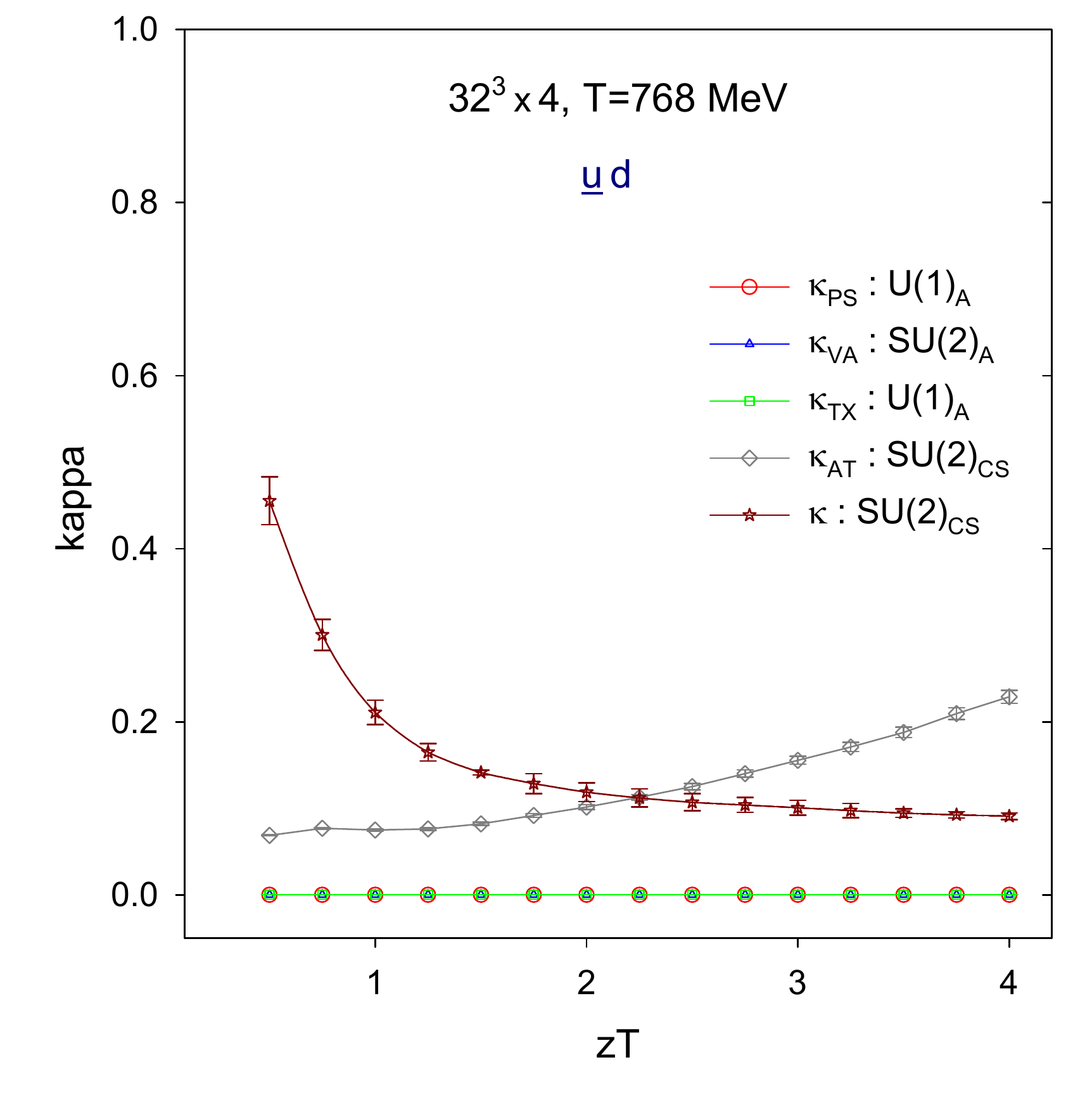}
\\
  \end{tabular}
\label{fig:K_ud_z}
\end{figure}

Next, we examine the symmetries in the spatial correlators  
with the symmetry-breaking parameters as defined by Eqs. (\ref{eq:k_PS_z}), (\ref{eq:k_TX_z}), 
(\ref{eq:k_VA_z}), (\ref{eq:k_AT_z}) and (\ref{eq:kappa_z}) in Sec. \ref{kappa}.
In Fig. \ref{fig:K_ud_z}, the symmetry-breaking parameters, $\kappa_{PS}$, 
$\kappa_{VA}$, $\kappa_{TX}$, $\kappa_{AT}$, and $\kappa$ corresponding 
to Fig. \ref{fig:Cz_ud} are plotted versus $zT$, for temperatures $T \sim 193-768$~MeV. 

For all six temperatures in the range $T \sim 193-768$~MeV, 
the $SU(2)_L \times SU(2)_R$ chiral symmetry is effectively restored 
with the maximum value of $ \kappa_{VA} $ equal to $8.7(4) \times 10^{-4} $ 
at $ T \sim 193 $~MeV and $zT = 0.75$. 

For the $U(1)_A$ symmetry, there are tiny breakings (especially at large $z$) at $T \sim 193 $~MeV
with the maximum value of $ \kappa_{TX} $ equal to $ 8.7(8) \times 10^{-3} $ at $zT=0.6875$, 
while that of $\kappa_{PS}$ is equal to $5.1(6) \times 10^{-2} $ at $zT=0.8125$.  
Therefore, it seems that $\kappa_{PS}$ and $\kappa_{TX}$ give incompatible answers 
at $T \sim 193$~MeV (similar to their counterparts in the $t$ correlators 
as shown in the top-right panel of Fig. \ref{fig:Ct_ud}), and also  
suggests that the effective restoration of $U(1)_A$ symmetry is likely to occur 
at temperatures higher than 193 MeV. 
To confirm or refute this, it is necessary to determine $\kappa_{PS}$ and $\kappa_{TX}$ 
in the continuum limit, which is beyond the scope of this paper.

\begin{figure}[!ht]
  \centering
  \caption{
    The $SU(2)_{CS}$ symmetry-breaking and -fading parameters ($\kappa_{AT}$, $\kappa$) 
    of $N_f=2+1+1$ lattice QCD at six temperatures in the range $ \sim 190-770$~MeV,  
    for $zT = (0.5, 1, 2)$.
  }
\begin{tabular}{@{}c@{}c@{}}
  \includegraphics[width=8.0cm,clip=true]{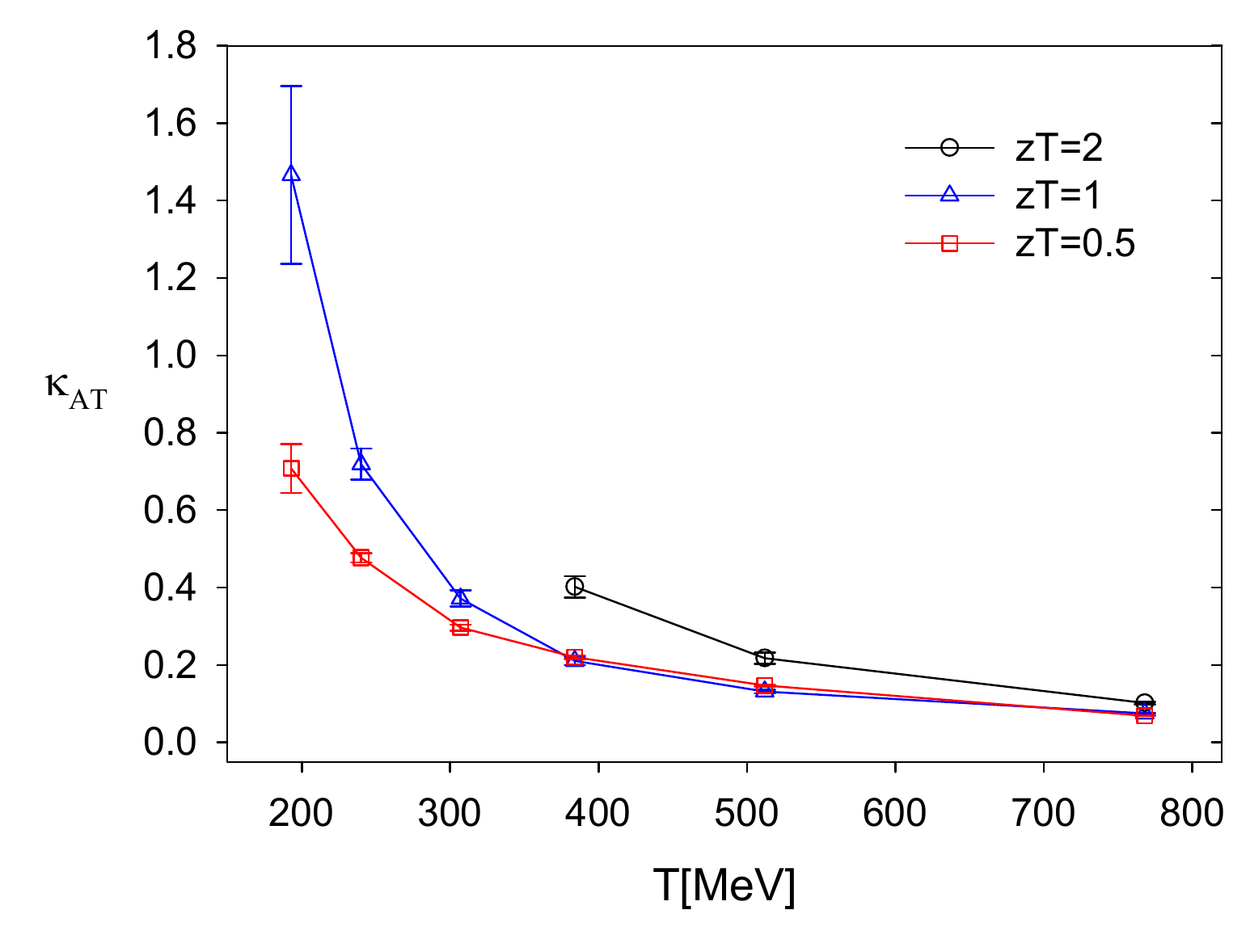}
&
  \includegraphics[width=8.0cm,clip=true]{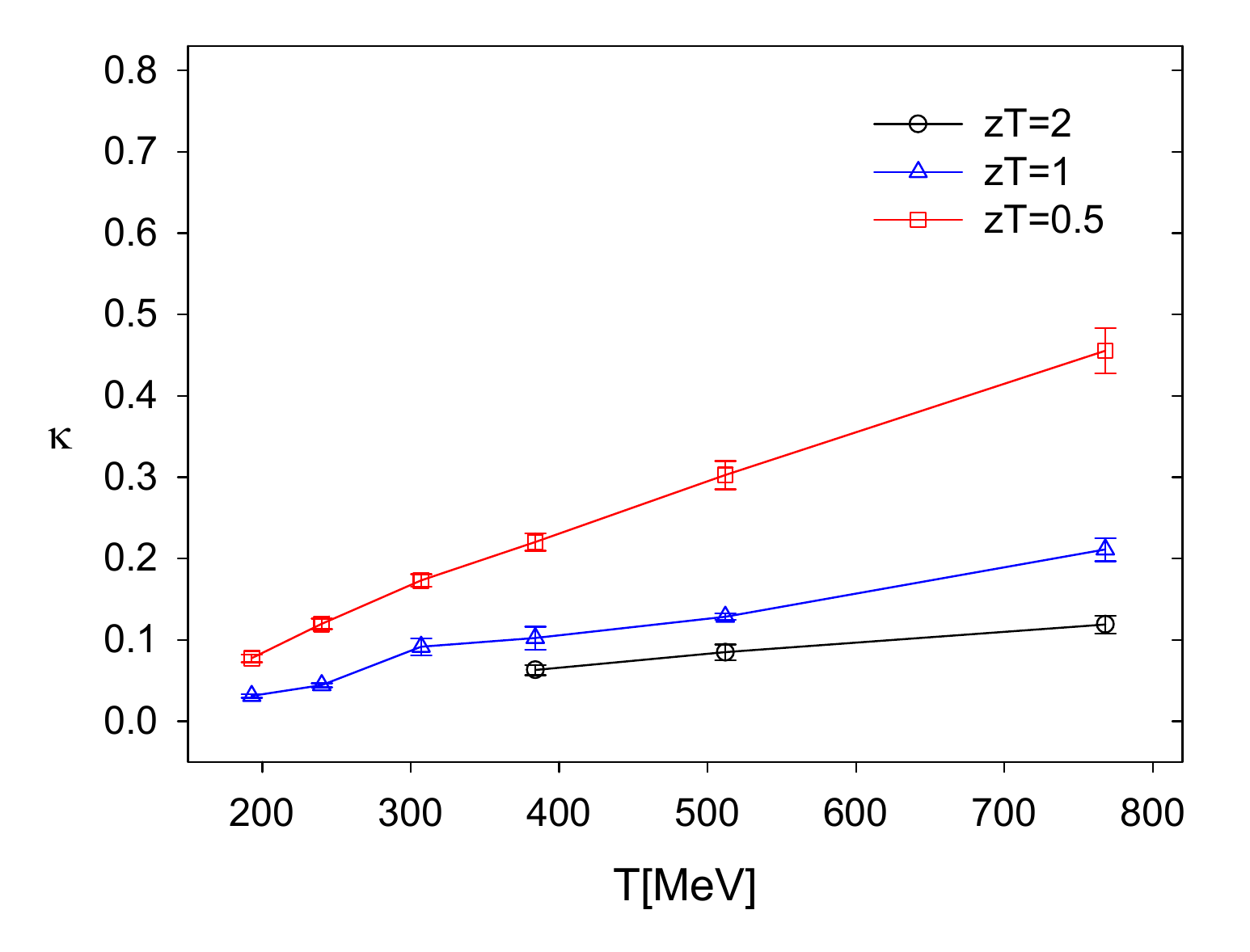}
\end{tabular}
  \label{fig:Kat_K_ud_z}
\end{figure}

For the $SU(2)_{CS}$ symmetry, the symmetry-breaking and -fading parameters 
$\kappa_{AT}(zT)$ and $\kappa(zT)$ are much larger than those 
($\kappa_{PS}$, $\kappa_{TX}$, and $\kappa_{VA}$)    
of $U(1)_A$ and $SU(2)_L \times SU(2)_R$ chiral symmetries, as shown in Fig. \ref{fig:K_ud_z}.

In Fig. \ref{fig:Kat_K_ud_z},  $\kappa_{AT}(zT)$ and $ \kappa(zT) $ 
are plotted versus the temperature $T$, for $zT = (0.5, 1, 2)$. 
In general, for any fixed $zT$, $\kappa_{AT}$ is a monotonic decreasing function of $ T $, 
while $\kappa$ is a monotonic increasing function of $T$.
 
Using the data of $\kappa_{AT}$ and $\kappa$ in Fig. \ref{fig:Kat_K_ud_z}  
and the criterion Eq. (\ref{eq:SU2_CS_crit_z}) for the emergence of 
approximate $SU(2)_{CS}$ symmetry,   
the ranges of temperatures satisfying Eq. (\ref{eq:SU2_CS_crit_z}) can be determined  
for any $zT$ and $\epsilon_{CS}$.  
In Table \ref{tab:Nf2p1p1_ud_z}, the ranges of temperatures satisfying 
Eq.~(\ref{eq:SU2_CS_crit_z}) with $\epsilon_{CS} = (0.20, 0.10, 0.05)$ 
are tabulated for $zT=(2.0, 1.0, 0.5)$.  
Note that for $zT=2$, the upper bounds of the windows,   
$T_x (> 770$~MeV) and $T_y (>770$~MeV) have not yet been determined, 
since the highest temperature in this study is $\sim 770$~MeV. 
In general, for any fixed $zT$, the window of temperatures is shrunk 
as $ \epsilon_{CS} $ is decreased [i.e., a more precise $SU(2)_{CS}$ symmetry].
At $\epsilon_{CS} = 0.10 $, the window is shrunk to zero for $zT = (2.0, 1.0, 0.5) $. 
In other words, the approximate $SU(2)_{CS}$ symmetry of the $z$ correlators of $\bar u \Gamma d$
in $N_f = 2+1+1$ lattice QCD cannot become a more precise symmetry with $ \epsilon_{CS} \le 0.10 $, 
unlike the $U(1)_A \times SU(2)_L \times SU(2)_R $ chiral symmetry, which is effectively restored 
as an exact symmetry for $T > T_1 \gtrsim T_c $. 
Consequently, in the range of temperatures with approximate $SU(2)_{CS} $ symmetry,  
even if the chromoelectic interactions may play a predominant role 
in binding $u$ and $d$ quarks into hadron-like objects, 
the role of chromomagnetic interactions in their bindings cannot be neglected.  

\begin{table}[h!]
\begin{center}
\caption{The approximate ranges of temperatures satisfying the criterion in 
         Eq.~(\ref{eq:SU2_CS_crit_z}) with $\epsilon_{CS} = (0.20, 0.15, 0.10) $, 
         for $zT=(2.0, 1.0, 0.5)$.  
In the second column, $T_x$ ($> 770$~MeV) and $T_y$ ($> 770$~MeV) have yet to be determined.} 
\setlength{\tabcolsep}{4pt}
\vspace{2mm}
\begin{tabular}{|cccc|}
\hline
  $\epsilon_{CS}$
  & $zT=2.0$
  & $zT=1.0$
  & $zT=0.5$ \\
\hline
\hline
0.20 &  $\sim 550$~MeV$-T_x$($> 770$~MeV)  & $\sim$ 380-730 MeV  &  NULL  \\
0.15 &  $\sim 660$~MeV$-T_y$($> 770$~MeV)  & $\sim$ 480-580 MeV  &  NULL  \\
0.10 &  NULL                               &  NULL               &  NULL  \\
\hline
\end{tabular}
\label{tab:Nf2p1p1_ud_z}
\end{center}
\end{table}

\begin{figure}[!ht]
  \centering
  \caption{The spatial $z$ correlators of $ \bar u \Gamma d $ meson interpolators
           constructed with the free-quark propagators.
  }
  \begin{tabular}{@{}c@{}c@{}}
  \includegraphics[width=7.0cm,clip=true]{Cz_l32t16_b620_ud_free.pdf}
&
  \includegraphics[width=7.0cm,clip=true]{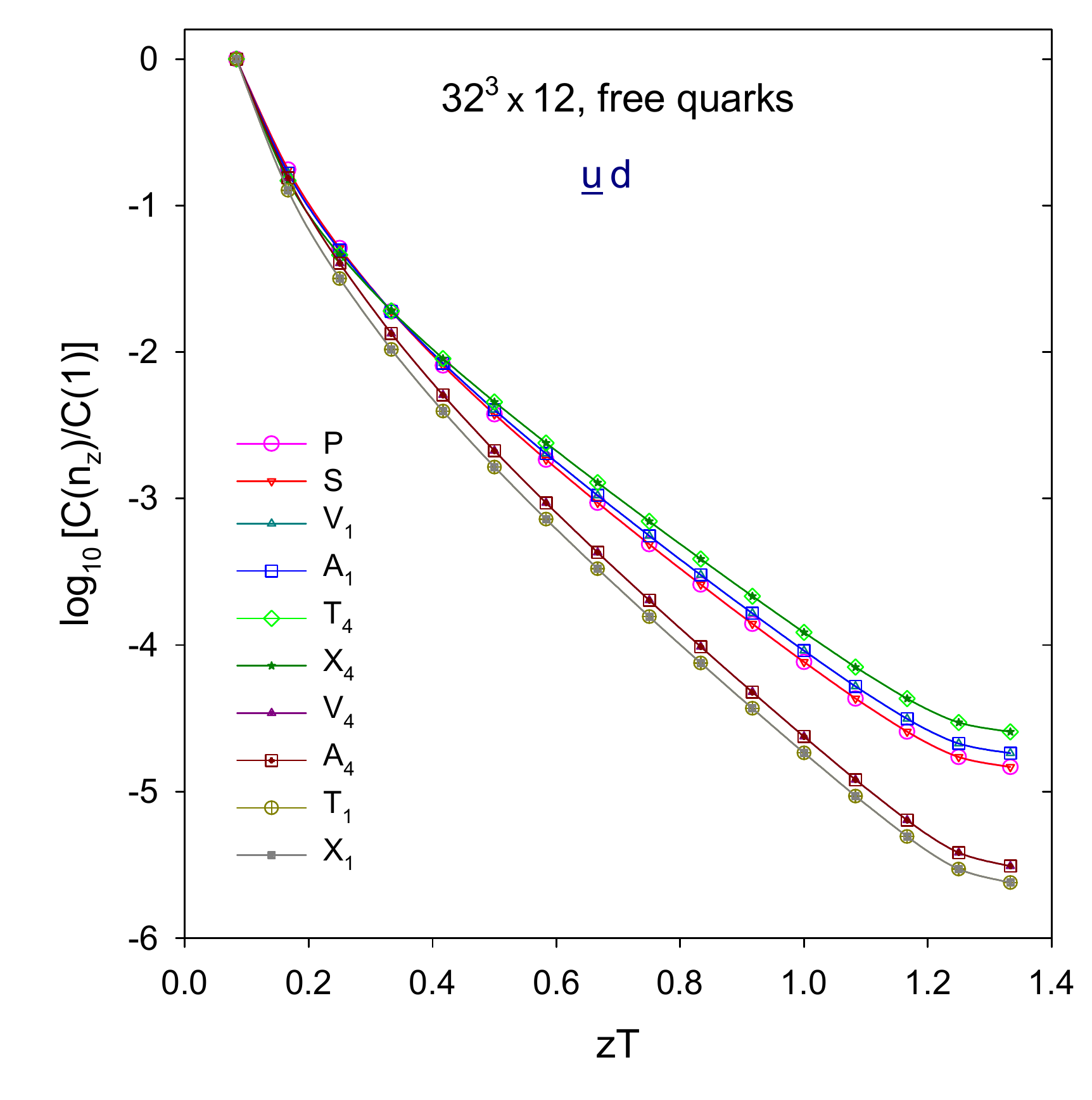}
\\
  \includegraphics[width=7.0cm,clip=true]{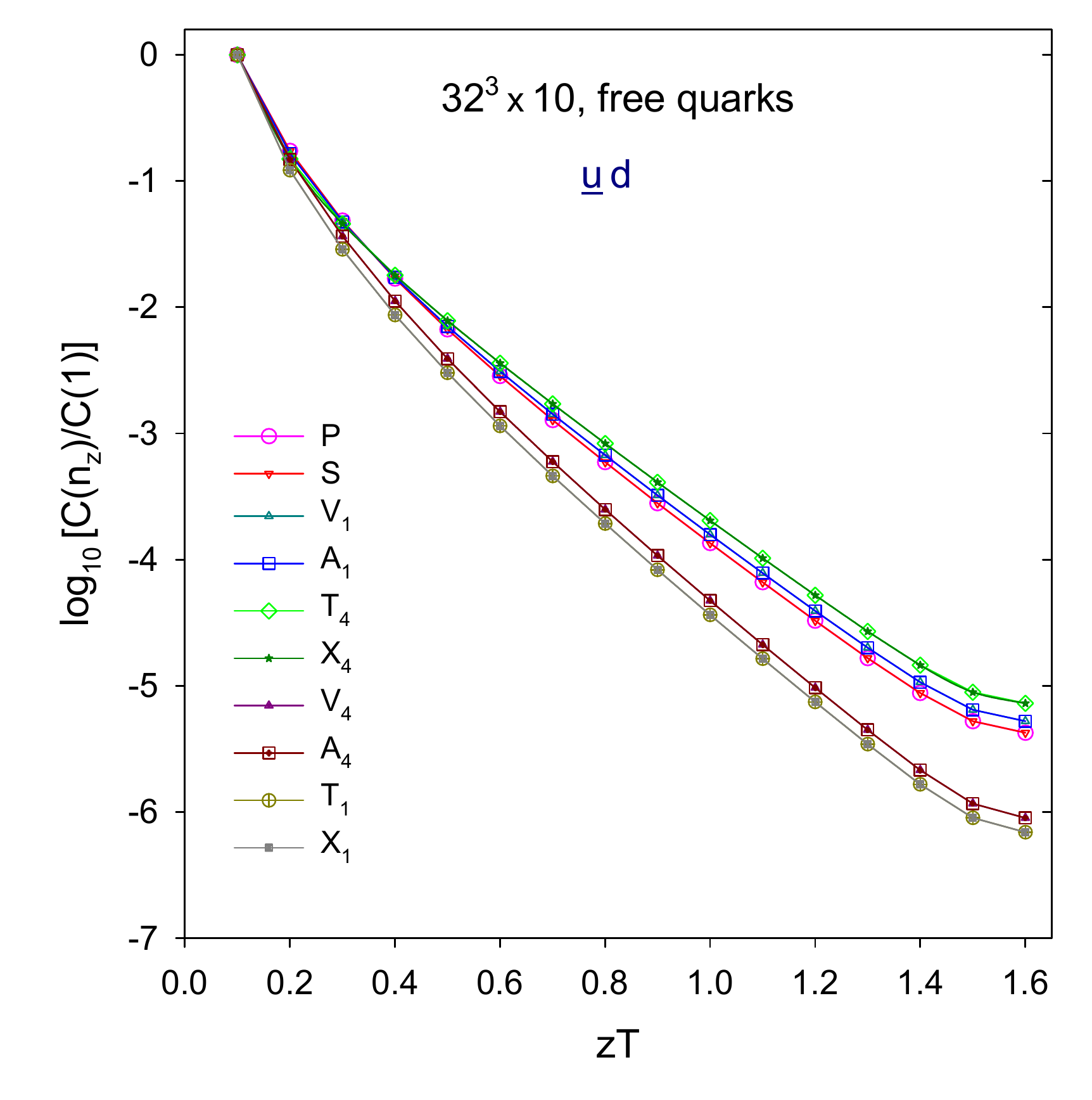}
&
  \includegraphics[width=7.0cm,clip=true]{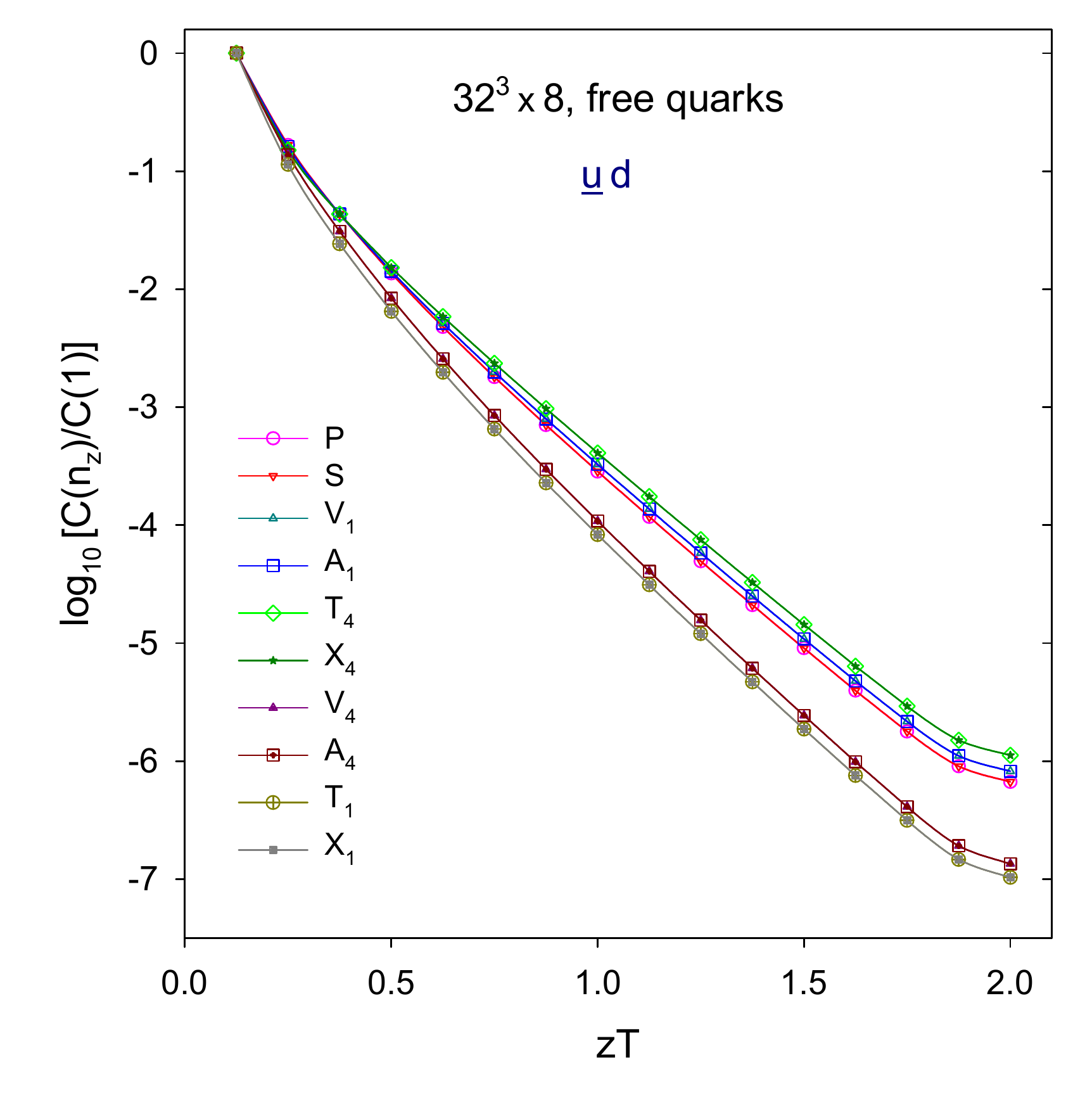}
\\
  \includegraphics[width=7.0cm,clip=true]{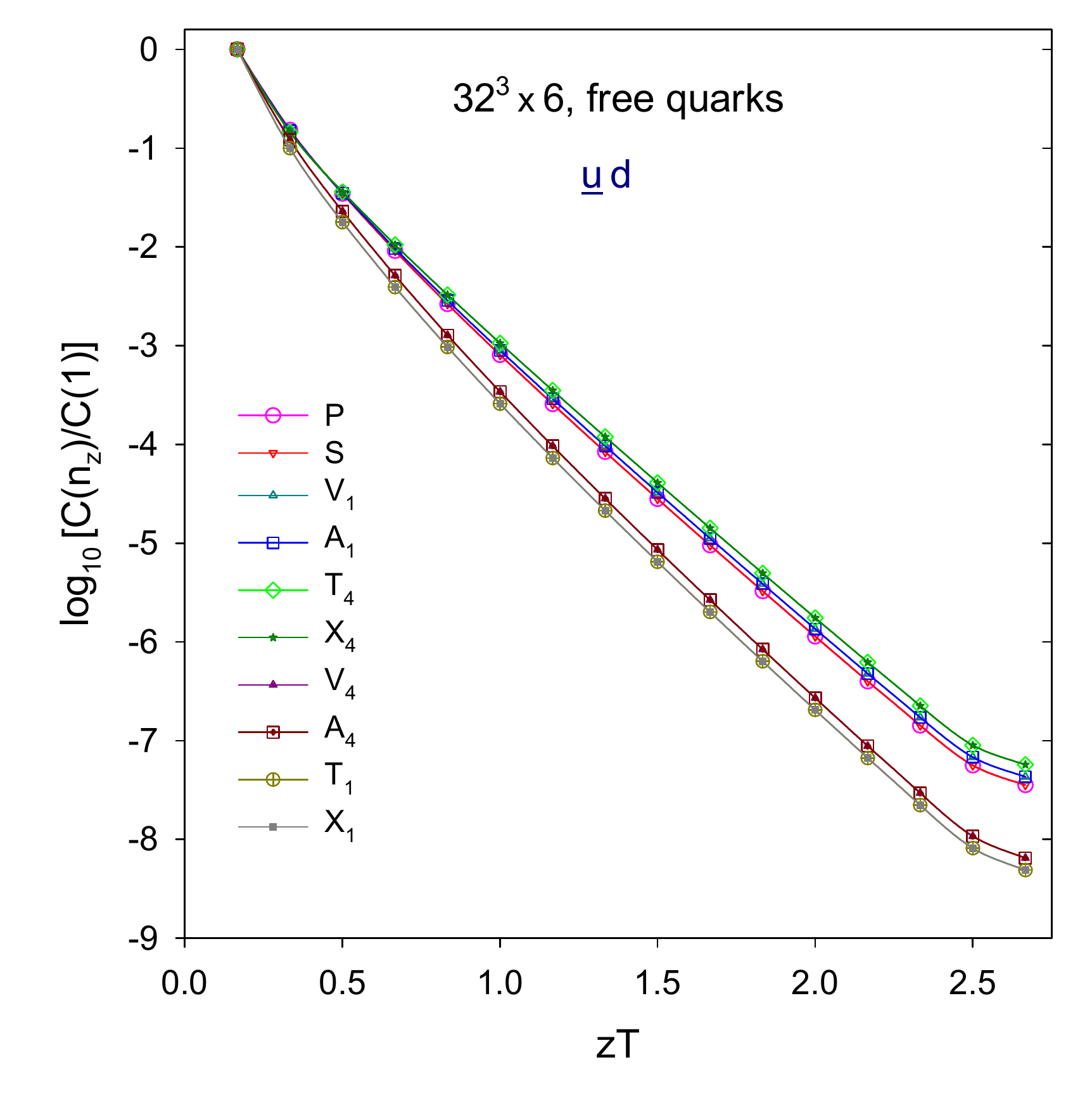}
&
  \includegraphics[width=7.0cm,clip=true]{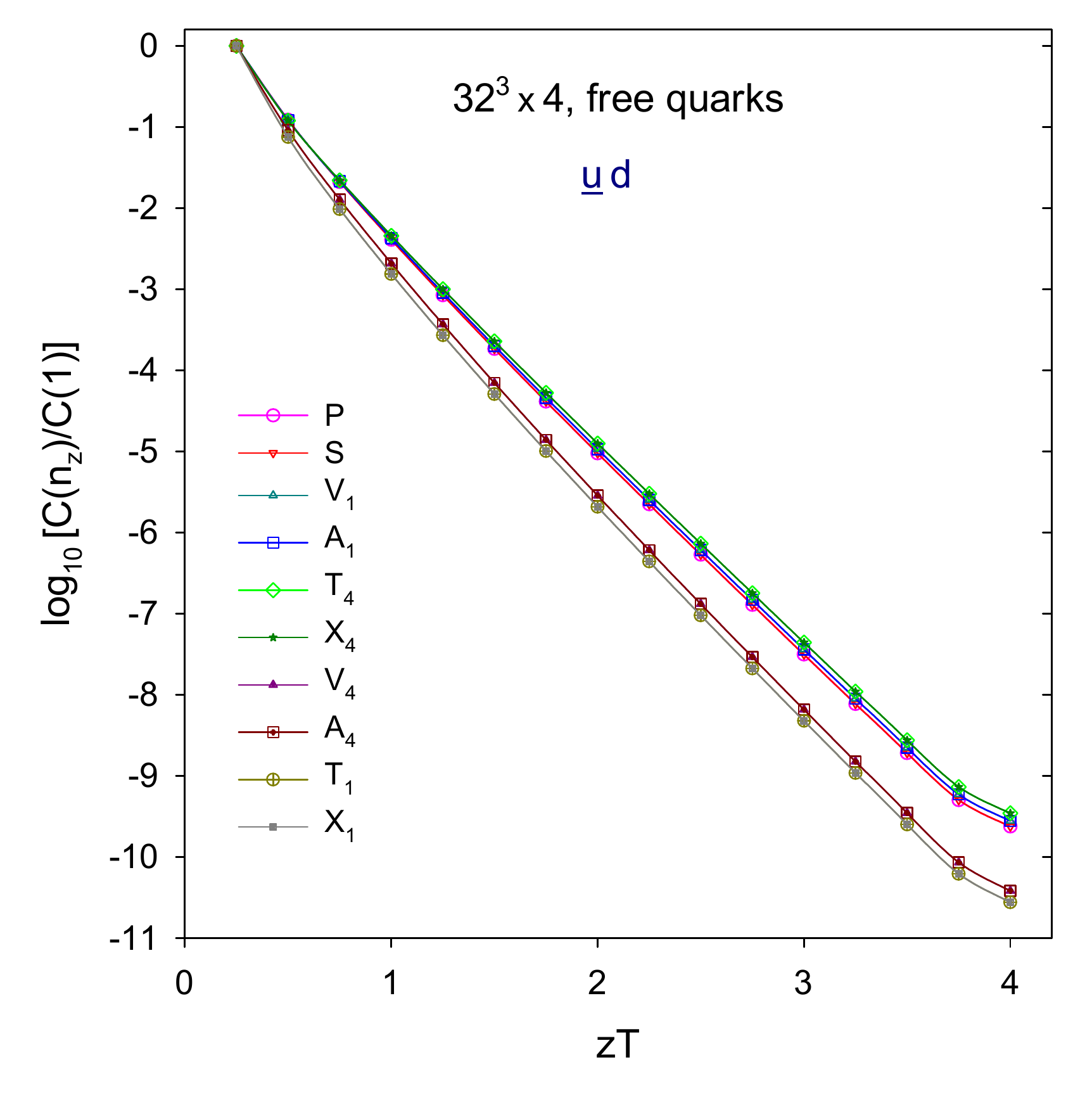}
\\
  \end{tabular}
\label{fig:Cz_ud_free}
\end{figure}

\begin{figure}[!ht]
  \centering
  \caption{
   The symmetry-breaking parameters of the $z$ correlators of $ \bar u \Gamma d $ with free quarks. 
  }
  \begin{tabular}{@{}c@{}c@{}}
  \includegraphics[width=7.0cm,clip=true]{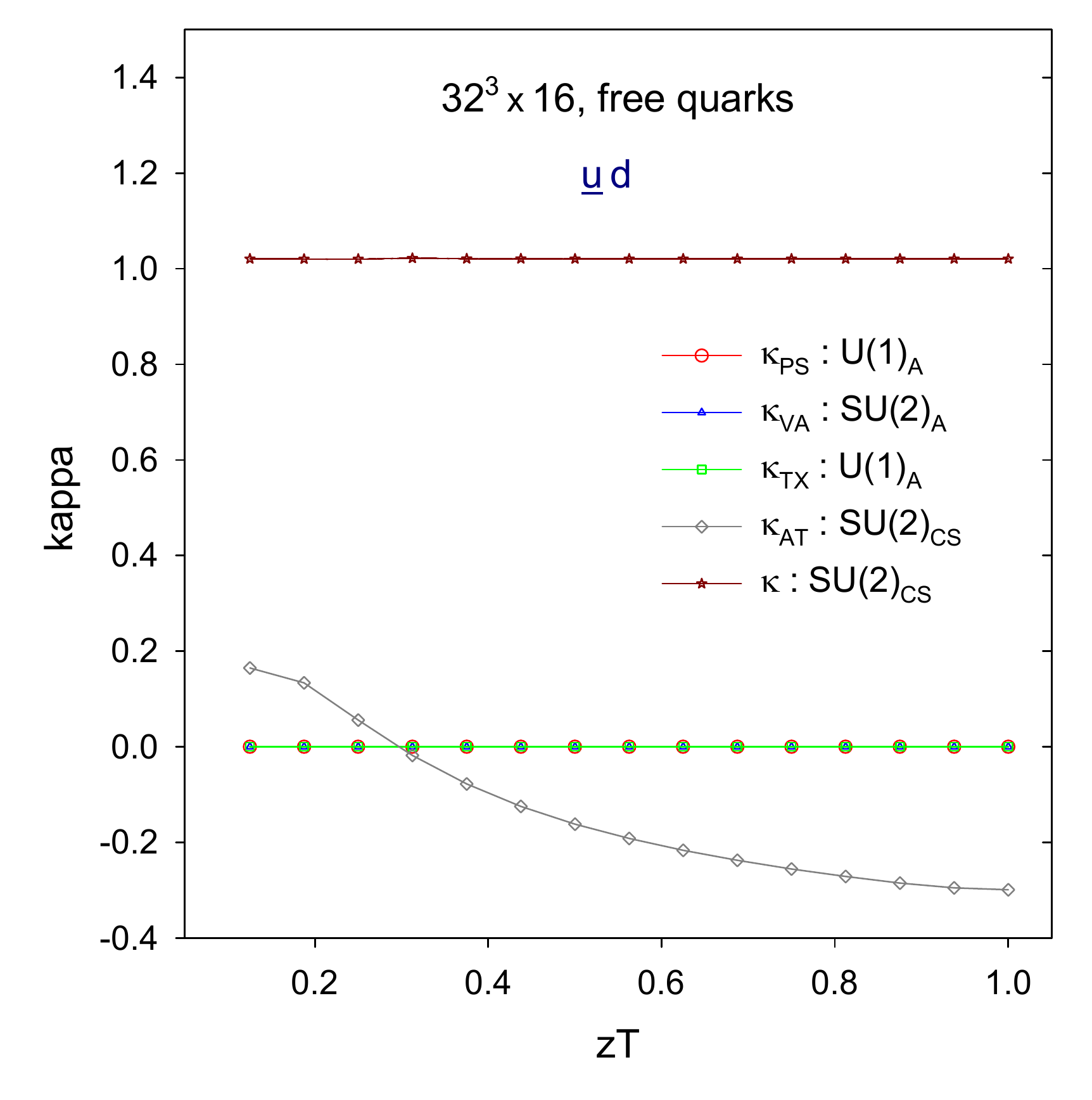}
&
  \includegraphics[width=7.0cm,clip=true]{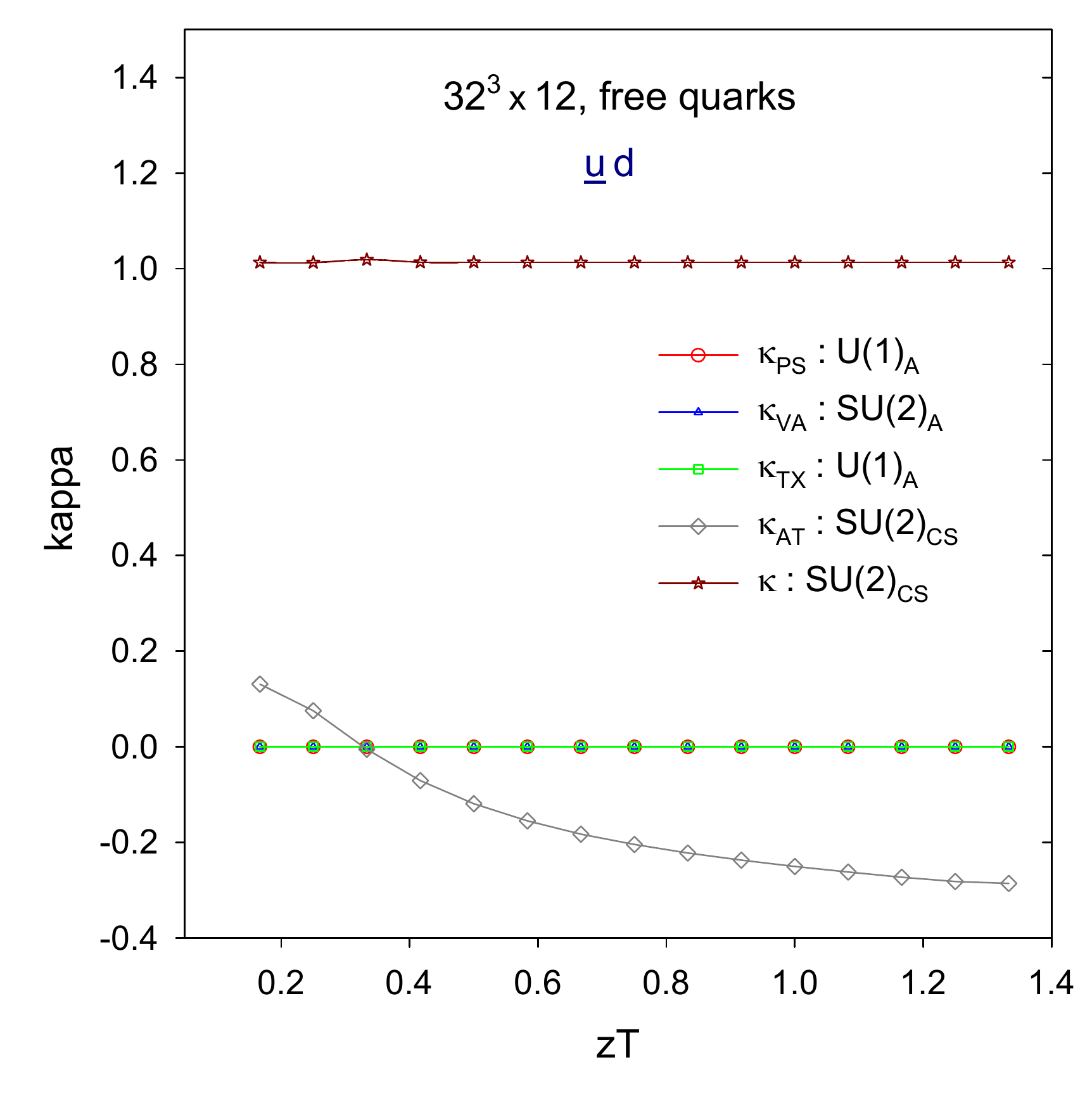}
\\
  \includegraphics[width=7.0cm,clip=true]{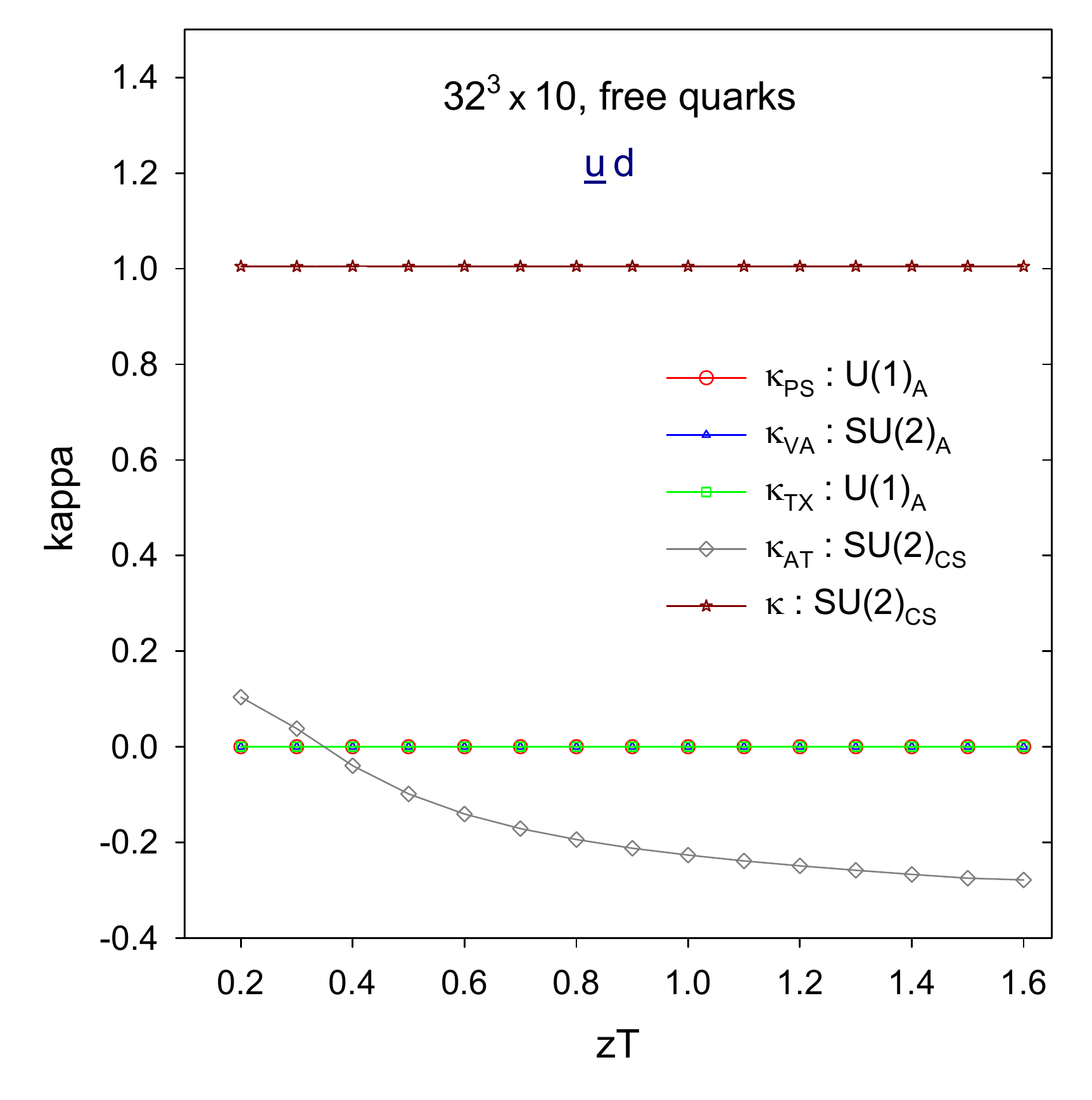}
&
  \includegraphics[width=7.0cm,clip=true]{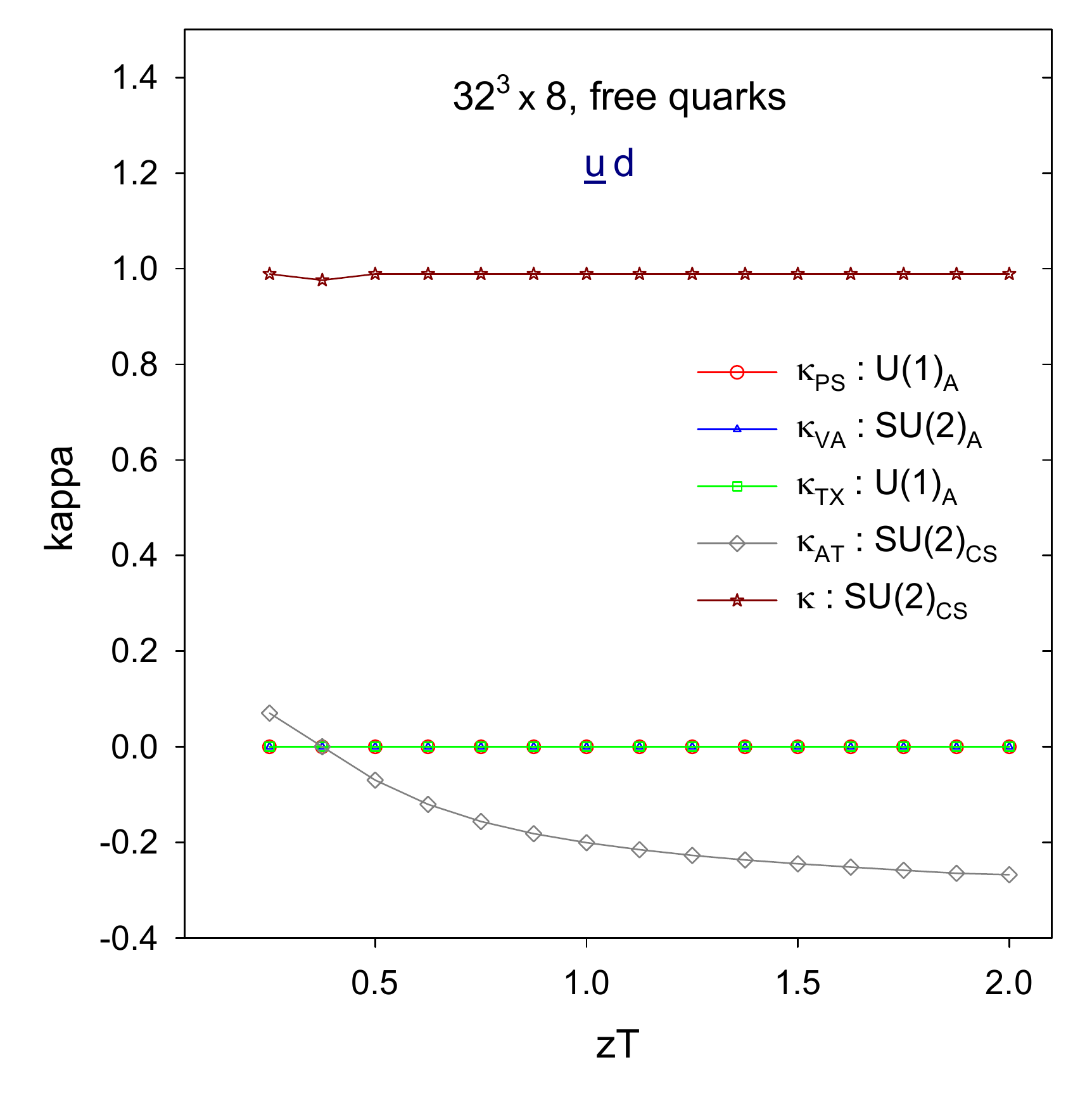}
\\
  \includegraphics[width=7.0cm,clip=true]{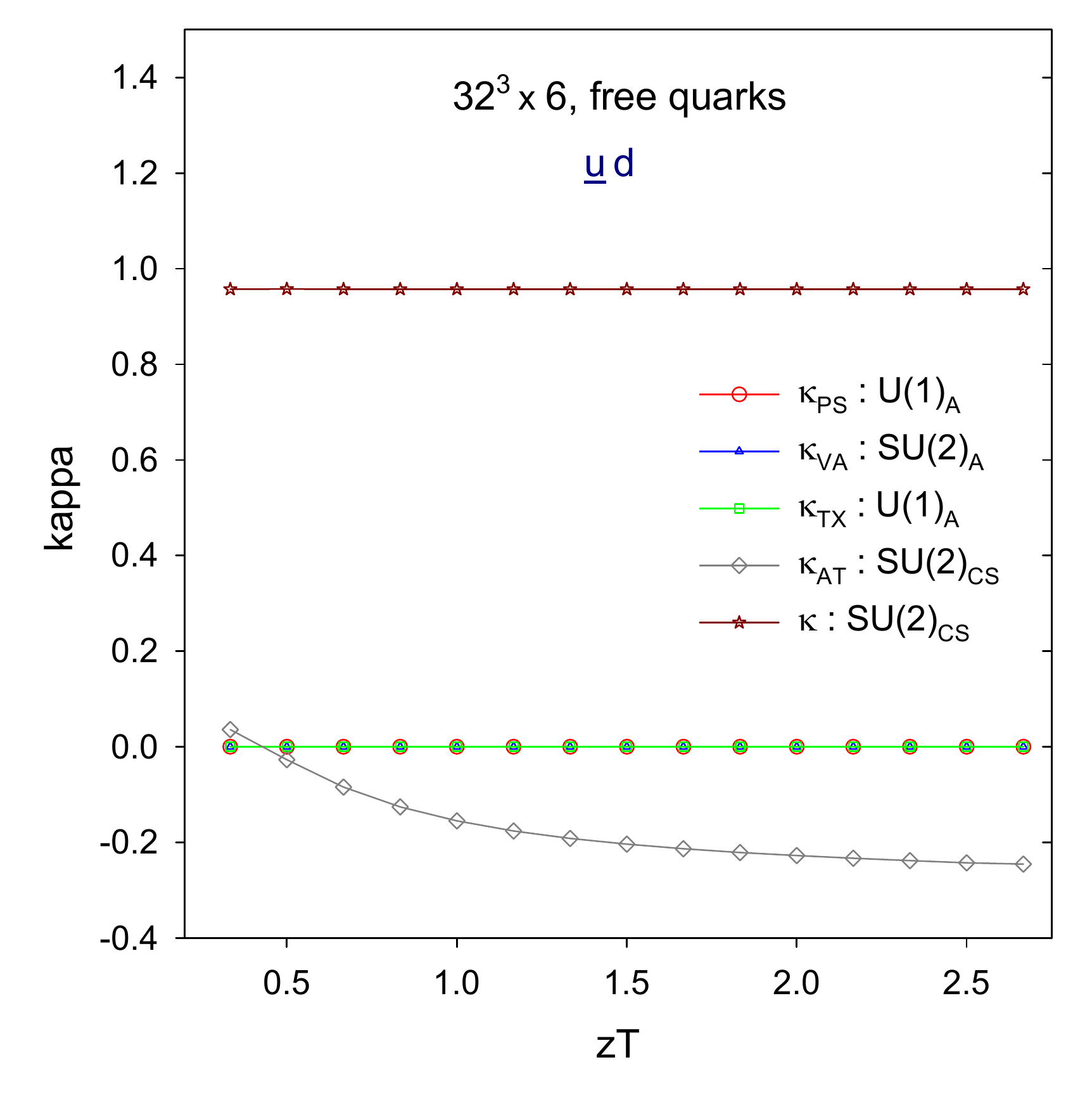}
&
  \includegraphics[width=7.0cm,clip=true]{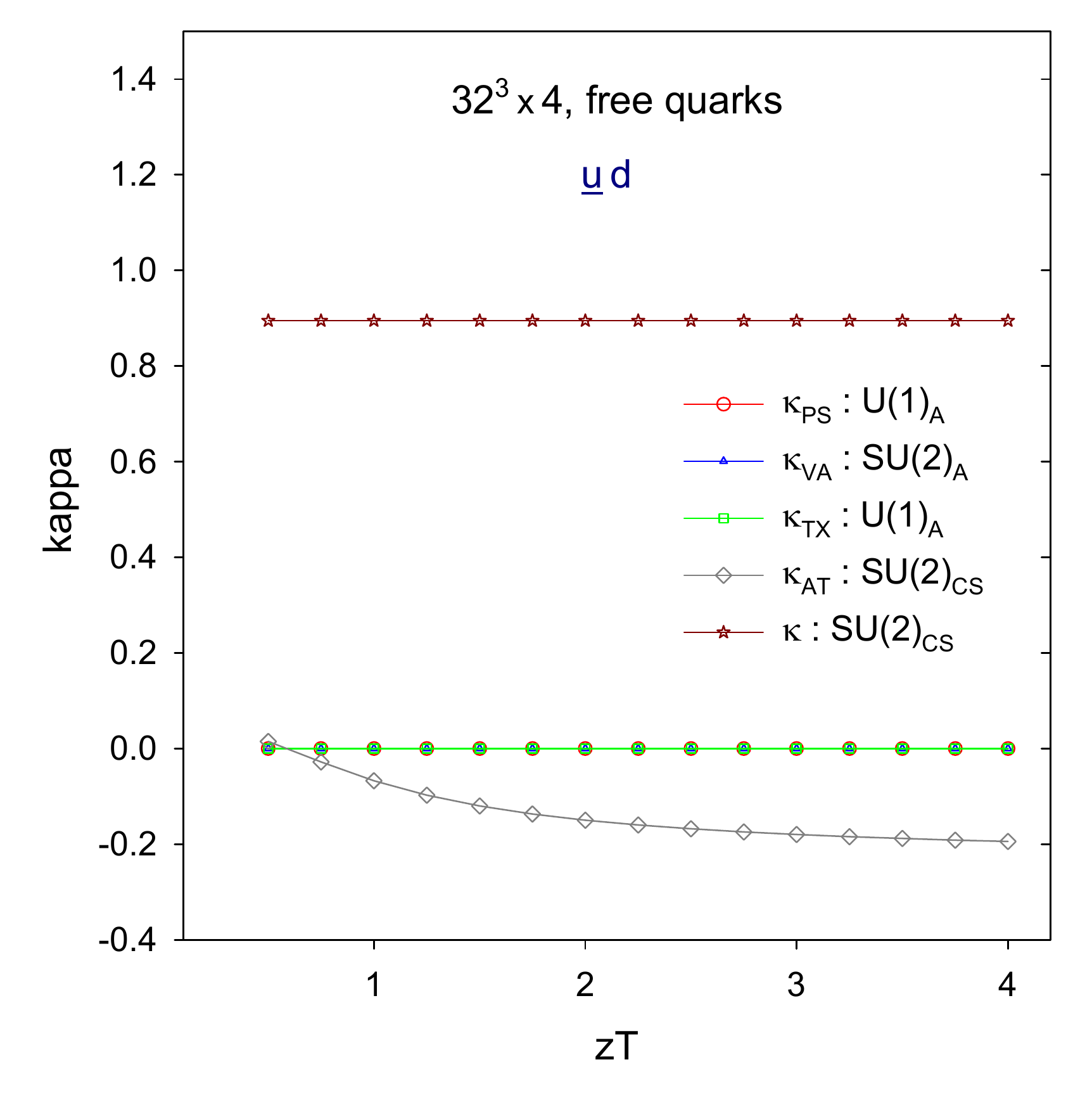}
\\
  \end{tabular}
\label{fig:K_ud_z_free}
\end{figure}

\subsection{Comparison with the noninteracting theory}
\label{free_z}

The spatial $z$ correlators of $\bar u \Gamma d$ constructed with 
free-quark propagators are plotted in Fig. \ref{fig:Cz_ud_free}.
The free-quark propagators are computed with the same set of boundary conditions 
(see Sec. \ref{issues}),
the same lattice size, and the same $u/d$ quark masses as those in $N_f=2+1+1$ lattice QCD, 
but with all link variables equal to the identity matrix.
Note that the lattice spacing $a$ and the temperature $T=1/(N_t a) $
are not defined for the free quarks. 
Thus, the label $zT$ of the horizontal axis in Fig. \ref{fig:Cz_ud_free}
should be regarded as $ zT = n_z/N_t $. 
In the following, the temperature $T$ for all quantities with free quarks is always understood 
to be the corresponding temperature $T=1/(N_t a) $ in $N_f = 2+1+1$ lattice QCD with the same $N_t$.

In Fig. \ref{fig:Cz_ud_free}, for all six lattice sizes $32^3 \times (16,12,10,8,6,4) $, 
the $U(1)_A \times SU(2)_L \times SU(2)_R$ chiral symmetry is almost exact in spite of the 
nonzero $u/d$ quark masses, as shown by the degeneracies $C_P(z) = C_S(z)$,  
$C_{T_k}(z) = C_{X_k}(z)$, and $ C_{V_k}(z) = C_{A_k}(z)$ for $k=1,2,4 $.  
Consequently, it appears that there are only five distinct $z$ correlators on each panel 
of Fig. \ref{fig:Cz_ud_free}. They appear in the order 
\bea 
\label{eq:Cz_ud_free}
C_{T_4, X_4}{\rm (free)} > C_{V_1, A_1}{\rm (free)} > C_{P, S}{\rm (free)} 
   > C_{V_4, A_4}{\rm (free)} > C_{T_1, X_1}{\rm (free)}, \ {\rm for} \ n_z \ge 7, 
\eea
which is different from that of $N_f = 2+1+1$ lattice QCD for $ T \sim 190-770$~MeV 
in Eq. (\ref{eq:Cz_ud}), i.e.,   
$$ C_{P, S} > C_{V_1, A_1} > C_{T_4, X_4} > C_{V_4, A_4} > C_{T_1, X_1}, \ {\rm for} \ n_z \ge 7, $$ 
where the latter is consistent with that of lattice QCD at $ T < T_c \sim 150$~MeV. 
Note that the orderings of $C_{P,S} $, $C_{V_1,A_1} $, and $C_{T_4, X_4}$ in 
Eq. (\ref{eq:Cz_ud_free}) are reversed from those in Eq. (\ref{eq:Cz_ud}). 

Next, we examine the symmetries in the $z$ correlators of free quarks  
with the symmetry-breaking parameters as defined in Sec. \ref{kappa}.

In Fig. \ref{fig:K_ud_z_free}, the symmetry-breaking parameters are plotted versus $zT = n_z/N_t$
for $N_t = (16, 12, 10, 8, 6, 4)$. 
For $U(1)_A$ and $SU(2)_L \times SU(2)_R$ chiral symmetries, 
$\kappa_{PS} \simeq \kappa_{TX} \simeq \kappa_{VA} < 10^{-7}$,  
which shows that the $U(1)_A \times SU(2)_L \times SU(2)_R$ chiral symmetry is almost exact 
in the noninteracting theory with free quarks, in spite of the nonzero $u/d$ quark masses.
For the $SU(2)_{CS}$ symmetry, the symmetry-breaking and -fading parameters
$\kappa_{AT}(zT)$ and $\kappa(zT)$ are much larger than those 
($\kappa_{PS}, \kappa_{TX}, \kappa_{VA}$)
of $U(1)_A$ and $SU(2)_L \times SU(2)_R$ chiral symmetries.

\begin{figure}[!ht]
  \centering
  \caption{
    The $SU(2)_{CS}$ symmetry-breaking and -fading parameters ($\kappa_{AT}$, $\kappa$) 
    of the spatial meson correlators with free quarks, versus the corresponding 
    temperature $T=1/(N_t a) $ in $N_f = 2+1+1$ lattice QCD with the same $N_t$,  
    for $zT = n_z/N_t = (0.5, 1, 2)$.
  }
\begin{tabular}{@{}c@{}c@{}}
  \includegraphics[width=8.0cm,clip=true]{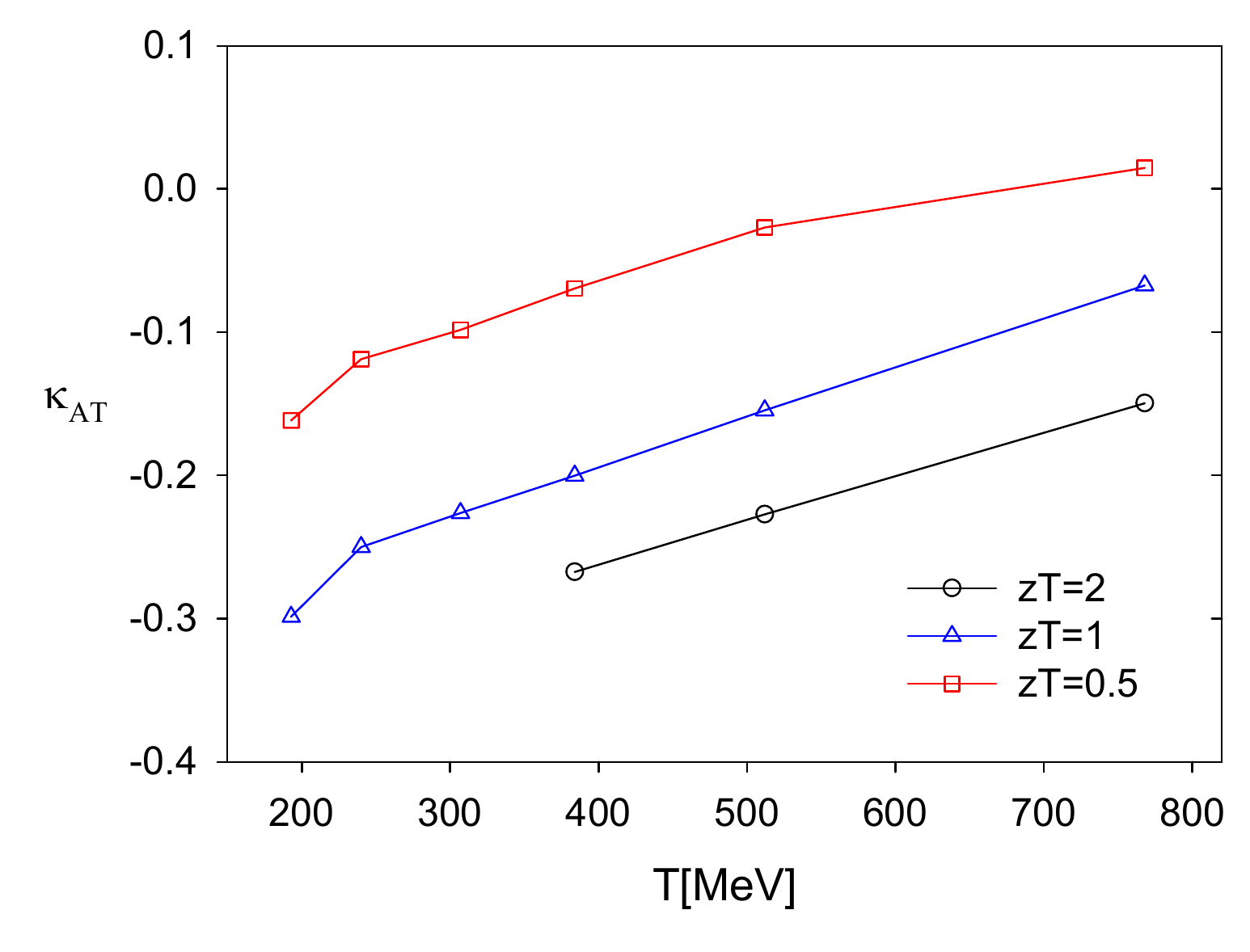}
&
  \includegraphics[width=8.0cm,clip=true]{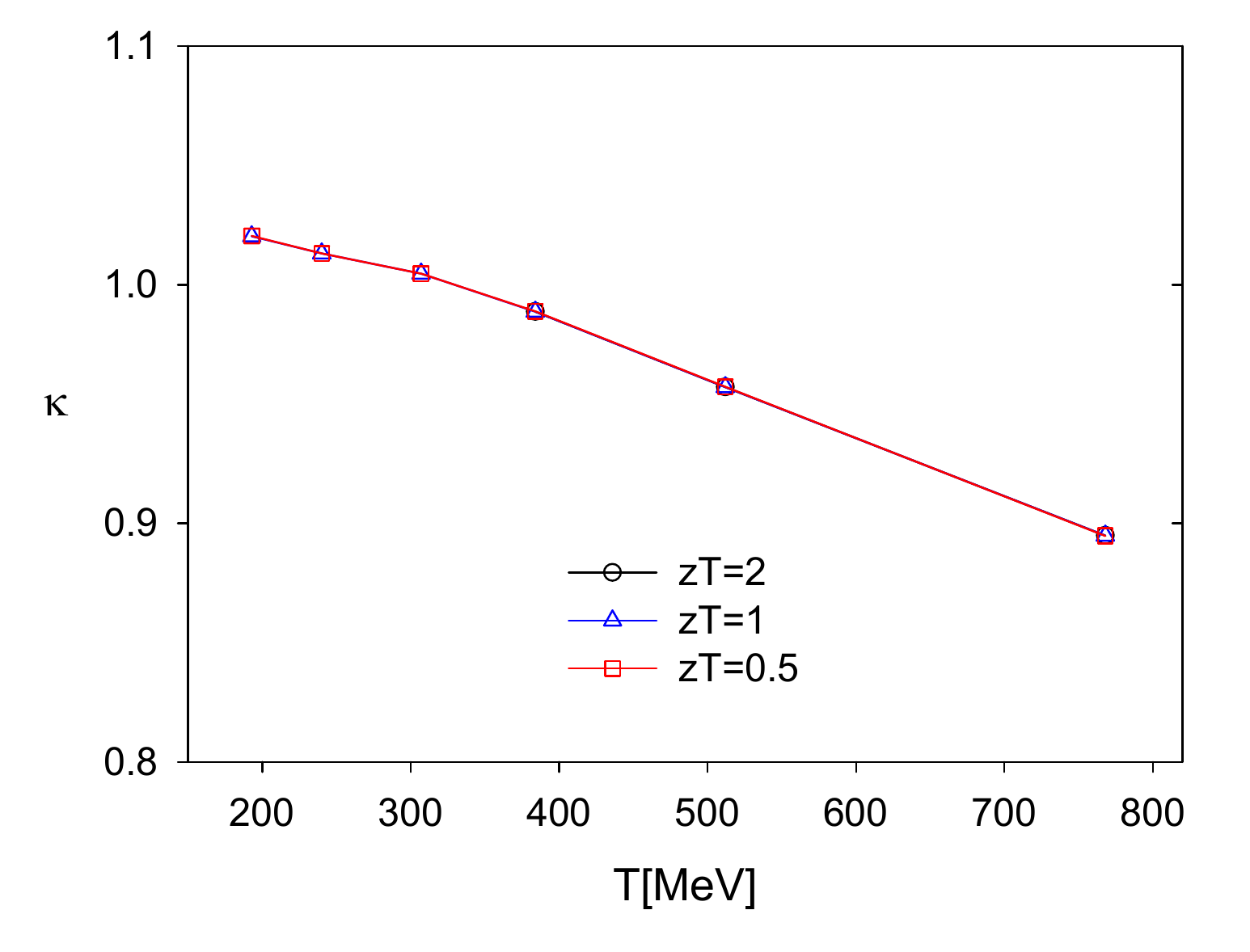}
\end{tabular}
  \label{fig:Kat_K_ud_z_free}
\end{figure}

In Fig. \ref{fig:Kat_K_ud_z_free}, 
the data of $\kappa_{AT}(zT) $ and $\kappa(zT)$ in Fig. \ref{fig:K_ud_z_free} 
of the noninteracting theory are plotted versus the corresponding temperature $T=1/(N_t a) $ 
in $N_f = 2+1+1$ lattice QCD with the same $N_t$, for $zT = n_z/N_t = (0.5, 1, 2)$.
In general, for any fixed $zT$, $|\kappa_{AT}| \lesssim 0.3 $ and $\kappa > 0.89 $ for any $T$. 
Obviously, there does not exist any window satisfying 
the criterion of Eq. (\ref{eq:SU2_CS_crit_z}) with $\epsilon_{CS} < 0.89 $.
Thus, the $SU(2)_{CS}$ symmetry does not emerge in the noninteracting theory on a lattice, 
in contrast to the $N_f=2+1+1$ lattice QCD at the physical point, 
with the emergence of approximate $SU(2)_{CS}$ symmetry in the windows, 
as tabulated in Table \ref{tab:Nf2p1p1_ud_z}.  
This implies that $u$ and $d$ quarks at these temperatures 
must be dynamically very different from the free or quasifree fermions.  
If the deconfined quarks in high-temperature QCD behave like free or quasifree quarks, then  
the $u$ and $d$ quarks in $N_f=2+1+1$ lattice QCD at the temperatures 
with approximate emergent $SU(2)_{CS}$ symmetry 
are likely to be confined inside hadron-like objects,  
which are predominantly bound by the chromoelectric interactions into color singlets.
Moreover, since $SU(2)_{CS}$ is a rather approximate emergent symmetry,           
the role of chromomagnetic interactions in forming these hadron-like objects cannot be neglected.

\subsection{Comparison with the $N_f=2$ lattice QCD}

In Ref.~\cite{Rohrhofer:2019qwq}, 
the symmetries of $z$ correlators of $\bar u \Gamma d$
were studied in $N_f=2$ lattice QCD with M\"obius domain-wall fermions,  
using nine ensembles of lattice sizes $ [32^3 \times (12,8,6,4) ] $ 
and lattice spacings $ \left[ a=(0.051, 0.065, 0.075, 0.096, 0.113)~{\rm fm} \right] $,  
covering the temperatures in the range $ \sim 220-960$~MeV. 

Comparing the $z$ correlators of $N_f=2+1+1 $ lattice QCD in 
Fig. \ref{fig:Cz_ud} with those of $N_f=2 $ lattice QCD in Fig. 1 of Ref. \cite{Rohrhofer:2019qwq},  
we see that in both cases, the order of Eq.~(\ref{eq:Cz_ud}) is satisfied.  
Also, the $U_A(1) $ and $SU(2)_L \times SU(2)_R $ chiral symmetries are effectively restored 
for all studied temperatures, in terms of the degeneracies $C_P(z) = C_S(z)$,  
$C_{T_k}(z) = C_{X_k}(z)$, and $ C_{V_k}(z) = C_{A_k}(z)$ for $k=1,2,4 $.   

For the $SU(2)_{CS} $ symmetry, its breaking in $N_f=2+1+1$ lattice QCD is larger than 
that in $N_f=2$ lattice QCD at the same temperature $T$. 
This can be seen by comparing the degeneracy  
in the multiplet $M_2 = (V_1, A_1, T_4, X_4)$ in Fig. \ref{fig:Cz_ud}  
with that in Fig. 1 of Ref. \cite{Rohrhofer:2019qwq}, 
and similarly for the multiplet $M_4 = (V_4, A_4, T_1, X_1)$.
Moreover, this can be seen by comparing the $SU(2)_{CS}$ symmetry-breaking 
and -fading parameters [$\kappa_{AT}(zT)$, $ \kappa(zT)$] 
between $N_f=2+1+1$ and $N_f=2$ lattice QCD.

\begin{figure}[!ht]
  \centering
  \caption{
    The $SU(2)_{CS}$ symmetry-breaking and -fading parameters ($\kappa_{AT}$, $|\kappa|$),   
    at $zT = 2$ for six temperatures $T \sim 260-960$~MeV in $N_f=2$ lattice QCD.
    The data points of $\kappa_{AT} = C_{A_1}/C_{T_4} - 1$ in the left panel are read off from      
    the ratio $ C_{A_1}/C_{T_4} $ shown in Figs. 3 and 4 of Ref.~\cite{Rohrhofer:2019qwq}, 
    while the right panel exactly matches the Fig. 5 of Ref.~\cite{Rohrhofer:2019qwq}.
  }
\begin{tabular}{@{}c@{}c@{}}
  \includegraphics[scale=0.5]{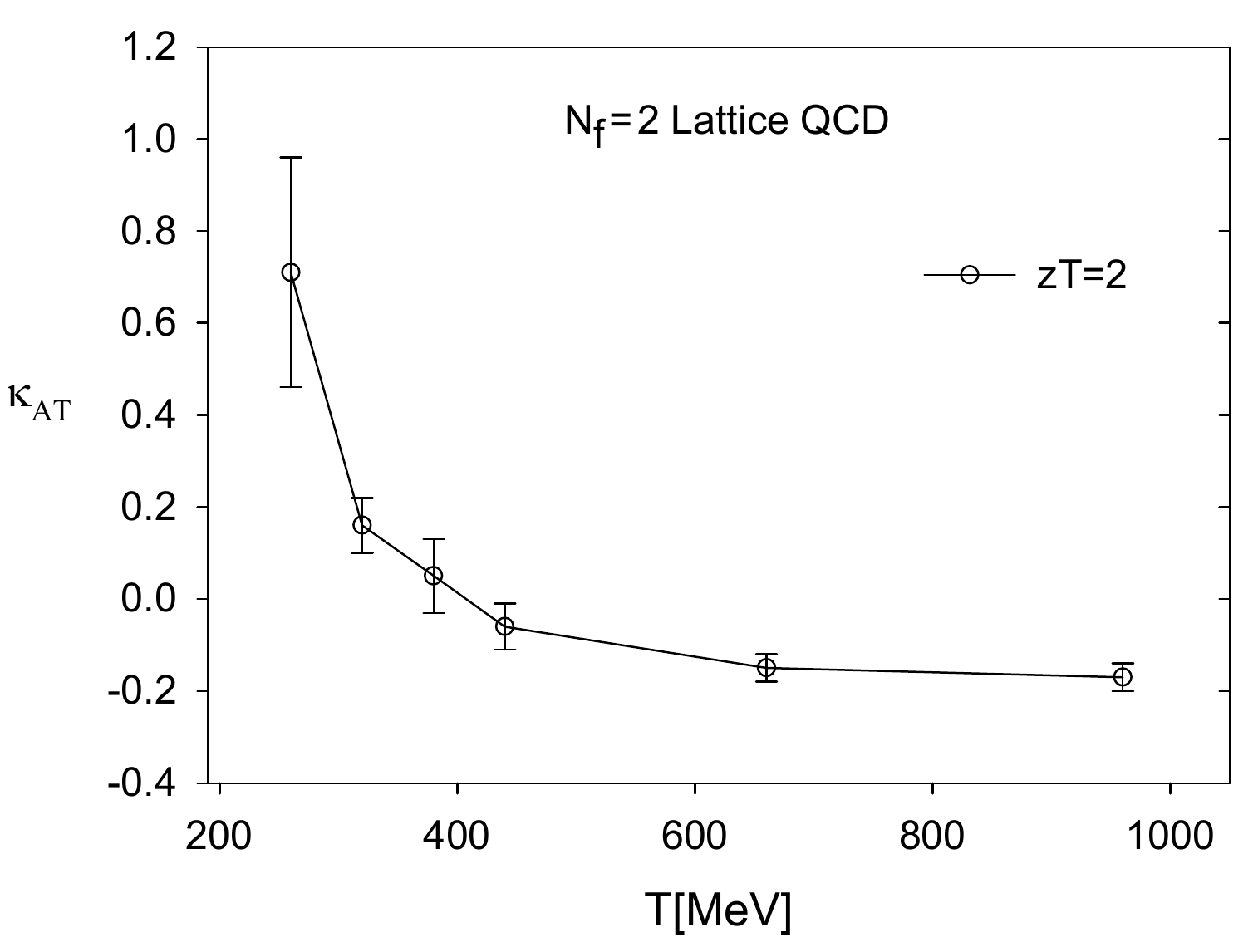}
&
  \raisebox{0.38cm}{\includegraphics[scale=0.6]{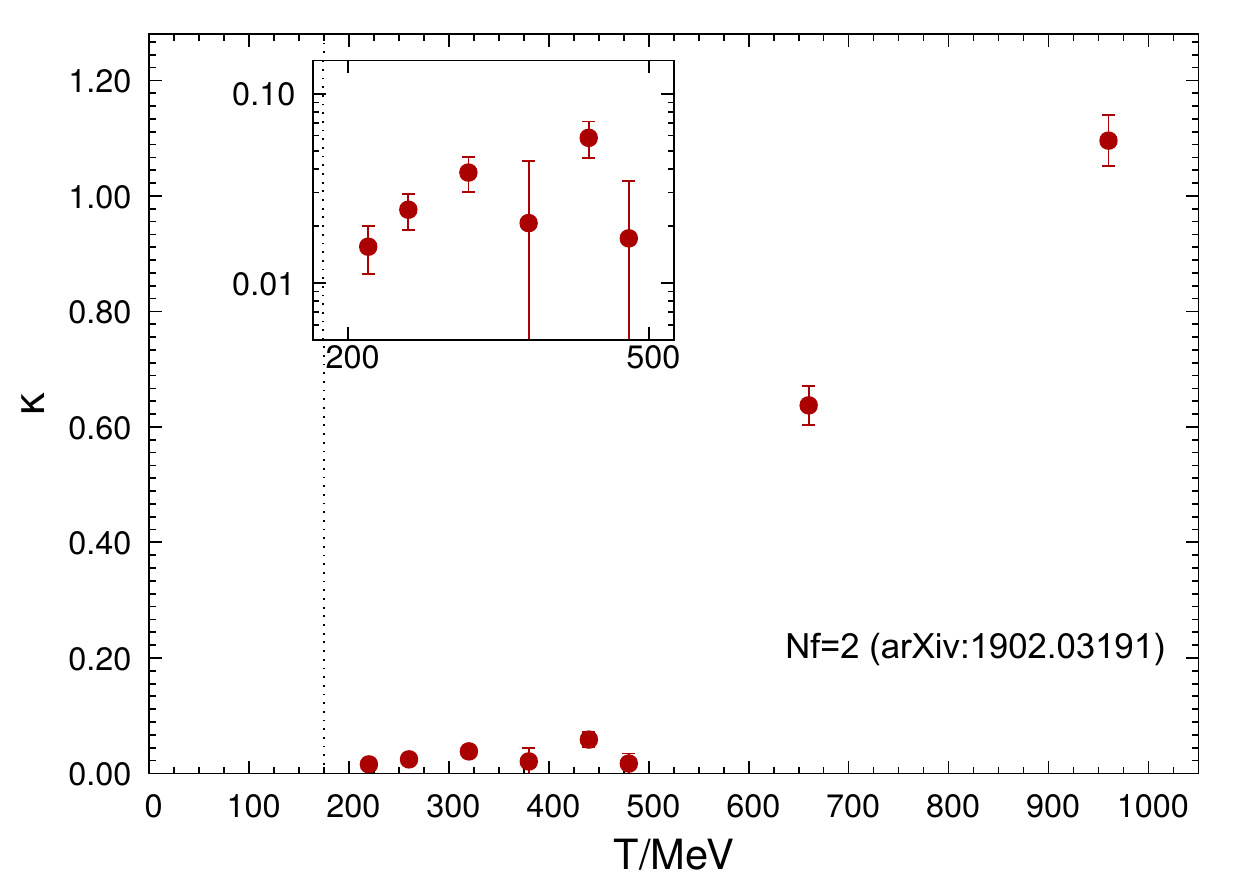}}
\end{tabular}
  \label{fig:Kat_K_Nf2_z}
\end{figure}

Reading off the ratio $C_{A_1}(zT)/C_{T_4}(zT)$ from Figs. 3 and 4 of Ref. \cite{Rohrhofer:2019qwq}, 
the value of $\kappa_{AT}(zT) = C_{A_1}(zT)/C_{T_4}(zT) - 1$ can be obtained 
for $N_f=2$ lattice QCD. At $zT=2$, the values of $\kappa_{AT}$ for six temperatures are plotted 
in the left panel of Fig. \ref{fig:Kat_K_Nf2_z},   
while those of $|\kappa|$ are shown in the right panel of Fig. \ref{fig:Kat_K_Nf2_z},   
which exactly matches the Fig. 5 of Ref. \cite{Rohrhofer:2019qwq}.

\begin{table}[h!]
\begin{center}
\caption{The ranges of temperatures satisfying the criterion in Eq. (\ref{eq:SU2_CS_crit_z}) 
with $\epsilon_{CS} = (0.20, 0.15, 0.10, 0.05, 0.01)$ at $zT=2.0$,  
for $N_f=2$ lattice QCD (Ref. \cite{Rohrhofer:2019qwq}) and $N_f=2+1+1$ lattice QCD (this work).
The third column ($N_f=2+1+1$) is taken from Table \ref{tab:Nf2p1p1_ud_z}, 
where $T_x$ ($> 770$~MeV) and $T_y$ ($> 770$~MeV) have yet to be determined.} 
\setlength{\tabcolsep}{4pt}
\vspace{2mm}
\begin{tabular}{|ccc|}
\hline
$\epsilon_{CS}$ & $N_f=2$ \cite{Rohrhofer:2019qwq} & $ N_f=2+1+1 $ (this work) \\
\hline
\hline
0.20 &  $\sim 320-500$~MeV   &  $\sim 550$~MeV$-T_x$($> 770$~MeV)  \\
0.15 &  $\sim 326-500$~MeV   &  $\sim 660$~MeV$-T_y$($> 770$~MeV)  \\
0.10 &  $\sim 350-500$~MeV   &  NULL               \\
0.05 &  $\sim 380-430$~MeV   &  NULL               \\
0.01 &  NULL                 &  NULL               \\
\hline
\end{tabular}
\label{tab:Nf2_Nf2p1p1_z}
\end{center}
\end{table}

Using the data of $\kappa_{AT}$ and $\kappa$ as shown in Fig. \ref{fig:Kat_K_Nf2_z} 
and the criterion in Eq. (\ref{eq:SU2_CS_crit_z}) for the emergence 
of approximate $SU(2)_{CS}$ symmetry,  
we obtain the ranges of temperatures satisfying Eq. (\ref{eq:SU2_CS_crit_z}) 
for $\epsilon_{CS} = (0.20, 0.15, 0.10, 0.05, 0.01)$,   
as tabulated in the second column of Table \ref{tab:Nf2_Nf2p1p1_z}.
For comparison, the corresponding results of $N_f=2+1+1$ lattice QCD are also
tabulated in the third column, which are taken from the second column 
of Table \ref{tab:Nf2p1p1_ud_z}.  

First, for a given $\epsilon_{CS}$,  
the lower bound of the window in $N_f=2+1+1$ lattice QCD is shifted to 
a higher temperature than that in $N_f=2$ lattice QCD. 
This is mainly due to the fact that the value of $\kappa_{AT}$ in the former 
is larger than that in the latter at the same temperature.
Thus, the former needs to go to a higher temperature in order to attain 
the same value of $\kappa_{AT}$. 

Second, for $N_f=2$ lattice QCD, the window satisfying the criterion (\ref{eq:SU2_CS_crit_z}) 
is shrunk as $ \epsilon_{CS}$ is decreased [i.e., a more precise $SU(2)_{CS}$ symmetry]. 
On the other hand, for $N_f = 2+1+1$ lattice QCD, since the upper bounds 
$T_x (> 770$~MeV) and $T_y (>770$~MeV) have yet to be determined, 
it is unclear whether the window is shrunked as $ \epsilon_{CS}$ is decreased from 0.20 to 0.15.
Since the window is shrunk to zero as $ \epsilon_{CS}$ is decreased from 0.15 to 0.10, 
we speculate that the window is also shrunk as $ \epsilon_{CS}$ is decreased from 0.20 to 0.15.

Third, the window in $N_f=2$ lattice QCD is nonzero even for $\epsilon$ is decreased to 0.05, 
while the window in $N_f=2+1+1$ lattice QCD has been shrunk to zero for $\epsilon_{CS} \le 0.10$.  
Finally, the window in $N_f=2$ lattice QCD is shrunked to zero as $\epsilon_{CS}$ 
is decreased to $ 0.01 $. 

Evidently, the $SU(2)_{CS}$ symmetry in $N_f=2+1+1$ lattice QCD 
is a more approximate emergent symmetry than that in $N_f=2$ lattice QCD.

\section{Conclusions and Outlook}
\label{conclusion}   

In this study, we have generated six gauge ensembles of $N_f=2+1+1$ lattice QCD with $(u/d,s,c)$ optimal 
domain-wall quarks at the physical point, on the $32^3 \times (16, 12, 10, 8, 6, 4)$ lattices 
with two lattice spacings $a \sim (0.064, 0.069)$~fm,  
for six temperatures in the range $\sim 190-770$~MeV, as summarized in Table \ref{tab:6_ensembles}.
The plan is to complete 17 gauge ensembles with 
three lattice spacings $a \sim (0.064, 0.069, 0.075)$~fm, which can be used to extract  
the continuum limit of the observables, for temperatures in the range $\sim 160-770$~MeV.  

Using six gauge ensembles, we computed the temporal and spatial correlators for the complete set of 
Dirac bilinears (scalar, pseudoscalar, vector, axial vector, tensor vector, 
and axial-tensor vector), and each for six combinations of quark flavors  
($\bar u d$, $\bar u s$, $\bar u c$, $\bar s c$, $\bar s s$, and $\bar c c $). 
In this paper, we focus on the meson correlators of $u$ and $d$ quarks,  
while those of other flavor combinations will be analyzed in a forthcoming paper \cite{chiu:2022ab}.
  
We examine the implications of these results for the effective restoration 
of the $SU(2)_L \times SU(2)_R $ and $U(1)_A$ chiral symmetries,  
as well as the emergence of approximate $SU(2)_{CS}$ chiral spin symmetry 
in $N_f=2+1+1$ lattice QCD, 
using the symmetry-breaking parameters $\kappa_{PS}$, $\kappa_{TX}$, $\kappa_{VA}$, 
and $(\kappa_{AT}, \kappa)$ as discussed in Sec. \ref{kappa}. 
The window of temperatures for the emergence of approximate $SU(2)_{CS}$ symmetry is determined 
for temporal and spatial correlators,  
according to the criteria in Eqs. (\ref{eq:SU2_CS_crit_t}) and (\ref{eq:SU2_CS_crit_z}), 
respectively. 
Comparing the windows in Table \ref{tab:Nf2p1p1_ud_t} (of the temporal correlators) 
with those in Table \ref{tab:Nf2p1p1_ud_z} (of the spatial correlators),  
we see that the former are nonzero for $\epsilon_{CS} $ down to 0.05 (at $tT=0.5$),  
while the later are shrunked to zero for $\epsilon_{CS} \le 0.10$ (at any $zT$). 
Theoretically, the temporal and spatial correlators have very different physical contents, 
e.g., the former are related to the thermal masses of the melting mesons, 
while the latter to the screening masses. Thus it is not surprising to see 
that the approximate $SU(2)$ symmetry emerges differently in these two sets of correlators.   

Comparing $N_f=2+1+1$ lattice QCD (in this work) 
with $N_f=2$ lattice QCD in Refs. \cite{Rohrhofer:2019qwq,Rohrhofer:2019qal}, 
we see that in both cases, the $U_A(1) $ and $SU(2)_L \times SU(2)_R $ chiral symmetries 
are effectively restored for all studied temperatures, 
in terms of the degeneracies $C_P(z) = C_S(z)$, $C_{T_k}(z) = C_{X_k}(z)$, 
and $ C_{V_k}(z) = C_{A_k}(z)$, for both spatial and time correlators.   
However, for the approximate $SU(2)_{CS}$ symmetry, 
it emerges differently in $N_f=2+1+1$ and $N_f=2$ lattice QCD, 
as shown in Fig. \ref{fig:K_compare_t} for the symmetry-fading parameter $\kappa$ 
of the temporal correlators, 
and by comparing Fig. \ref{fig:Kat_K_ud_z} with Fig. \ref{fig:Kat_K_Nf2_z} 
for the $SU(2)_{CS}$ symmetry-breaking and -fading parameters $(\kappa_{AT}, \kappa)$ 
of the spatial correlators.   
In general, the $SU(2)_{CS} $ symmetry breaking in $N_f=2+1+1$ lattice QCD is larger than 
that in $N_f=2$ lattice QCD at the same temperature $T$, for both spatial and temporal correlators. 
Comparing the windows for the emergence of approximate $SU(2)_{CS}$ symmetry   
as tabulated in Table \ref{tab:Nf2_Nf2p1p1_z} for $zT=2.0$, 
we see that the window of $N_f=2+1+1$ lattice QCD is shrunked to zero for $\epsilon_{CS} \le 0.10 $, 
while that of $N_f=2$ lattice QCD is nonzero as $\epsilon_{CS} $ is decreased to 0.05, 
then finally it is shrunked to zero for $\epsilon_{CS} \le 0.01 $. 

Since both $N_f=2$ and $N_f=2+1+1$ lattice results have not been extrapolated 
to the continuum, there are discrepancies due to the discretization uncertainties. 
Moreover, even in the continuum limit, there are 
discrepancies between $N_f=2+1+1$ and $N_f=2$ QCD 
due to the quantum fluctuations of heavy $c$ and $s$ quarks, which are present in the former  
but absent in the latter. This can be seen explicitly 
from the quantum expectation value of the meson correlation function of $u$ and $d$ quarks 
in $N_f=2+1+1$ lattice QCD with exact chiral symmetry,
\small
\bea
\label{eq:C_Nf2p1p1}
C_\Gamma(t,\vec{x}) 
&=& \frac{1}{Z} 
\int [dU] e^{-A_g(U)} 
\hspace{-2mm} \prod_{f=u,d,s,c} \hspace{-2mm} \det\left[ (D_c + m_f) (\Id + r D_c)^{-1} \right] \ 
\tr\left[ \Gamma (D_c + m_u)^{-1}_{x,0} \Gamma (D_c + m_d)^{-1}_{0,x} \right]; \nn 
Z &=& \int [dU] e^{-A_g(U)} 
\hspace{-2mm} \prod_{f=u,d,s,c} \hspace{-2mm} \det\left[ (D_c + m_f) (\Id + r D_c)^{-1} \right],  
\eea 
\normalsize
where $A_g(U)$ is the gauge action at temperature $T=1/(N_t a)$, 
$D_c$ is the chirally symmetric Dirac operator \cite{Chiu:1998gp}, 
and $(D_c + m_f)^{-1}$ is the valence quark propagator \cite{Chiu:1998eu}.
Moreover, the explicit breakings of $U(1)_A$, $SU(2)_L \times SU(2)_R$ and $SU(2)_{CS} $ symmetries
due to the quark masses of $s$ and $c$ heavy quarks are much larger than those of $u$ and $d$ light 
quarks. The former enters Eq. (\ref{eq:C_Nf2p1p1}) only through the quark determinants, 
while the latter also enters the meson correlator of each configuration 
through the $u/d$ quark propagators.

In physical reality, it is necessary to incorporate the $b$-quark determinant 
in (\ref{eq:C_Nf2p1p1}), i.e., to perform HMC simulations of $N_f=2+1+1+1$ lattice QCD 
with $(u/d, s, c, b)$ quarks \cite{Chiu:2020tml}. 
This gives more diverse quantum fluctuations than those in (\ref{eq:C_Nf2p1p1}).
Moreover, since the $b$ quark is much heavier than $(u,d,s,c)$ quarks, 
its explicit breakings of $U(1)_A$, $SU(2)_L \times SU(2)_R$ and $SU(2)_{CS} $ symmetries 
must be much larger than those due to $(u,d,s,c)$ quarks. 
Consequently, the effective restoration of $U_A(1) $ and $SU(2)_L \times SU(2)_R $ chiral symmetries 
in $N_f = 2+1+1+1$ lattice QCD would occur at different temperatures from those in 
$N_f=2+1+1$ lattice QCD.  Moreover, for the emergence of approximate $SU(2)_{CS}$ symmetry  
with a fixed $\epsilon_{CS}$ in the criteria of 
Eqs. (\ref{eq:SU2_CS_crit_z}) or (\ref{eq:SU2_CS_crit_t}), 
the lower bound of the window in $N_f=2+1+1+1$ lattice QCD
is likely to occur at a higher temperature than that in $N_f=2+1+1$ lattice QCD. 
Also, as $\epsilon_{CS}$ is decreased, the window of $N_f=2+1+1+1$ lattice QCD 
would have been shrunked to zero while the window of $N_f=2+1+1$ lattice QCD is still nonzero. 
The above speculations are based on the scenario of going from $N_f=2$ to $N_f=2+1+1$ lattice QCD 
as shown in Table \ref{tab:Nf2_Nf2p1p1_z}. 
Our worry is that the $SU(2)_{CS}$ symmetry might not emerge in 
lattice QCD with physical $(u, d, s, c, b)$ quarks, say, for $\epsilon_{CS} < 0.5 $ 
in the criteria of Eqs. (\ref{eq:SU2_CS_crit_z}) and (\ref{eq:SU2_CS_crit_t}).   

Comparing $N_f = 2+1+1$ lattice QCD at the physical point with 
the noninteracting theory on the lattice, 
we see that $u$ and $d$ quarks behave dynamically very differently 
from the free (and quasifree) fermions, 
since the $SU(2)_{CS}$ symmetry does not emerge in the latter, 
in contrast to the former with the approximate emergent $SU(2)_{CS}$ symmetry 
in the windows as tabulated in Tables \ref{tab:Nf2p1p1_ud_t} and \ref{tab:Nf2p1p1_ud_z}.  
If the deconfined quarks in high-temperature QCD behave like free or quasifree fermions, 
then the $u$ and $d$ quarks in $N_f = 2+1+1$ lattice QCD at the temperatures with 
approximate emergent $SU(2)_{CS}$ symmetry are likely to be confined inside hadron-like objects, 
which are predominantly bound by the chromoelectric interactions into color singlets. 
Nevertheless, the role of chromomagnetic interactions in forming these hadron-like objects 
cannot be neglected, since the emergent $SU(2)_{CS}$ symmetry is not an exact symmetry. 
It is interesting to find out the relationship between the degree of dominance of 
the chromoelectric interactions in these hadron-like objects and 
the $ \epsilon_{CS} $ in the criteria of 
Eqs. (\ref{eq:SU2_CS_crit_t}) and (\ref{eq:SU2_CS_crit_z}). 

To clarify the nature of these meson-like objects, it is necessary 
to examine the spectral functions of the $J=1$ mesons (i.e., $V_k$, $A_k$, $T_k$, and $X_k$)     
which are relevant to the $SU(2)_{CS}$ symmetry. 
If bound-state peaks exist in the spectral functions of the $J=1$ mesons,   
in the window $(T_{cs}, T_f)$ of the emergence of approximate $SU(2)_{CS}$ symmetry,    
and also the widths of these peaks gradually broaden,  
and the peaks eventually disappear as $ T \to T_f$, similar to what has been observed 
in the spectral function of the $J=0$ mesons $(P,S)$ 
for $N_f = 2$ lattice QCD \cite{Lowdon:2022xcl}, then the degrees of freedom in the $J=1$ mesons  
can be asserted to be color-singlet (melting) mesons rather than deconfined quarks and gluons. 
To this end, it is necessary to generalize the approach of Refs. \cite{Bros:1992ey,Bros:2001zs}
for $J=0$ mesons to $J=1$ mesons. Also, the spatial correlators of 
$J=1$ mesons are required to be evaluated to high precision even at large distances,  
without the contamination of unphysical meson states,  
such that the damping factor $D_{m,\beta}(\vec{u})$ \cite{Bros:2001zs}
of each $J=1$ meson channel can be extracted reliably. 
The proposed prescription in Sec. \ref{issues} provides a viable way to attain this 
goal$-$that is, to compute two sets of quark propagators with periodic and antiperiodic
boundary conditions in the $z$ direction, while their boundary conditions in $(x,y,t)$ directions 
are the same [i.e., periodic in the $(x,y)$ directions, and antiperiodic in the $t$ direction].  
Then, each set of quark propagators are used to construct the $z$ correlators independently, 
and finally taking the average of these two spatial $z$ correlators. 
Finally, there is another viable prescription for eliminating 
the contribution of the unphysical meson states, as follows.  
First, the backward ($-\hat{z}$) running quark propagator is eliminated for each configuration 
by averaging two quark propagators with periodic and antiperiodic boundary conditions 
in the $z$-direction. Then, the resulting quark propagator is used for constructing 
the $z$ correlators of this configuration.
Consequently, the $z$ correlators are free of backward-propagating meson states
as well as the unphysical meson states, and they behave like $ \sim e^{-M z} $ rather than 
$ \sim \cosh [ M( L_z/2  - z)] $. The advantage of the new prescription is that 
the effective mass $M^{\rm eff}_\Gamma(z) = \ln [ C_\Gamma(n_z)/C_\Gamma(n_z + 1) ] $ 
has a longer plateau than that of the proposed prescription in Sec. \ref{issues}, 
which is essential for the determination of screening mass reliably. 
Once two sets of quark propagators with periodic and antiperiodic
boundary conditions in the $z$ direction are computed, then the $z$ correlators of these 
two prescriptions can be constructed respectively.

\vfill
\section*{Acknowledgements}

The author is grateful to Academia Sinica Grid Computing Centre 
and National Center for High Performance Computing for the computer time and facilities.
This work is supported by the National Science and Technology Council
(Grants No.~108-2112-M-003-005, No.~109-2112-M-003-006, and No.~110-2112-M-003-009), 
and Academia Sinica Grid Computing Centre (Grant No.~AS-CFII-112-103).



\begin{thebibliography}{99}


\bibitem{Detar:1987kae}
C.~E.~DeTar and J.~B.~Kogut,
``The Hadronic Spectrum of the Quark Plasma,''
\href{https://doi.org/doi:10.1103/PhysRevLett.59.399}{\blue Phys. Rev. Lett. \textbf{59}, 399 (1987)};
``Measuring the Hadronic Spectrum of the Quark Plasma,''
\href{https://doi.org/doi:10.1103/PhysRevD.36.2828}{\blue Phys. Rev. D \textbf{36}, 2828 (1987)}

\bibitem{Bazavov:2019www}
A.~Bazavov, S.~Dentinger, H.~T.~Ding, P.~Hegde, O.~Kaczmarek, F.~Karsch, E.~Laermann, A.~Lahiri, S.~Mukherjee and H.~Ohno, \textit{et al.}
``Meson screening masses in (2+1)-flavor QCD,''
\href{https://doi.org/doi:10.1103/PhysRevD.100.094510}{\blue Phys. Rev. D \textbf{100}, no.9, 094510 (2019)}
[arXiv:1908.09552 [hep-lat]].

\bibitem{Glozman:2014mka}
L.~Y.~Glozman,
``SU(4) symmetry of the dynamical QCD string and genesis of hadron spectra,
\href{https://doi.org/doi:10.1140/epja/i2015-15027-x}{\blue Eur. Phys. J. A \textbf{51}, no.3, 27 (2015)}
[arXiv:1407.2798 [hep-ph]].

\bibitem{Glozman:2015qva}
L.~Y.~Glozman and M.~Pak,
``Exploring a new SU(4) symmetry of meson interpolators,''
\href{https://doi.org/doi:10.1103/PhysRevD.92.016001}{\blue Phys. Rev. D \textbf{92}, no.1, 016001 (2015)}
[arXiv:1504.02323 [hep-lat]].

\bibitem{Rohrhofer:2019qwq}
C.~Rohrhofer, Y.~Aoki, G.~Cossu, H.~Fukaya, C.~Gattringer, L.~Y.~Glozman, S.~Hashimoto, C.~B.~Lang and S.~Prelovsek,
``Symmetries of spatial meson correlators in high-temperature QCD,''
\href{https://doi.org/10.1103/PhysRevD.100.014502}{\blue Phys. Rev. D \textbf{100}, no.1, 014502 (2019)}
[arXiv:1902.03191 [hep-lat]].


\bibitem{Rohrhofer:2019qal}
C.~Rohrhofer, Y.~Aoki, L.~Y.~Glozman and S.~Hashimoto,
``Chiral-spin symmetry of the meson spectral function above $T_c$,''
\href{https://doi.org/10.1016/j.physletb.2020.135245}{\blue Phys. Lett. B \textbf{802}, 135245 (2020)}
[arXiv:1909.00927 [hep-lat]].

\bibitem{chiu:2022ab}
T.~W.~Chiu, in preparation.


\bibitem{Chiu:2002ir}
  T.~W.~Chiu,
``Optimal domain wall fermions,''
\href{http://doi.org/10.1103/PhysRevLett.90.071601}{\blue Phys.\ Rev.\ Lett.\  {\bf 90}, 071601 (2003)}
  [hep-lat/0209153]; 
``Domain-Wall Fermion with $ R_5 $ Symmetry,''
\href{https://doi.org/10.1016/j.physletb.2015.03.036}{\blue Phys.\ Lett.\ B {\bf 744}, 95 (2015)}
  [arXiv:1503.01750 [hep-lat]].


\bibitem{Chiu:2011bm}
  T.~W.~Chiu, T.~H.~Hsieh, Y.~Y.~Mao [TWQCD Collaboration],
``Pseudoscalar Meson in Two Flavors QCD with the Optimal Domain-Wall Fermion,''
\href{https://doi.org/10.1016/j.physletb.2012.09.067}{\blue Phys.\ Lett.\ B {\bf 717}, 420 (2012)}
  [arXiv:1109.3675 [hep-lat]].

\bibitem{Chen:2014hyy}
  Y.~C.~Chen, T.~W.~Chiu [TWQCD Collaboration],
``Exact Pseudofermion Action for Monte Carlo Simulation of Domain-Wall Fermion,''
\href{https://doi.org/10.1016/j.physletb.2014.09.016}{\blue Phys.\ Lett.\ B {\bf 738}, 55 (2014)}
  [arXiv:1403.1683 [hep-lat]].


\bibitem{Chen:2022fid}
Y.~C.~Chen, T.~W.~Chiu and T.~H.~Hsieh [TWQCD Collaboration], 
``Topological susceptibility in finite temperature QCD with physical (u/d,s,c) domain-wall quarks,''
\href{https://doi.org/10.1103/PhysRevD.106.074501}{\blue Phys. Rev. D \textbf{106}, no.7, 074501 (2022)}
[arXiv:2204.01556 [hep-lat]].


\bibitem{Narayanan:2006rf}
  R.~Narayanan and H.~Neuberger,
``Infinite N phase transitions in continuum Wilson loop operators,''
\href{https://doi.org/10.1088/1126-6708/2006/03/064}{\blue JHEP {\bf 0603}, 064 (2006)}
  [hep-th/0601210].

\bibitem{Luscher:2010iy}
  M.~Luscher,
``Properties and uses of the Wilson flow in lattice QCD,''
\href{https://doi.org/10.1007/JHEP08(2010)071}{\blue JHEP {\bf 1008}, 071 (2010)}; 
 Erratum: [\href{https://doi.org/10.1007/JHEP03(2014)092}{\blue JHEP {\bf 1403}, 092 (2014)}]
  [arXiv:1006.4518 [hep-lat]].


\bibitem{Bazavov:2015yea}
  A.~Bazavov {\it et al.} [MILC Collaboration],
``Gradient flow and scale setting on MILC HISQ ensembles,''
\href{https://doi.org/10.1103/PhysRevD.93.094510}{\blue Phys.\ Rev.\ D {\bf 93}, no. 9, 094510 (2016)}
  [arXiv:1503.02769 [hep-lat]].



\bibitem{Chen:2012jya}
  Y.~C.~Chen, T.~W.~Chiu [TWQCD Collaboration],
``Chiral Symmetry and the Residual Mass in Lattice QCD with the Optimal Domain-Wall Fermion,''
\href{https://doi.org/10.1103/PhysRevD.86.094508}{\blue Phys.\ Rev.\ D {\bf 86}, 094508 (2012)}
  [arXiv:1205.6151 [hep-lat]].


\bibitem{Glozman:2020qvo}
L.~Y.~Glozman and C.~B.~Lang,
``A finite box as a tool to distinguish free quarks from confinement at high temperatures,''
\href{https://doi.org/10.1140/epja/s10050-021-00494-9}{\blue Eur. Phys. J. A \textbf{57}, no.6, 182 (2021)}
[arXiv:2007.10942 [hep-lat]].


\bibitem{Chiu:1998gp}
T.~W.~Chiu and S.~V.~Zenkin,
``On solutions of the Ginsparg-Wilson relation,''
\href{https://doi.org/10.1103/PhysRevD.59.074501}{\blue Phys. Rev. D \textbf{59}, 074501 (1999)}
[arXiv:hep-lat/9806019 [hep-lat]].

\bibitem{Chiu:1998eu}
T.~W.~Chiu,
``GW fermion propagators and chiral condensate,''
\href{https://doi:10.1103/PhysRevD.60.034503}{\blue Phys. Rev. D \textbf{60}, 034503 (1999)}
[arXiv:hep-lat/9810052 [hep-lat]].


\bibitem{Chiu:2020tml}
T.~W.~Chiu,
``Beauty mesons in $N_f$=2+1+1+1 lattice QCD with exact chiral symmetry,''
\href{https://doi.org/10.1103/PhysRevD.102.034510}{\blue Phys. Rev. D \textbf{102}, no.3, 034510 (2020)}
[arXiv:2004.02142 [hep-lat]].


\bibitem{Lowdon:2022xcl}
P.~Lowdon and O.~Philipsen,
``Pion spectral properties above the chiral crossover of QCD,''
\href{https://doi.org/10.1007/JHEP10(2022)161}{\blue JHEP \textbf{10}, 161 (2022)}
[arXiv:2207.14718 [hep-lat]].


\bibitem{Bros:1992ey}
J.~Bros and D.~Buchholz,
``Particles and propagators in relativistic thermo field theory,''
\href{https://doi.org/10.1007/BF01565114}{\blue Z. Phys. C \textbf{55}, 509-514 (1992)}

\bibitem{Bros:2001zs}
J.~Bros and D.~Buchholz,
``Asymptotic dynamics of thermal quantum fields,''
\href{https://doi.org/10.1016/S0550-3213(02)00059-7}{\blue Nucl. Phys. B \textbf{627}, 289-310 (2002)}
[arXiv:hep-ph/0109136 [hep-ph]].




\end{thebibliography}
\end{document}